\documentclass[10pt,numbers,sort&compress]{elsarticle}
\usepackage[a4paper, left=1.in, right=1.in, top=1.2in, bottom=1.2in]{geometry}

\usepackage{bbm}
\usepackage{graphicx}
\usepackage{amsmath}
\usepackage{amssymb}
\usepackage{xcolor}
\usepackage{enumitem} 
\usepackage{subcaption} 
\usepackage[table]{xcolor} 

\usepackage{booktabs}
\usepackage{array}
\usepackage{adjustbox}

\newcolumntype{P}[1]{>{\raggedright\arraybackslash}p{#1}}

\usepackage{hyperref}
\hypersetup{
        colorlinks=true, 
        linkcolor=blue,  
        filecolor=blue,  
        urlcolor=blue,   
        citecolor=blue   
    }
\usepackage{url}
\usepackage{cleveref}
\usepackage{wrapfig}
\usepackage{pythonhighlight}
\usepackage{enumitem}

\crefname{appendix}{}{}

\journal{Fusion Engineering and Design}

\begin{document}

\begin{frontmatter}

\title{The Data Fusion Labeler (dFL): Challenges and Solutions to Data Harmonization, Labeling, and Provenance in Fusion Energy}

\author[1,4]{Craig Michoski\corref{cor1}}
\author[1]{Matthew Waller}
\author[2]{Brian Sammuli}
\author[2]{Zeyu Li}
\author[1]{Tapan Ganatma Nakkina}
\author[2]{Raffi Nazikian}
\author[2]{Sterling Smith}
\author[2]{David Orozco}
\author[1]{Dongyang Kuang}
\author[3]{Martin Foltin}
\author[2]{Erik Olofsson}
\author[1]{Mike Fredrickson}
\author[1]{Jerry Louis-Jeune}
\author[1,4]{David R. Hatch}
\author[1,4]{Todd A. Oliver}
\author[2]{Mitchell Clark}
\author[1]{Steph-Yves Louis}
\cortext[cor1]{Corresponding author. Email: michoski@sophel.io}
\address[1]{Sophelio, Austin, TX USA}
\address[2]{General Atomics, San Diego, CA, USA}
\address[3]{Hewlett Packard Enterprise, Palo Alto, CA, USA}
\address[4]{University of Texas, Austin, TX, USA}

\begin{abstract}
Fusion energy research increasingly depends on the ability to integrate heterogeneous, multimodal datasets from high-resolution diagnostics, control systems, and multiscale simulations. The sheer volume and complexity of these datasets demand the development of new tools capable of systematically harmonizing and extracting knowledge across diverse modalities. The  \texorpdfstring{\textbf{Data Fusion Labeler}}{Data Fusion Labeler} (dFL) is introduced as a unified workflow instrument that performs uncertainty-aware data harmonization, schema-compliant data fusion, and provenance-rich manual and automated labeling at scale. By embedding alignment, normalization, and labeling within a reproducible, operator-order-aware framework, dFL reduces time-to-analysis by greater than 50X (e.g., enabling $>$200 shots/hour to be consistently labeled rather than a handful per day), enhances label (and subsequently training) quality, and enables cross-device comparability. Case studies from DIII-D demonstrate its application to automated ELM detection and confinement regime classification, illustrating its potential as a core component of data-driven discovery, model validation, and real-time control in future burning plasma devices.
\end{abstract}


\end{frontmatter}


\section{Introduction}

\begin{figure}
    \centering
    \includegraphics[width=0.95\linewidth]{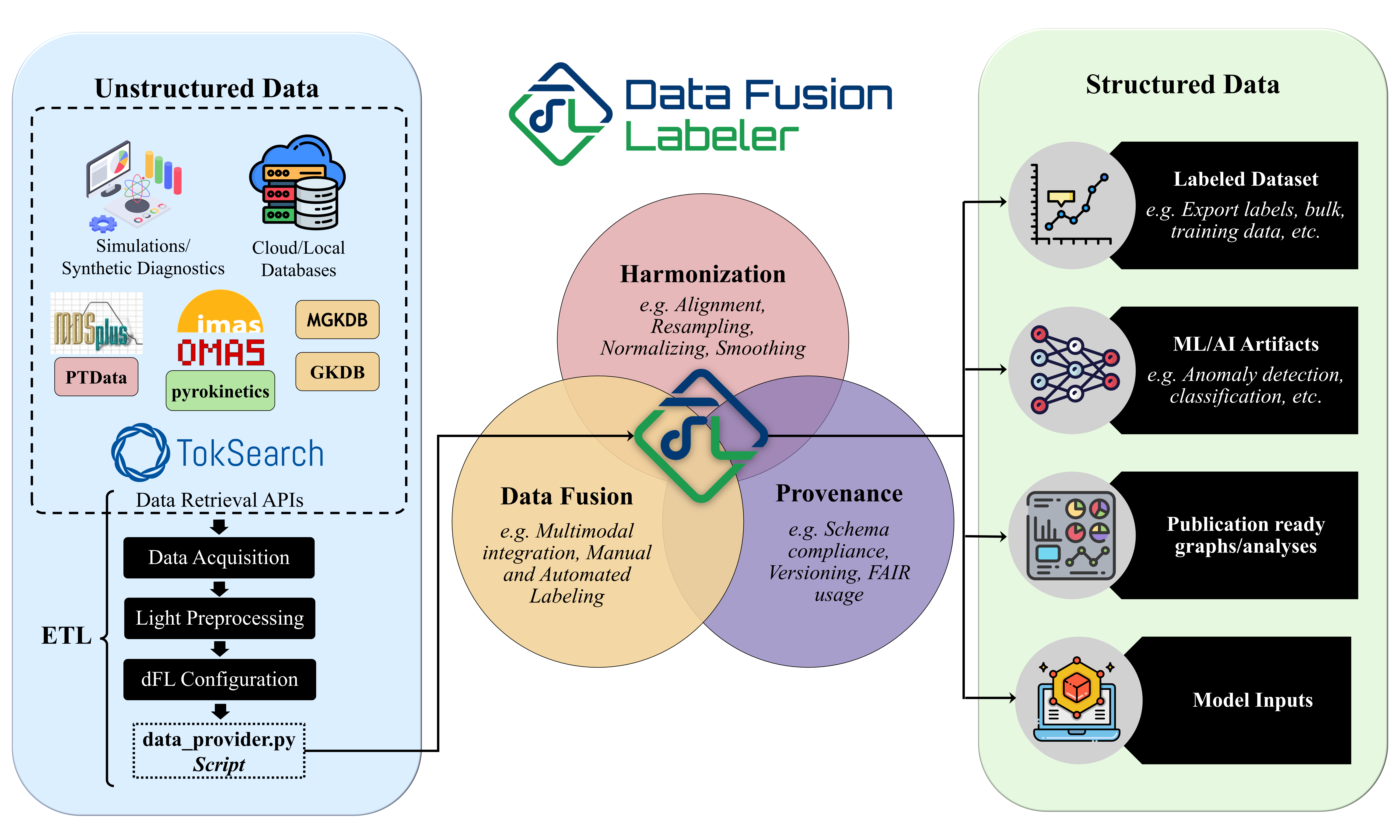}
    \caption{Overview of the Data Fusion Labeler (dFL) workflow.}
    \label{fig:dFL}
\end{figure}

Fusion energy devices and their high-fidelity simulations now produce petabyte-scale data streams spanning up to nine orders of magnitude in time (from microseconds to tens of seconds) and covering spatial scales from the ion gyroradius ($\rho_i \sim 1\ \mathrm{mm} - 1\ \mathrm{cm}$) to the full device ($R \sim 1 - 10\ \mathrm{m}$). These streams arise from heterogeneous sources, such as kilohertz profile diagnostics (Thomson scattering, charge-exchange recombination spectroscopy), megahertz magnetic coils and fast cameras in the edge/SOL \citep{zweben2017invited}, neutron and bolometry systems, real-time control telemetry, and synthetic diagnostics embedded in extended MHD and gyrokinetic codes \citep{jenko2000gene,Kotschenreuther1995}. Managing and interpreting such multimodal data motivates dedicated tools like the \textbf{Data Fusion Labeler} (dFL) (viz. Figure \ref{fig:dFL}), which provides harmonization, fusion, and provenance tracking across this diverse landscape.

While extreme in scale, these challenges are not unique to fusion. Comparable issues arise in domains such as brain--computer interfaces (BCI), manufacturing, robotics, biomedical data, finance, weather forecasting,  marketing analytics, etc., where multimodal, asynchronous, and noisy time-series data must be aligned, fused, and interpreted under uncertainty. In all such contexts, the central barriers include the following:  

\begin{itemize}
    \item \textbf{Multimodality and heterogeneity:} diverse data sources often operate at different sampling rates, in different units, and with distinct metadata conventions, making cross-signal comparisons non-trivial. This is common in domains ranging from multimodal medical imaging to financial time series, and in fusion research arises from disparate diagnostics such as magnetics, Thomson scattering, and neutron detection \citep{strait2008chapter,ponce2010thomson,bertalot2012fusion,petty2008sizing,dumont2021magnetic,wehner2016predictive}.  
    \item \textbf{Imbalanced and noisy datasets:} rare but high-impact events---whether seizures in EEG, extreme weather events, fraud in transaction streams, or plasma disruptions in fusion devices \citep{vega2022disruption,buttery2000neoclassical,ham2020filamentary}---occur against a backdrop of far more common stable conditions. Noise and uncertainty from calibration errors, environmental interference, or diagnostic degradation further complicate inference \citep{michoski2024gaussian,liu2024plasma}.  
    \item \textbf{Sparse and incomplete labeling:} expert-generated annotations are costly and often incomplete, forcing workflows into semi- or weakly-supervised regimes. This holds across many fields---clinical medicine, finance, or plasma physics---where the volume of unlabeled data vastly exceeds curated subsets \citep{anirudh20232022,pavone2023machine,hatfield2021data,wang2023scientific}.  
    \item \textbf{Dynamic and nonstationary behavior:} the underlying processes evolve over time, often with changing statistics and regime shifts. Examples include market volatility, neural adaptation, climate variability, or plasma instabilities and control interventions in fusion \citep{ryzhkov2023magneto,conway2021geodesic,hoi2021online,gheibi2021applying}.  
    \item \textbf{Integration of multiscale simulations and models:} many domains must reconcile simulations or models that span different spatial or temporal scales and mathematical representations, whether in weather forecasting, structural engineering, or multiscale plasma simulations from MHD to gyrokinetics \citep{churchill2020machine,hatch2022reduced}.  
\end{itemize}

In fusion energy these barriers are often compounded  by temporal imbalances (megahertz magnetics versus kilohertz profiles), missing data \citep{osman2018survey}, and the need for principled handling of imbalanced classes without distorting statistical calibration in hazard-function and survival modeling \citep{olofsson2018event,Olofsson_2025}. Collectively, they mirror the ``long tail'' of problems in other data-rich sciences: sparse critical events, heterogeneous modalities, dynamic systems, and costly human labels.

By contrast, the inherent alignment of photographic, video, and textual data has allowed those modalities to flourish within diffusion and transformer architectures, accelerating progress across computer vision and natural language processing. The time-series modalities that dominate fusion and other physical sciences lack this natural alignment, and thus have not yet fully benefited from comparable advances. The \textbf{Data Fusion Labeler} (dFL) (outlined in Figure \ref{fig:dFL}) is designed to close this gap, bringing the harmonization, fusion, and structured labeling required to elevate time-series data to the same precipice of advancement now transforming other fields.

Converting these diverse, asynchronous, and noise-burdened signals into plasma physics insight, reliable control signals, and design constraints hinges on three interlocking pillars:  

\begin{enumerate}[label=\textbf{\Roman*.}]
    \item \textbf{Data harmonization} \citep{Cheng2024}: systematic curation, standardization, alignment, and uncertainty-aware representation that render disparate diagnostics and model outputs \emph{commensurable} in units, coordinates, sampling, and metadata;  
    \item \textbf{Data fusion} \citep{Castanedo2013}: the mathematically principled combination of harmonized sources to produce estimates and decisions with lower bias, tighter credible intervals, and improved robustness relative to any constituent alone; and  
    \item \textbf{Data provenance} \citep{Herschel2017}: complete, machine-actionable records of origin, processing history, and versioning that enable reproducibility, auditability, and FAIR \citep{Wilkinson2016} reuse across devices and campaigns \citep{Wilkinson2016,Peng2011}.  
\end{enumerate}

In magnetic confinement experiments, harmonization can be exemplified by the sub-millisecond synchronization of edge fast-camera movies with Mirnov coil signals, a prerequisite for uncovering causal relationships between visible turbulence structures and magnetic perturbations. Data fusion can be illustrated by combining Thomson scattering and charge-exchange recombination spectroscopy, whose complementary sensitivities to electrons and ions enable construction of self-consistent pressure and rotation profiles unattainable from either diagnostic alone. Data provenance is captured by preserving the exact EFIT version, coil set, and preprocessing filters used in equilibrium reconstruction, since even minor differences in these choices can lead to divergent $q$-profiles and stability assessments.  

In data engineering terms, harmonization and data fusion extend the classical ETL paradigm \citep{Souibgui2019}: harmonization can be viewed as a physics-aware specialization of ``Transform,'' ensuring cross-diagnostic commensurability, while data fusion operates downstream of ETL, synthesizing harmonized inputs into higher-level constructs with reduced uncertainty and enhanced robustness. This perspective is consistent with the Joint Directors of Laboratories (JDL) hierarchy, which treats source preprocessing and alignment as ``Level 0'' \citep{White1991DataFusionLexicon}---a \emph{prerequisite} for higher-level estimation, identification, and decision-making \citep{Hall1997}.  

In magnetic confinement fusion, this principle is operationalized by the ITER-driven \emph{Integrated Modelling \& Analysis Suite (IMAS)}, which provides a machine-independent schema for experimental and synthetic data, enabling toolchains to interoperate without bespoke converters \citep{Pinches2020}. Workflow environments such as \textsc{OMFIT} then leverage IMAS/IDS structures to compose equilibrium reconstruction, transport solvers, stability analysis, and turbulence models in a single, provenance-aware graph \citep{Meneghini2015}.  

Harmonization stands as the foundational step before any attempt at learning or control. Without it, downstream analyses risk being dominated by idiosyncrasies---unit drift, timestamp skew, schema mismatches, signal polarity inversions, undocumented calibration changes, diagnostic preprocessing artifacts, or missing metadata---rather than by the underlying physics. The study of edge turbulence and pedestal dynamics, for example, demands sub-millisecond temporal registration between fast imaging and magnetic probes to uncover true causal structure \citep{zweben2017invited}. Similarly, the integration of gyrokinetic surrogates such as TGLF or GENE outputs with profile diagnostics requires consistent geometry, coordinate conventions, and uncertainty representations \citep{Staebler2007,jenko2000gene}. Harmonization is therefore not a superficial preprocessing task; it is an \emph{information-preserving filter} that renders multi-diagnostic inference and control both tractable and defensible.

With commensurability established, \emph{data fusion} methods ranging from classical Bayesian estimators, to modern representation-learning techniques, to sophisticated automated labeling and feature extraction can be brought to bear. By exploiting complementary sensitivities across diagnostics, these methods reduce variance, resolve ambiguities, enhance robustness to noise and diagnostic failure, and improve reliability under non-ideal conditions. In disruption prediction, for instance, the fusion of magnetic, radiative, and density signals has outperformed single-channel predictors, enabling generalization across a range of operational scenarios \citep{KatesHarbeck2019}. Similar benefits are observed in event detection tasks such as identifying ELMs, sawteeth, or ITB formation, where combining profile evolution, fluctuation spectra, and actuator traces leads to sharper recall and precision.

Underlying both data harmonization and data fusion is the necessity of rigorous data provenance. The reproducibility of scientific claims and the portability of trained models depend on transparent lineage: knowing what data were used, how they were transformed, by whom, with which code and parameters, and against which schema version. The FAIR principles codify these requirements \citep{Wilkinson2016}, and the iterative nature of campaign analysis demands that labels, features, and models be versioned alongside the raw signals. Without such practices, reported performance cannot be independently verified, nor can models be deployed with confidence as devices and diagnostics evolve \citep{Peng2011}.  

In this work, we introduce the \textbf{Data Fusion Labeler} (dFL) as a unifying workflow instrument that sits at the interface between data harmonization, data fusion, and data provenance. dFL (\emph{i}) ingests multi-diagnostic and multimodal experimental and synthetic data in (optionally) IMAS/OMAS-compatible layouts \citep{Pinches2020,Meneghini2015,xu2025summary} processed through uncertainty aware alignment; (\emph{ii}) performs filling, resampling, normalization, smoothing, and visualization at multiple resolutions; (\emph{iii}) supports manual, statistical, physics-informed, simulation-driven, and classifier-based \emph{labeling and autolabeling} to convert raw streams into structured event corpora; and (\emph{iv}) captures complete, machine-readable provenance for every label and export, enabling exact replay and cross-facility reuse. In practice, dFL reduces time-to-analysis from weeks to hours, turns ad-hoc scripts into documented transformations, and converts fragile, person-specific workflows into durable, shareable artifacts. By standardizing label semantics and embedding them in harmonized, provenance-rich contexts, dFL directly addresses the data scarcity and class-imbalance challenges that pervade disruption prediction, confinement-regime identification, and MHD mode classification \citep{deVries2011,KatesHarbeck2019}.  

This paper formalizes the architectural requirements for label-centric fusion workflows and presents the design of dFL to meet them. We (1) articulate harmonization constraints induced by fusion physics and diagnostic practice; (2) detail a provenance model aligned with FAIR principles; (3) describe manual and automated labeling modalities, including statistics-based change detection, physics-informed rules, simulation-driven autolabeling, and supervised classifiers; and (4) demonstrate that rigorous harmonization and provenance multiply the physics value extracted by downstream analysis, modeling, and control. Throughout, we ground the discussion in fusion-specific literature and tools (IMAS/OMFIT, turbulence and transport models, high-speed edge diagnostics) \citep{Pinches2020,Meneghini2015,jenko2000gene,Staebler2007,zweben2017invited}, while also highlighting analogies to other high-stakes data-rich fields (BCI, finance, weather), emphasizing that the challenges and solutions embodied in dFL are broadly relevant.  

\section{Challenges of Data Fusion in Fusion Energy Data}\label{data}

Fusion energy research faces an especially demanding set of challenges when it comes to the fusion of heterogeneous data sources. Tokamaks and stellarators rely on a diverse set of diagnostics, operational systems, and simulation frameworks, producing massive multimodal datasets that must be harmonized, fused, and labeled under uncertainty. While many of these issues---such as multimodality, imbalance, noise, sparse labeling, and nonstationarity---are also encountered in other data-intensive fields (e.g., medicine, finance, or weather forecasting), they appear in fusion at extreme scale and with unique physical constraints. Addressing them is not optional: they determine the reliability of physics inference, the safety of machine operation, and the reproducibility of results across campaigns and devices.

\subsection{Multimodal, Heterogeneous, Imbalanced, Noisy Datasets}

In fusion energy research, particularly in the operation and diagnostics of tokamaks and other fusion plasma devices, data from various sources are collected to monitor the system's behavior and ensure safe and efficient operations. These datasets are inherently multimodal, arising from the combination of diagnostic instruments such as magnetic coils \citep{strait2008chapter}, Thomson scattering \citep{ponce2010thomson}, and neutron detectors \citep{bertalot2012fusion}, operational data like plasma parameters \citep{petty2008sizing} and external heating systems \citep{dumont2021magnetic}, as well as simulation and model data in areas like model predictive control (MPC) \citep{wehner2016predictive}. These data sources often differ in their resolution, format, and sampling frequency, leading to a heterogeneous dataset that must be reconciled for comprehensive analysis, and especially for ML/AI data-enabled pipelines. Additionally, due to the dynamic and high-energy environment in a tokamak, the data can exhibit high dimensionality, with a wide range of variables describing the plasma state, control systems, and performance metrics. The integration of such diverse data types presents significant challenges for modeling and analysis, requiring sophisticated algorithms capable of handling multiple modalities simultaneously.

Data fusion has emerged as a pivotal strategy to address these challenges by combining and integrating information from diverse sources into a unified, enriched dataset. This process facilitates the resolution of heterogeneity by aligning and harmonizing data formats, resolutions, and sampling frequencies across modalities. For example, data fusion can effectively reconcile high-frequency magnetic sensor data with low-frequency Thomson scattering measurements, enabling synchronized and comprehensive analysis. By leveraging advanced methods, such as multimodal deep learning \citep{wang2024multi} and Bayesian frameworks \citep{oliver2024automated}, researchers can unify disparate data types, enhancing the interpretability of complex plasma behavior and improving the robustness of downstream models.

A particularly pressing, though often under-emphasized issue in fusion energy research is the prevalence of imbalanced data. Many fusion diagnostics and operational datasets exhibit significant disparities in the frequency and distribution of events of interest. For instance, high-disruption scenarios in tokamaks \citep{vega2022disruption}---critical for understanding and preventing damage to plasma-facing components---occur far less frequently than stable operating conditions. Similarly, certain anomalous behaviors, such as neoclassical tearing modes \citep{buttery2000neoclassical} or edge-localized modes \citep{ham2020filamentary}, are relatively rare (or sparsely contained in the diagnostic data stream) but hold immense importance for both scientific understanding and operational safety. This imbalance skews the datasets, resulting in challenges for predictive modeling, as standard machine learning algorithms often become biased toward the majority class, failing to adequately capture rare but critical phenomena. The imbalance also complicates the selection of appropriate evaluation metrics, requiring the use of metrics like precision, recall, and F1-score \citep{grandini2020metrics} to ensure meaningful model assessment.

Data fusion techniques are particularly advantageous in addressing imbalanced datasets. By integrating data from multiple diagnostics, fusion methodologies can amplify the representation of under-represented events through the synthesis of complementary information. For instance, correlating rare disruptions captured in magnetic diagnostics with soft X-ray observations can enhance the detection and characterization of such events, mitigating the effects of imbalance. Additionally, data fusion supports robust feature extraction across modalities, which can improve the sensitivity of models to rare but critical phenomena.

Fusion energy datasets are further compounded by temporal imbalances, where data collected at different timescales contribute unevenly to the overall dataset. For example, while magnetic diagnostics may sample at megahertz frequencies \citep{zweben2017invited}, other diagnostics, such as Thomson scattering systems, operate at much lower frequencies. This discrepancy can lead to an over-representation of certain data modalities in modeling processes, while others are under utilized despite their potential importance.

Additionally, fusion datasets often contain significant noise and/or uncertainties \citep{michoski2024gaussian, liu2024plasma} arising from a variety of sources, including sensor calibration errors, electromagnetic interference, and signal degradation over time. Noise can obscure critical patterns, particularly in diagnostics requiring high sensitivity, such as those detecting soft X-rays or fast ion populations. Distinguishing meaningful signals from noisy data is particularly challenging in imbalanced datasets, as rare events are more likely to be masked by random fluctuations or measurement errors. Missing values---whether due to diagnostic malfunctions, misalignments in data collection, or physical limitations of the instrumentation---further exacerbate these issues, introducing gaps that must be addressed through imputation \citep{osman2018survey} or alternative strategies.

To address these challenges, the fusion data community has increasingly focused on developing robust data handling and analysis pipelines. Techniques such as synthetic oversampling (e.g., SMOTE) \citep{lin2021review} and statistically robust data resampling methodologies \citep{faghihi2020moment} are being explored to mitigate the impact of class imbalance. Advanced data fusion techniques, such as multimodal deep learning \citep{wang2024multi} and Bayesian frameworks \citep{oliver2024automated}, are being employed to reconcile heterogeneous datasets, enabling the extraction of meaningful insights across disparate diagnostic systems. These methods not only resolve heterogeneity but also bolster the ability to manage noisy and imbalanced data effectively. Additionally, methods for denoising data, including wavelet transforms \citep{lu2024neutron,farge2015wavelet} and advanced filtering techniques \citep{anirudh20232022}, are becoming essential for preprocessing, particularly in high-noise environments.  However, it is important to note that not all imbalance should be corrected through oversampling or reweighting. In some cases---particularly those that are explicitly statistical in their workflow---preserving the original distribution is essential for maintaining statistical calibration of model outputs. For example, in hazard function learning, naive rebalancing can distort the interpretation of survival probabilities unless special precautions are taken \citep{olofsson2018event,Olofsson_2025}. More broadly, there are differing views on whether imbalance should always be ``fixed,'' with some arguing that careful metric selection and calibration can be preferable to data-level interventions.

\subsection{Sparse and Incomplete Data Labeling in Fusion Diagnostics}

In addition to the challenges posed by heterogeneous and noisy data, another significant issue in fusion energy research is the sparse and incomplete nature of data labeling \citep{anirudh20232022}. Many fusion diagnostics, especially those employed in tokamaks, operate in real-time and are designed to monitor a variety of plasma parameters over extended periods. However, annotating this data with meaningful labels and metatags---such as classifying plasma stability modes, identifying anomalous events, or categorizing different operational regimes---is often sparse due to the complexity of the plasma behavior and the high cost of manual labeling. In many cases, only a small fraction of the collected data is annotated, leading to the problem of semi-supervised or unsupervised learning, where models must make predictions with limited labeled data. This challenge is compounded by the fact that many labels are incomplete, especially when diagnostics fail or data gaps occur due to technical issues or extreme plasma conditions. To overcome this, advanced techniques in active learning \citep{pavone2023machine}, transfer learning \citep{hatfield2021data}, and weak supervision \citep{wang2023scientific} are required to efficiently handle sparse and incomplete labels while maintaining model accuracy.

\subsection{Dynamic and Nonstationary Nature of Fusion Devices}

Fusion systems, particularly tokamaks, exhibit highly dynamic and nonstationary behavior, further complicating data analysis. The physical processes within a tokamak \citep{ryzhkov2023magneto,conway2021geodesic}, such as plasma instabilities, heating, and magnetic confinement, evolve over time and are influenced by a wide range of external and internal factors. These processes often do not follow fixed statistical distributions or patterns, and the underlying system behavior can change rapidly in response to operational adjustments, perturbations, or environmental influences. As a result, fusion data are often nonstationary, meaning that statistical properties such as mean, variance, and correlations can vary over time. This nonstationarity can lead to difficulties in developing predictive models that generalize well across different operational phases or fusion events. Moreover, the dynamic nature of the system means that models must continuously adapt to changing conditions, requiring the use of online learning techniques \citep{hoi2021online} or adaptive algorithms \citep{gheibi2021applying} capable of handling time-varying data. The challenge of capturing both short-term fluctuations and long-term trends in such dynamic environments necessitates sophisticated modeling techniques, such as recurrent neural networks (RNNs), autoregressive approaches, or other time-series forecasting methods, which are designed to accommodate the evolving characteristics of the system.

\subsection{Multiscale Simulation Data in Fusion Energy}

Another significant challenge in fusion energy research arises from the integration of simulation data generated by multiscale pipelines, which model fusion systems at different spatial and temporal resolutions \citep{churchill2020machine,hatch2022reduced}. These simulation models span a broad hierarchy: large-scale macroscopic descriptions of plasma dynamics, such as magnetohydrodynamic (MHD) models, reduced and extended MHD formulations that capture two-fluid or Hall physics \citep{jardin2010review, charidakos2014action, sovinec2004nimrod}, and kinetic models that resolve particle distribution functions in velocity space to address turbulence, fast-ion dynamics, and wave-particle interactions \citep{bernard2024edge, picls2024pcls}. In addition, multiscale efforts extend beyond core plasma physics to encompass kinetic-material interaction models for plasma-material interfaces, plasma facing components (PFCs), and edge/SOL dynamics \citep{nordlund2014pwimodeling, lasa2024multiscale}, as well as coupled models for subsystems such as fueling, exhaust, and impurity transport \citep{wiesen2017edge, ROBBE2023112229}.

The data produced at each scale often resides in distinct numerical or mathematical representations: structured or unstructured grids for continuum MHD and fluid models, marker distributions or velocity-space discretization for kinetic solvers, and surface or mesh-based data for material response simulations.  Moreover, the geometries employed can vary widely; global core plasma models may impose cylindrical or toroidal symmetry, while localized edge or wall-interaction simulations require realistic three-dimensional geometry with surface roughness or material microstructure resolved.

This variability in representations complicates integration across scales and hinders the derivation of comprehensive insights from the full simulation hierarchy. Bridging these gaps requires the development of robust data fusion techniques that can: (\emph{i}) map information across scales and between disparate numerical representations, (\emph{ii}) reconcile differences in resolution, uncertainty, and accuracy, and (\emph{iii}) incorporate both experimental and simulated data streams to improve predictive capability. Such harmonized workflows are essential not only for theory-experiment comparison, but also for developing reduced models suitable for control and decision-making in real-time tokamak operation.

\medskip


\section{dFL System Overview}\label{sec:dFL_overview}

The \textbf{Data Fusion Labeler} (dFL) provides a modular, extensible framework for ingesting, harmonizing/fusing, and labeling multimodal fusion energy datasets, as shown graphically in Figure \ref{fig:dFL}. Data ingestion is managed through the \texttt{data\_provider} script, which serves as the primary interface between raw archives and the dFL functionalities. This mechanism enables users to construct \emph{custom ingestion pipelines}, including tailored graphing and analysis tools, custom independent variable formats, user-defined normalization schemes, specialized smoothers, customized labels and autolabelers, or domain-specific filters and feature maps. For low-level access, the \texttt{Fetch Data} functionality allows users to override default ingestion and implement arbitrary preprocessing pipelines prior to rendering signals in the dFL interface, ensuring maximal flexibility when integrating novel diagnostics or experimental workflows.  The full \texttt{data coordinator} API, and extensive examples and demos are provided in the \href{http://dfl.sophelio.io/documentation}{dFL documentation}

Beyond extensibility, dFL ships with native support for a full suite of preprocessing operations required for fusion data pipelines, including resampling, normalization, smoothing, trimming, interpolation, etc. These operators are integrated into the graphical interface, ensuring that users can both prototype and deploy harmonization strategies without leaving the dFL environment. All operations can be made fully provenance-tracked, guaranteeing that every transformation---whether native or user-defined---is recorded for reproducibility and downstream audit.

\begin{wraptable}{l}{0.68\textwidth}
\centering
\footnotesize
\renewcommand{\arraystretch}{1.15} 
\setlength{\tabcolsep}{5pt}        
\textbf{Nomenclature Key for DIII-D TokSearch Point Names} \\[6pt] 
\begin{tabular}{|p{1.7cm}|p{6.5cm}|}
\hline
\rowcolor[gray]{0.9}
\textbf{DIII-D Signal Name} & \textbf{Physical Measurement Description} \\
\hline
\texttt{ip} & Plasma current, representing the total toroidal current flowing through the confined plasma (typically in megaamperes, MA). \\
\hline
\texttt{magnetics amplitudes} & Amplitudes of toroidal mode components (\( n \)) of the poloidal magnetic field (\( B_{p} \)), obtained from \texttt{modespec}/\texttt{modespyec} analysis of toroidally distributed magnetic probes; typically expressed in Gauss or mT. \\
\hline
\texttt{pinj} & Total neutral beam injection (NBI) power delivered to the plasma, summed over all active injectors (in megawatts, MW). \\
\hline
\texttt{wnt} & Average of the measured electron density and temperature pedestal width, with the unit in $\psi_N$. \\
\hline
\texttt{tinj} & Neutral beam injection torque, quantifying the total angular momentum input to the plasma from the NBI system (in $\mathrm{N\,m}$). \\
\hline
\texttt{echpwrc} & Electron cyclotron heating (ECH) power coupled to the plasma, as measured by the real-time ECH power controller (in MW). \\
\hline
\texttt{betan} & Normalized plasma beta, $\beta_N = \langle \beta \rangle / (I_p / a B_T)$, denoting the ratio of plasma to magnetic pressure normalized to plasma current and magnetic field. \\
\hline
\texttt{density} & Line-averaged electron density from interferometry diagnostics (in units of $10^{19}\ \mathrm{m^{-3}}$). \\
\hline
\texttt{Pedestal ratio} & Ratio of the measured pedestal width to the conventional scaling law based on KBM estimation \citep{Snyder2011}, serving as an indicator of pedestal structure and stability. \\
\hline
\texttt{fs0*} & Filterscope channels measuring line-integrated photon flux from D$_\alpha$ and impurity emissions near the plasma edge (in $\mathrm{photons\ s^{-1}\ cm^{-2}\ sr^{-1}}$). \\
\hline
\end{tabular}
\label{tab:nomenclature_toksearch}
\vspace{-29pt} 
\end{wraptable}

In practice, this architecture allows dFL to serve as both a turnkey environment and a development platform: users can rely on the included preprocessing library for common workflows, or seamlessly inject custom operators at the ingestion level to meet specialized needs. This dual capability makes dFL well-suited for experimental campaigns, simulation archives, and cross-device data fusion tasks, where flexibility, reproducibility, and performance are equally essential.

\section{Harmonization \& Data Fusion}\label{sec:harmonization}

In fusion energy research, particularly for tokamak experiments and simulations, data harmonization is often undervalued compared to direct physics analysis. Yet it is foundational: without rigorous harmonization, the quality, accuracy, reproducibility, and physical coherence of downstream results are compromised. By enforcing consistency across heterogeneous diagnostics and simulation outputs, harmonization transforms disparate signals into a common, physically meaningful framework suitable for reliable inference and control. Equally, harmonization is the prerequisite for \emph{data fusion}: once signals are co-registered and rendered commensurable, one can combine information across diagnostics in statistically principled ways (e.g., inverse-variance weighting, information-form Kalman updates, and multi-output Gaussian-process co-kriging) to produce estimates with reduced posterior variance relative to any single channel.

Key harmonization tasks include synchronizing signals in time and space, reducing datasets to relevant intervals, addressing gaps or missing values, and applying resampling, smoothing, and normalization to ensure comparability without erasing important physics. The sequencing of these operations is critical, as many are non-commutative and can yield different results if applied in different orders. A principled harmonization workflow preserves essential information, minimizes artifacts, and prepares data for advanced analysis, modeling, and multi-device data fusion. In practice, harmonization defines the likelihood terms and error models that downstream fusion methods rely on, e.g., transforming raw measurements into a common state-space with known noise covariance enables Bayesian fusion \citep{Khaleghi2013Multisensor}.

In fact---as is already well-established in other data-heavy engineering and scientific disciplines---the single most pragmatic lever to accelerate learning, prediction, and control in fusion-energy science resides not in ever more elaborate model architectures, but rather in improving the {\it information substrate} upon which those models operate. When heterogeneous diagnostics and simulation outputs are harmonized and fused into a coherent representation, one effectively drives the underlying state-space manifold to {\it lower informational entropy}---yielding a dataset that is more compact, better aligned with the physics, and less burdened by spurious variation. Symbolically:
\[
\mathrm{Better\ Data} \; \Rightarrow\;\mathrm{Lower\ Entropy\ Manifold}
\;\Rightarrow\;\mathrm{Simpler\ Model}
\;\Rightarrow\;\mathrm{Higher\ Generalizability.}
\]
By reducing the variance and bias introduced by mis-synchronization, inconsistent units, missing provenance, or unlabeled modes of operation---and by enforcing a physically consistent {\it operator ordering} in harmonization pipelines, where trimming, filling, resampling, smoothing, and normalization are applied in a reproducible and non-commutative sequence---one ensures that transformations preserve causal and phase relationships among signals.  This disciplined treatment of operator composition transforms harmonization from a cosmetic preprocessing step into a mathematically coherent filter, one that preserves invariants of the underlying dynamics and prevents artificial distortion of cross-diagnostic correlations.  In doing so, even modest modeling approaches can generalize across campaign, device, and regime boundaries with greater stability and interpretability. Empirical evidence from multimodal fusion studies confirms that improvements in input coherence and alignment frequently yield greater downstream gains than incremental model refinement or neural lift alone~\citep{Kline2022MultimodalFusion, Cheng2024DataHarmonization, Nan2022DataHarmonisation}. In the fusion context, this realization underscores why the architecture of the Data Fusion Labeler (dFL) is focused first on harmonization and fusion, rather than pushing the envelope of operator complexity.


The dFL workflow is designed to provide users with the ability to design and customize their own harmonization strategies.  While a large number of native operations are already fully features, dFL is designed so that custom harmonization and visualization strategies can be easily and readily integrated into the existing dFL GUI by simple python scripting.  As a consequence, the harmonization functions included in Sections \ref{gap-fill}--\ref{normalize} should be thought of as options, not restrictions.  Some simple examples of implementing custom harmonization operations are shown in the included appendices, and extensive examples are detailed in the \href{http://dfl.sophelio.io/documentation}{dFL documentation}.  In addition, dFL support the ability to customize the \texttt{Fetch Data} option, that allows users to perform any preprocessing steps they like before loading the data into the dFL GUI using the same simple \texttt{data coordinator} API (additional details and examples of this ingestion-level customization approach can be found in \ref{AppE} and are extensively detailed in the \href{http://dfl.sophelio.io/documentation}{dFL documentation}). Beyond harmonization, dFL’s labeling and export stages can be used to preserve per-signal uncertainties and masks so that downstream \emph{feature-, score-, or decision-level data fusion} (e.g., stacking \citep{Hamda2023DSFusionIoT}, Dempster-Shafer combinations \citep{Koksalmis2020SensorFusionDS}, or log-odds aggregation \citep{Hanea2021ExpertsAggregation}) can correctly weight sources and propagate uncertainty \citep{Nazari2022DecisionFusion}. \\

\subsection{Basic Data Feature Support}\label{basic}

This subsection outlines the essential data handling features in dFL, such as dynamic visualization-driven resampling, data filling, and data trimming, that together ensure data alignment, consistency, and clarity across diverse fusion datasets. It also emphasizes the importance of interactive navigation tools, which allow users to zoom, label, and explore the data at different resolutions without compromising accuracy or performance. These steps further determine the statistical structure (e.g., sampling operators, effective bandwidths, and masks) on which multi-rate data fusion and cross-diagnostic inference depend.

\subsubsection{Data Alignment} 
Data alignment is a foundational prerequisite for any meaningful integration of multimodal fusion datasets. In both experimental and simulation contexts, diagnostics operate on disparate temporal grids, spatial coordinate systems, and reference frames, often with differing latencies and acquisition rates. Without precise alignment---both temporally and spatially---comparisons between diagnostics can introduce phase errors, distort cross-correlations, and obscure causal relationships between measured quantities. In downstream ML/AI or model-based workflows, such misalignments can propagate as systematic biases, degrading predictive accuracy and interpretability. Robust alignment procedures, therefore, are not merely a convenience but a requirement for preserving the physical coherence of the data, enabling valid feature extraction, and supporting reliable fusion of heterogeneous sources into a unified analytic framework.  From a formal data fusion perspective, alignment supplies a common state \(x(t)\) and observation maps \(y_i(t_i)=H_i x(t)+\epsilon_i\) with known \(t\mapsto t_i\); without this, likelihoods across \(i\) cannot be coherently multiplied nor can cross-covariances be estimated.

\begin{wrapfigure}{r}{0.57\textwidth} \centering \vspace{-10pt} \includegraphics[width=0.55\textwidth]{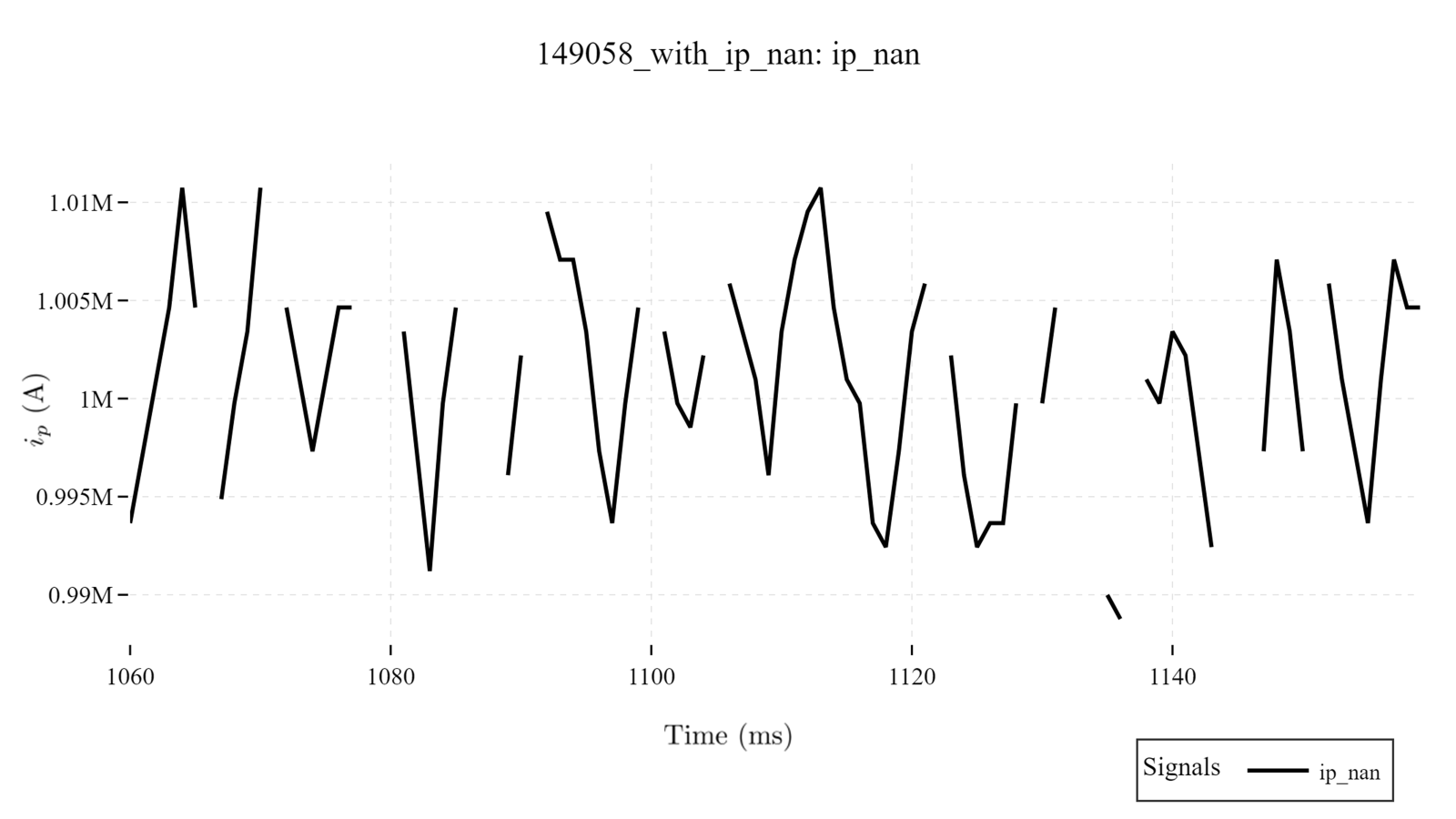} \\ \includegraphics[width=0.55\textwidth]{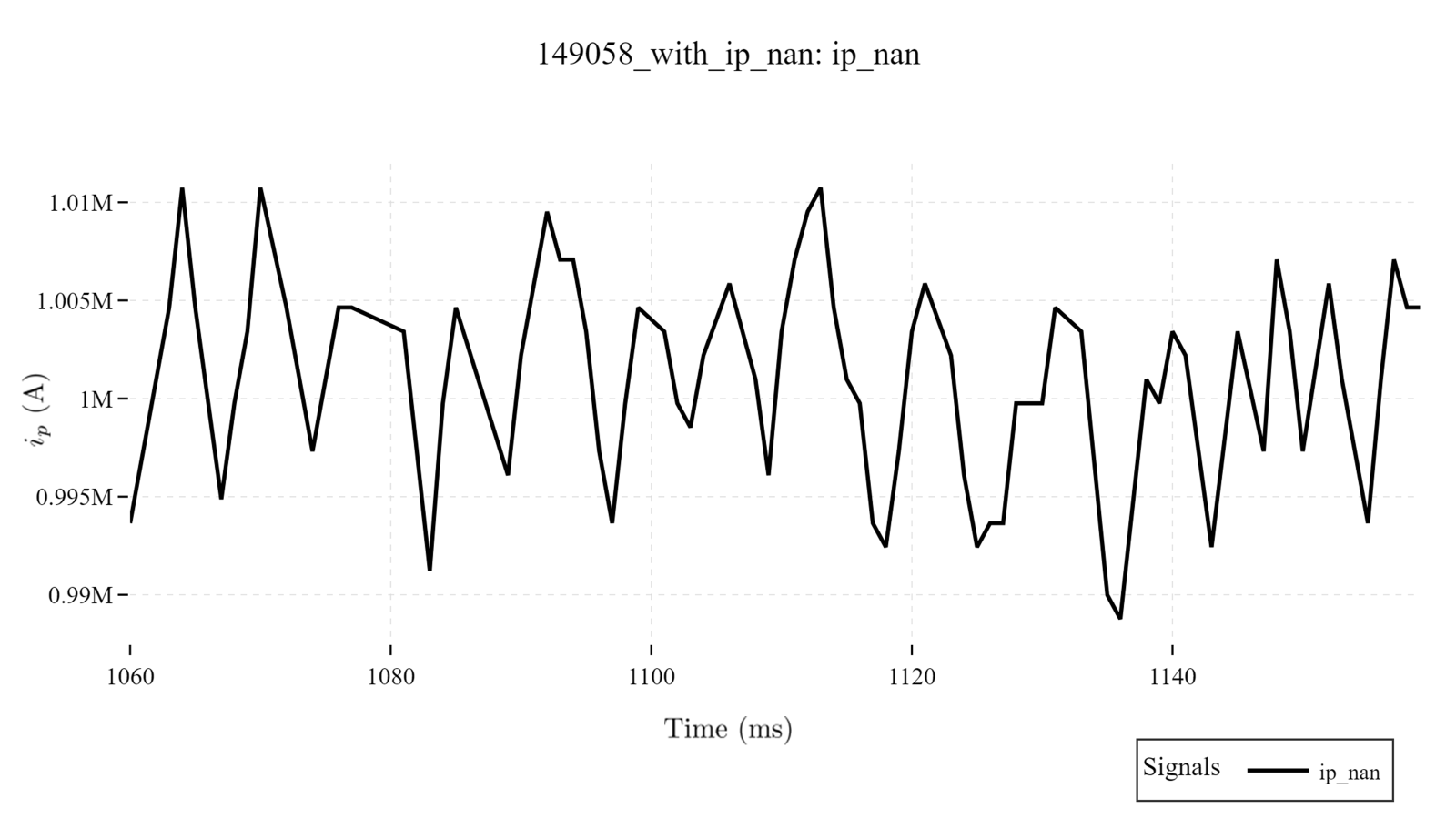}
\vspace{-10pt} 
\caption{An image showing the fill capability on a signal with NaNs scattered throughout. The top image is the NaN-riddled plasma current $i_p$ signal before fill, and the bottom image is after dFL fill is applied. } \label{fig:fill} \end{wrapfigure} As a rule, dFL generally deals with data alignment at the data ingestion level, ensuring that temporal and spatial indices from disparate sources are reconciled as early as possible in the pipeline.  By enforcing alignment at ingestion, dFL prevents the accumulation of phase mismatches and indexing errors that would otherwise compound through downstream preprocessing steps such as smoothing, normalization, or feature extraction. However, even with such safeguards, certain \emph{instrumental effects}---most notably digitizer drift, ADC baseline wander, or timing jitter---can subtly distort the apparent temporal coherence between diagnostics. These drifts, arising from hardware instabilities rather than physical dynamics, manifest as slow offsets or clock skews that shift relative signal phases over the duration of a discharge. Because their magnitude and character depend on device-specific electronics and environmental conditions, they typically require \emph{case-by-case calibration or detrending}, often through baseline correction or reference-pulse tracking. dFL treats these as calibration anomalies external to the alignment framework itself but provides mechanisms for their metadata tagging, ingestion realignment, and/or post-hoc correction. This layered approach ensures that subsequent operations (whether for visualization, model training, or real-time control) operate on datasets that are both temporally reconciled and physically trustworthy, preserving the statistical integrity and interpretability of the fused result.

It is also worth noting that dFL can handle raw pulse-counting data via utilizing custom filters/graphs in \texttt{fetch data} (within the \texttt{data provider}). That is, \texttt{fetch data} can be used to ingest list-mode or pre-binned counts with live time, apply dead-time/pile-up corrections, convert to rates with uncertainties, optionally variance-stabilize, align by bin edges, and use Poisson-aware/heteroscedastic losses. Nevertheless care must be taken when utilizing dFL's built-in operations on raw pulse-counting data, as several default operators assume fixed-grid, roughly constant-variance noise, where treating counts as continuous (e.g., generic smoothing or $z$-scores) can bias both rates as well as fusion weights.

The data presented in this paper integrates with TokSearch \citep{Sammuli_2018,toksearch_docs,Sammuli2024}, a high-performance toolkit developed for parallel access to heterogeneous fusion data. TokSearch is a high-performance query and retrieval engine integrated into dFL as a backend data provider, which is widely used to prepare experimental and simulation archives for AI/ML workflows.  Within dFL, it provides the data access backbone, ensuring that labeling operates on consistently aligned inputs (as the data has been pre-processed for alignment in TokSearch).  The important and subtle topic of physical units and unit-handling in dFL is discussed in some detail in \ref{units}, and the additional alignment considerations from resampling are addressed in Section \ref{resample}.

\subsubsection{Data Trimming}
\label{data-trim}

When large heterogeneous diagnostics (e.g. such as magnetic probes at 2~MHz, Thomson scattering at a kilohertz, synthetic turbulence fields on adaptive meshes, etc.) are loaded \emph{en masse}, the cardinality of samples ($N$) quickly exceeds what a browser, a GPU, or even a high-bandwidth PCIe bus can ingest without stalling.  A judicious \emph{trim}, applied \emph{before} gap filling or resampling, collapses the problem from $\mathcal{O}(N)$ I/O and memory traffic to $\mathcal{O}(n)$, where $n\!\ll\!N$ is the subset aligned with the scientific question at hand. Trimming can also be utilized to localize data fusion to regimes where the underlying process is, e.g., approximately stationary, improving the validity of stationary noise assumptions in weighted least squares and spectral fusion, etc.

Conceptually, trimming is nothing more than restricting the domain of a signal: $x(t)\;\mapsto\;x(t)\,\mathbbm{1}_{[t_s,\,t_e]}$. Practically, the operation re-indexes metadata, rewrites chunk boundaries, and releases memory pages---whether GPU-resident (VRAM) or system-resident (virtual memory)---thereby accelerating every downstream transform (e.g., FFT, filtering, feature extraction, etc.) while also guarding against aliasing that would arise if one down-sampled first. By reducing the number of resident buffers, trimming alleviates pressure across the memory hierarchy: GPU workloads gain additional VRAM headroom, while CPU-based workflows avoid excessive paging and cache thrashing in system RAM. Because no samples are overwritten, the cut is reversible: clearing the window restores the full record, preserving provenance and reproducibility. Analysts are encouraged to trim iteratively, i.e., begin with a coarse window to maintain GUI responsiveness, refine as patterns emerge, and, if adjacent segments must later be concatenated, preserve a sliver of overlap ($\approx$5\%) to mitigate edge effects in convolutional or recurrent models. This strategy echoes long-standing advice from digital signal processing and tidy-data theory: reduce data volume before heavy computation, and delineate observational units prior to statistical modeling~\citep{SmithDSP,WickhamTidy}. 

\subsubsection{Data Fill, Leakage, \& Causality}\label{gap-fill}

In many practical circumstances, data sequences contain scattered null entries. Missing samples (NaNs, sensor dropouts, empty rows) disrupt statistical analyses and machine-learning pipelines, etc.  The dFL pipeline therefore supplies three canonical remedies: (1) \textbf{linear interpolation} estimates interior gaps from adjacent points, producing smooth first-order continuity but leaving edge NaNs untouched; (2) \textbf{constant edge extension} copies the first or last valid value outward, guaranteeing a complete vector for real-time or frequency-domain algorithms, albeit with flat segments at the boundaries; and (3) \textbf{hybrid fill} applies edge extension at the ends and linear interpolation inside, yielding a fully populated record with minimal distortion. An example of how the fill looks is provided in Fig.~\ref{fig:fill}.  Further algorithmic details can be found in \citep{vanBuuren2018flexible,pandasInterp,SciPyInterp} and the \href{http://dfl.sophelio.io/documentation}{dFL documentation}.  More advanced data imputation strategies are currently under development, and are planned for dFL automatic updates.  It should also be noted that users may also easily implement their own fill strategies into the GUI (see the \href{http://dfl.sophelio.io/documentation}{dFL documentation}), or perform them using \texttt{Fetch Data} in the \texttt{data coordinator} API (see \ref{AppE}).\\

\noindent \emph{On Causality and Data Leakage:}  A further concern in harmonization pipelines (particular in real-time control or MPC) is the risk of \emph{data leakage}, where labels or derived features inadvertently incorporate information that would not be causally available at the time of prediction. For example, if gap-filling or normalization procedures draw on post-disruption signals (e.g. feature normalization), or if labels are constructed using smoothed quantities that extend into the future of the prediction window, then downstream ML models may achieve artificially inflated performance. Causal constraints are equally essential for sequential data fusion techniques (e.g., Kalman filtering), where future-dependent preprocessing violates the Markov property and yields non-realizable filters. The dFL framework is designed to mitigate this by allowing causal constraints to be imposed during labeling and/or preprocessing, such as restricting fill operations to past-only data, or by explicitly tracking provenance so that analysts can audit whether any step introduces information from outside the causal horizon. Ensuring causal consistency is therefore essential for both scientific validity and the credibility of ML-driven inference in fusion workflows, and should be kept in mind for all subsequent harmonization and labeling steps. 

\subsubsection{Visualization, Navigation, Analysis, and Dynamic Resampling}\label{vis}

Interactive visualization is central to effective data exploration, physical intuitive building, and labeling in dFL, particularly when working with large, high-dimensional time-series or spatial datasets. The interface leverages Plotly's built-in navigation tools for \emph{panning}, \emph{zooming}, \emph{resetting axes}, \emph{interacting} with and \emph{selecting} different signals, and \emph{autoscaling}, enabling users to adjust their view of the data in real time. These interactions are tightly coupled with \textit{dynamic resampling} to ensure that both the visualized data and its associated labels remain clear, accurate, and responsive at all scales.

Labeling tasks such as marking plasma states, identifying anomalies, or tagging regions of interest, often require changing the granularity of the displayed data. When a user zooms in on a temporal or spatial region, the system resamples the data to a finer resolution, revealing more detail and enabling precise placement of labels and annotations. Without such resampling, labels can become misaligned, overlap, or lose contextual meaning. Conversely, when zooming out, the system reduces resolution to maintain visual clarity, avoid overplotting, and ensure smooth interface performance. In dFL, the default visual resampling method is based on the MinMax LTTB algorithm \citep{van2023minmaxlttb}, which efficiently preserves salient features of the signal while adapting the level of detail to the current view. 

\begin{figure} \centering \includegraphics[width=0.90\linewidth]{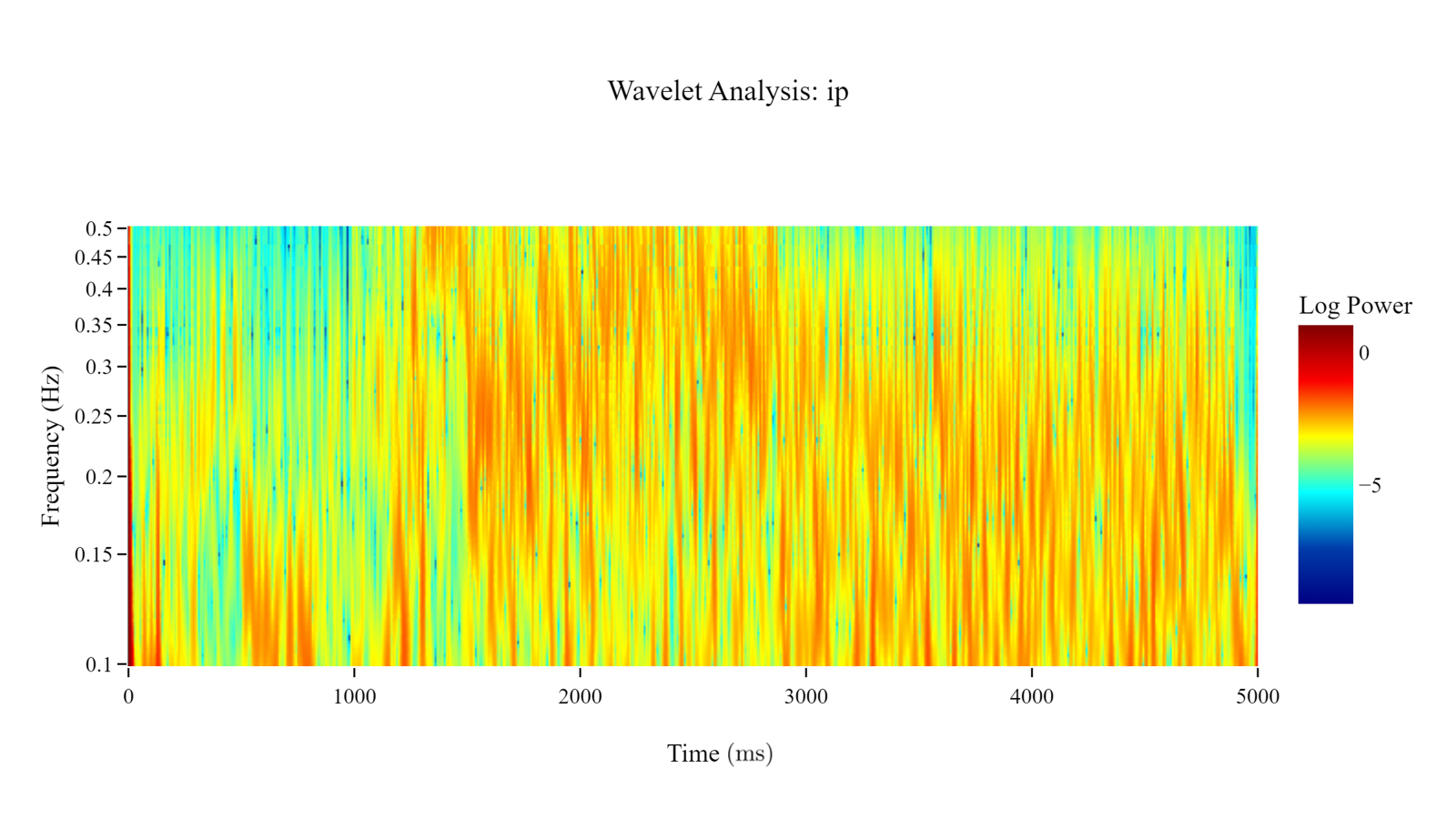} 
\caption{Here we show one of the natively supported frequency space graphs, a wavelet transform, with \texttt{Window Size}=256, \texttt{Overlap}=50\%, and \texttt{Morlet Width Param}=5, for the plasma current $i_{p}$ (Shot \#149058).} \label{fig:wavelet} 
\end{figure}

Beyond just navigation, dFL offers a diverse suite of interactive graphing and analysis tools for deep insight generation. Standard visualizations include time series plots with interactive signal toggling, multiple correlation plot types (scatter, heatmap, pairwise grids) supporting Pearson, Spearman, and Kendall methods, distribution plots (histograms and kernel density estimates), statistical analysis plots (rolling $z$-scores, CUSUM charts, moving-average confidence bands), and time-frequency visualizations such as spectrograms, short-time Fourier transforms (STFT), and continuous wavelet transforms (CWT) (see Fig.~\ref{fig:wavelet}, for example). All plots support parameter tuning, overlays, and result export, with behaviors optimized for responsiveness even on large datasets.  Full descriptions of all natively supported graphs types, along with definitions of all parameter settings, can be found in the full \href{http://dfl.sophelio.io/documentation}{dFL documentation}. 

\begin{figure} \centering \includegraphics[width=0.85\linewidth]{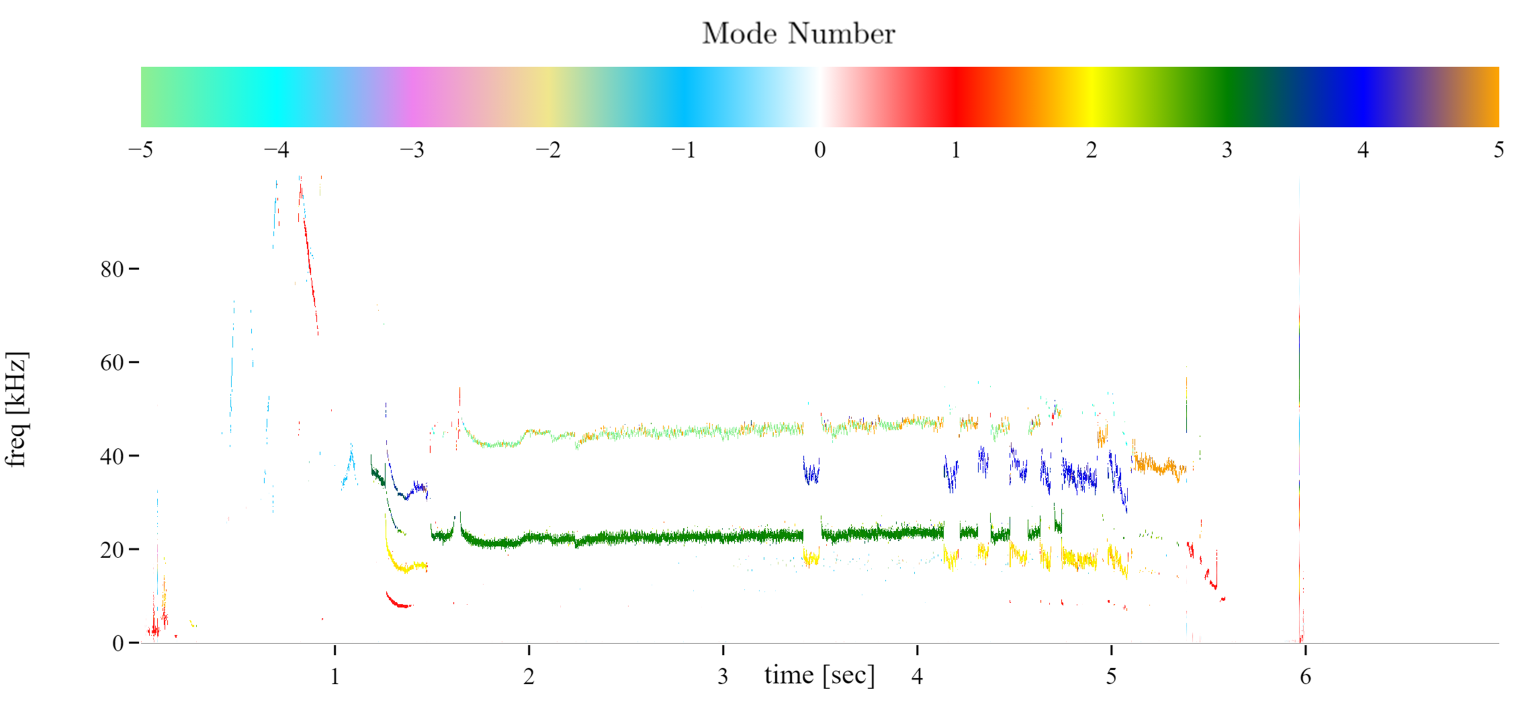} \caption{Here we show a custom graph type incorporated using the python-based Modespyec program for magnetic mode analysis.  The toroidal modes are set by contour colors (Shot \#149091).} \label{fig:modespec} \end{figure}

The system is also designed for extensibility. Custom graph types can be readily integrated by writing lightweight Python backends using the dFL \texttt{custom grapher dictionary} in the \texttt{data coordinator} API, which can easily leverage libraries such as \texttt{matplotlib} or \texttt{seaborn}, if so desired. This architecture allows advanced users to tailor visualizations to domain-specific needs, incorporate specialized statistical metrics, and adapt styles for publication-ready output. Detailed descriptions, parameterization guidelines, and implementation examples for all supported visualization types are provided in the full \href{http://dfl.sophelio.io/documentation}{dFL documentation}, and an example script is provided in \ref{AppC}.  Custom visual layers can be used to compute data fusion-relevant summaries (e.g., per-mode SNR or posterior weights), then incorporated into the dFL GUI, and their lineage exported. Figure \ref{fig:modespec} shows an example of a highly customized  Modespyec graphing tool integrated and available in dFL, with details provided in the \href{http://dfl.sophelio.io/documentation}{dFL documentation}. 

\subsection{Data Resampling}\label{resample}

Resampling plays a critical role in aligning and adjusting datasets from different sources, particularly when integrating data from fusion diagnostics and simulations. Fusion diagnostics, which monitor real-time plasma behavior, often collect data at varying sampling rates to capture transient phenomena such as plasma instabilities or rapid changes in magnetic fields.  \begin{wrapfigure}{l}{0.57\textwidth} \centering \vspace{-10pt} 
\includegraphics[width=0.55\textwidth]{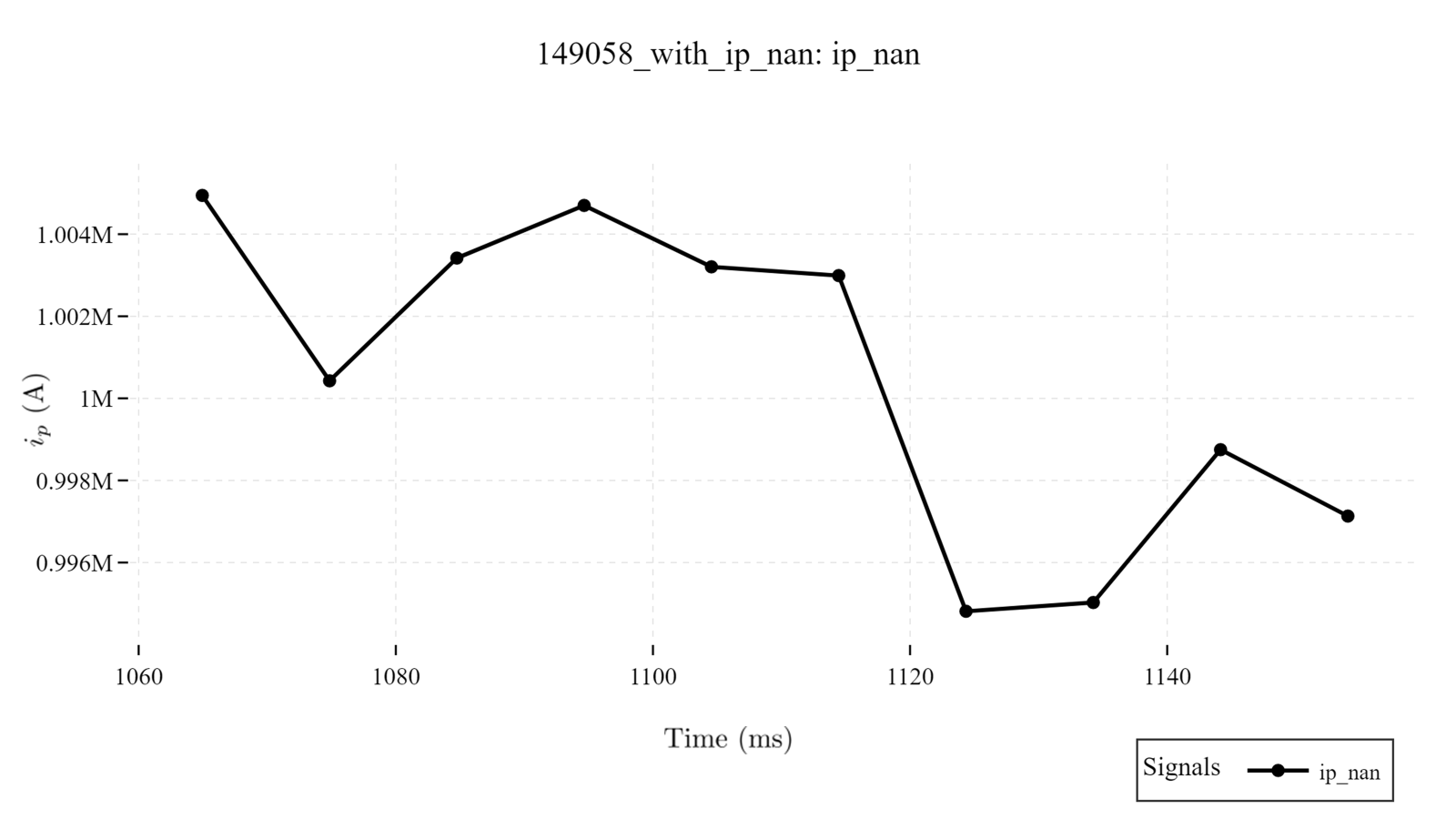} \\ \includegraphics[width=0.55\textwidth]
{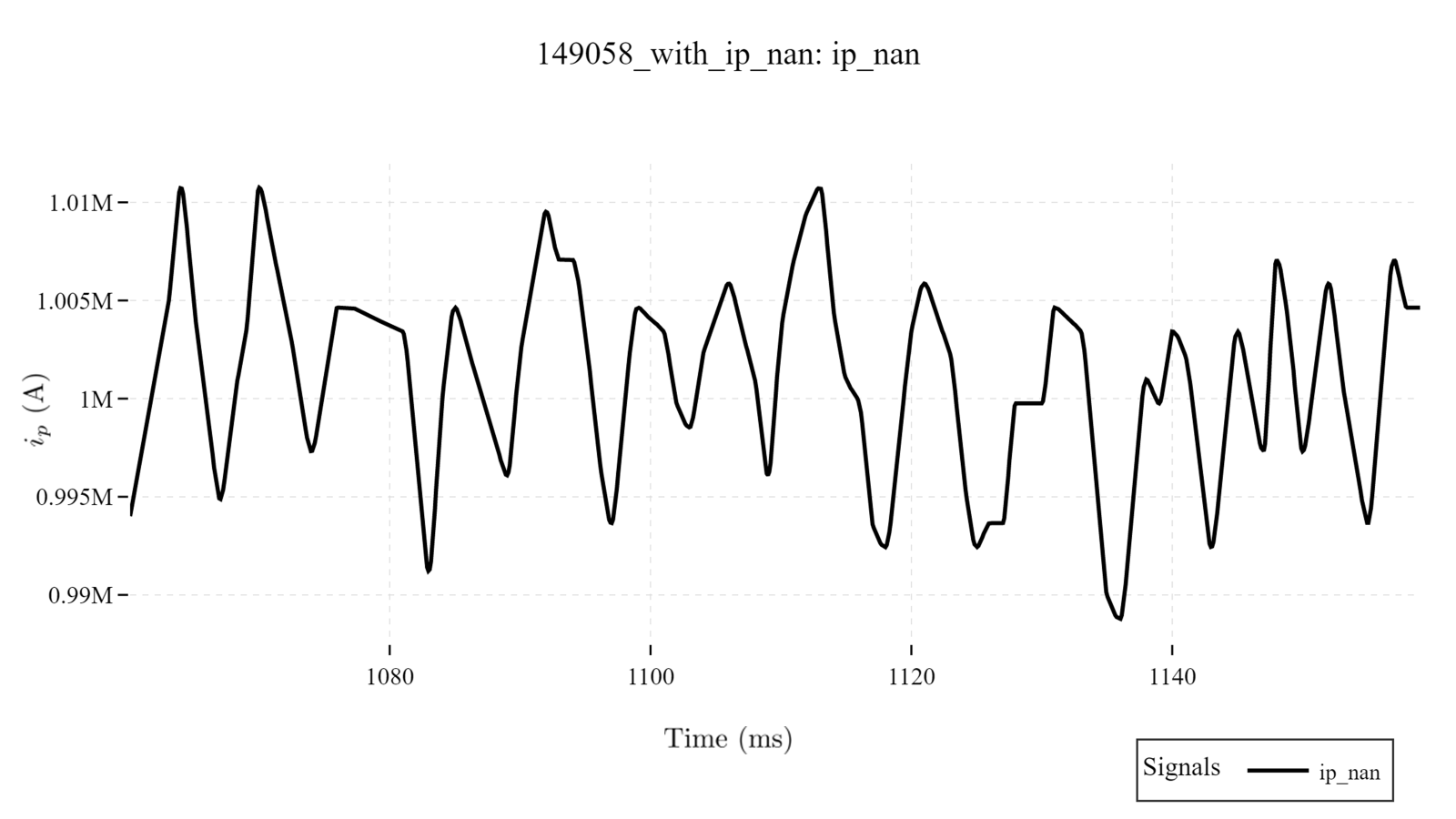}
\vspace{-10pt}
\caption{The same 100 point data segment shown in figure \ref{fig:fill}, first filled, then on top downsampled to 10 points using importance downsampling preserving the first two central moments, and on bottom upsampled to 1000 points using Mono-PCHIP.} \label{fig:resample} \end{wrapfigure} However, simulations, which model long-term behavior or large-scale dynamics, tend to produce results at often either lower (or at times, much higher) temporal resolutions than the corresponding experimental information.  To make meaningful comparisons between these data types, resampling techniques must be employed to adjust the sampling rate of one, and often both datasets, ensuring that they are aligned temporally on regular numerically stable grids. This may involve downsampling  high-frequency experimental data to match the lower resolution of simulation outputs or upsampling the simulation results to match the more granular experimental measurements, and any mixture of the two between different diagnostics and or simulations frameworks, etc.

The resampling process must be executed carefully to avoid pitfalls such as \textit{aliasing}, where high-frequency components of the data are incorrectly represented when downsampled, or the loss of important granular features that can occur when upsampling. To mitigate these issues resampling techniques, such as, e.g., linear, spline, or cubic interpolation, can be employed to ensure that key characteristics of the data are preserved. For example, cubic interpolation is often used when upsampling to maintain smooth transitions between data points and to prevent sharp discontinuities that could arise from simpler methods like linear interpolation.  While resampling ensures consistency in time-scale, it also introduces challenges in balancing computational efficiency, data fidelity, and data consistency, particularly when dealing with the large, multimodal datasets found in fusion energy.
Critically, downsampling before data fusion should include matched anti-alias filtering so that information from a high-rate channel is not spuriously injected at frequencies unsupported by lower-rate channels, which would otherwise bias frequency-domain data fusion (e.g. coherence-weighted estimators).

dFL natively supports five upsamplers and five downsamplers.  Most of these sampling techniques are conventional and implemented using well-established techniques.  An example of the resampling is provided in Figure \ref{fig:resample}.  Note, additional details and references to all the resampling techniques are provided in the online \href{http://dfl.sophelio.io/documentation}{dFL documentation} and in \ref{AppF} and \ref{AppG}.  It is worth noting that in certain instances, specific properties of the signal need to be preserved (e.g. spectral properties), in which case custom resampling might be required.  Options for integrating user-defined resampling methods into the dFL GUI are included in the \href{http://dfl.sophelio.io/documentation}{dFL documentation} (e.g., any DSP tool from \texttt{scipy.signal} can be integrated with a few lines of code), as are preprocessing examples using \texttt{Data Fetch} to load the data into dFL, as shown in \ref{AppE}.

\subsection{Data Smoothing}

Smoothing is a foundational operation in the broader process of data harmonization, serving as a bridge between raw, irregular, or noisy measurements and the coherent, analysis-ready representations required for high-fidelity interpretation. Far from being a cosmetic transformation, smoothing plays a decisive role in suppressing spurious high-frequency components often arising from sensor limitations, electromagnetic interference, numerical discretization artifacts, or signal degradation that do not correspond to the underlying physical phenomena. In fusion energy research, particularly in the interpretation of tokamak diagnostic signals, these noise sources can mask or distort the signatures of genuine plasma behavior, leading to mis-characterization of events or erroneous parameter estimation.  In addition, it should be duly noted that generally smoothing, like down-sampling, generates information loss and is non-invertible. As a consequence, in formal data fusion, for example, smoothing alters both  the autocovariances and cross-covariances that many fusion techniques exploit.  This is another reasons why dFL records and tracks smoothing kernels so that downstream estimators can adjust \(R_i\) (and the covariance between signals \(R_{ij}\)) accordingly.

Because of this nuance the importance and delicacy of smoothing is sometimes underestimated in discrete data workflows, where the emphasis instead often shifts prematurely toward feature extraction, modeling, or machine learning. Without proper smoothing, however, the downstream stages of data harmonization risk propagating noise as if it were signal, reducing predictive accuracy and obscuring physically meaningful correlations. For example, in spectral data fusion (e.g., coherence-weighted mode tracking), over-smoothing can attenuate narrow-band features and degrade fusion weights tied to cross-power estimates.  In this sense, smoothing is not merely a preparatory step, but a necessary safeguard against the contamination of the scientific inference pipeline.  

\begin{wrapfigure}{r}{0.57\textwidth} \centering \vspace{-10pt} 
\includegraphics[width=0.55\textwidth]{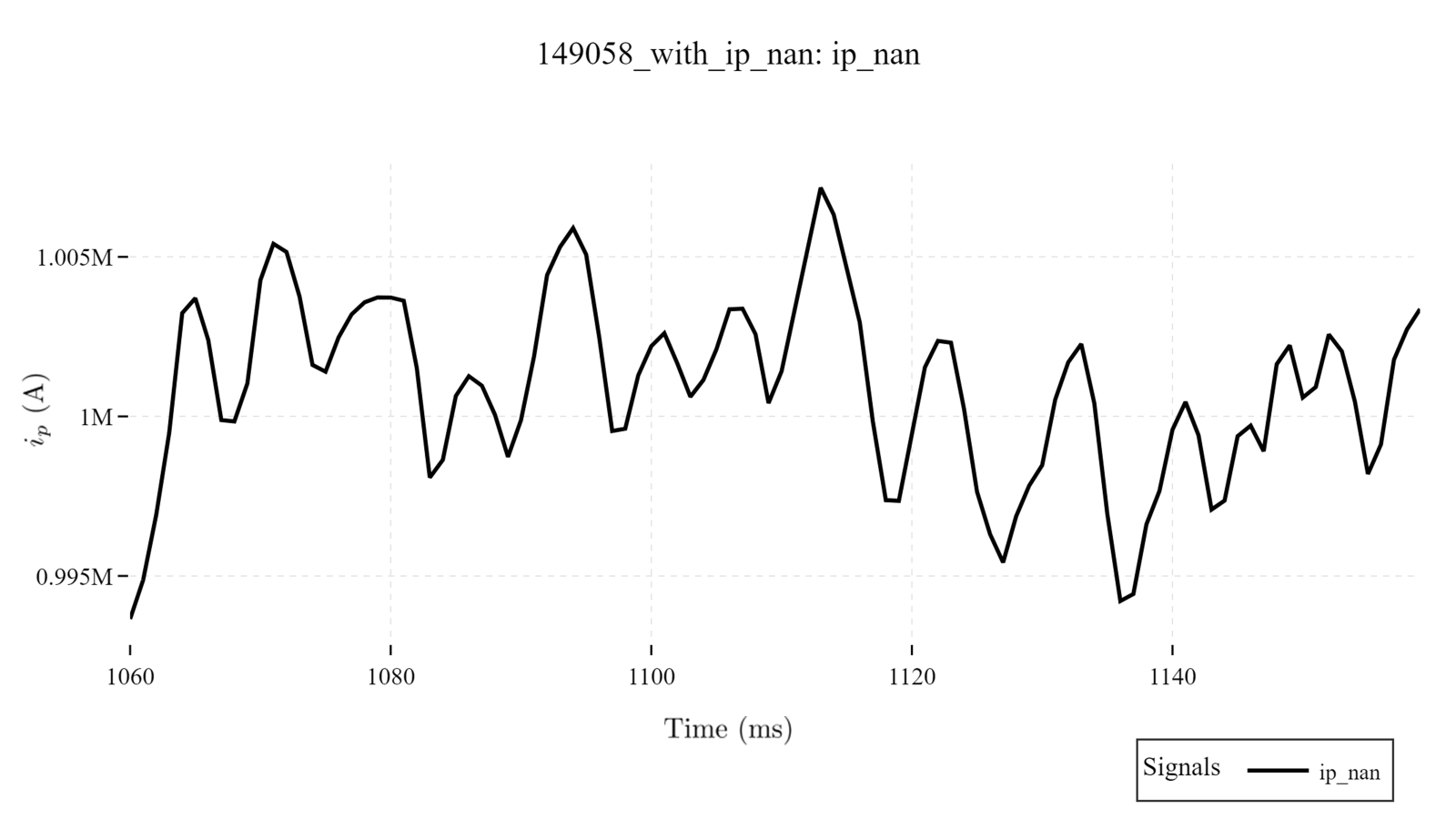} \caption{The same 100 point data segment shown in figures \ref{fig:fill}-\ref{fig:resample}, but here smoothed after fill using EMA with a span of 5.} \label{fig:smooth} \vspace{-10pt} 
\end{wrapfigure}The choice of smoothing method is highly context-dependent: it must respect both the statistical structure of the data and the physical context of the measurement process. An effective smoothing strategy should remove noise without erasing sharp but genuine features such as mode transitions, instability onsets, or abrupt control interventions. Striking this balance requires a principled understanding of the dataset’s spectral content, temporal or spatial resolution, and the downstream tasks it must support.

It is also important to distinguish smoothing from resampling, though certain resampling operators (see Section~\ref{basic}) incorporate implicit smoothing to suppress aliasing. Within dFL, smoothing is treated as a first-class, explicit operation whose parameters can be tuned independently of resampling, ensuring the flexibility to adapt to heterogeneous sources and irregular sampling grids typical of multimodal fusion datasets. The nuances of where smoothing belongs in the data processing pipeline---especially in the context of the order of operations---is addressed in Section~\ref{OOO}.   Moreover, building in custom smoothing and/or resampling into the dFL GUI can sometimes be ideal in order to preserve domain-specific features, e.g., spectral information.  Details of how to incorporate new smoothing/resampling algorithms can be found in the \href{http://dfl.sophelio.io/documentation}{dFL documentation}, and instructions on how to preprocess datasets using the \texttt{fetch data} functionality in the  dFL \texttt{data coordinator} API is show in \ref{AppE}.   Examples and definitions of natively supported smoothing algorithms in dFL are listed in \ref{AppH}. All smoothing parameters are exported so that data fusion methods can reconstitute the effective transfer function applied to each channel.

\subsection{Data Normalization}\label{normalize}

Normalization is another foundational operation in the data fusion and harmonization pipeline, ensuring that heterogeneous measurements can be meaningfully compared, integrated, and analyzed without scale-induced biases. In multimodal fusion datasets, diagnostic variables span orders of magnitude in both units and absolute values, ranging from electron temperatures in the keV regime, to magnetic field strengths in milliteslas or teslas, to particle densities exceeding $10^{20}\ \mathrm{m}^{-3}$. Left unnormalized, such disparities distort statistical measures, inflate condition numbers in numerical algorithms, and can cause optimization procedures in numerical models to converge poorly or toward biased minima.  Normalization also improves numerical stability in data fusion solvers that invert normal equations (or accumulate information matrices), reducing sensitivity to poor conditioning that can otherwise misallocate weights across modalities.

\begin{wrapfigure}{l}{0.57\textwidth} \centering 
\vspace{-10pt} 
\includegraphics[width=0.55\textwidth]{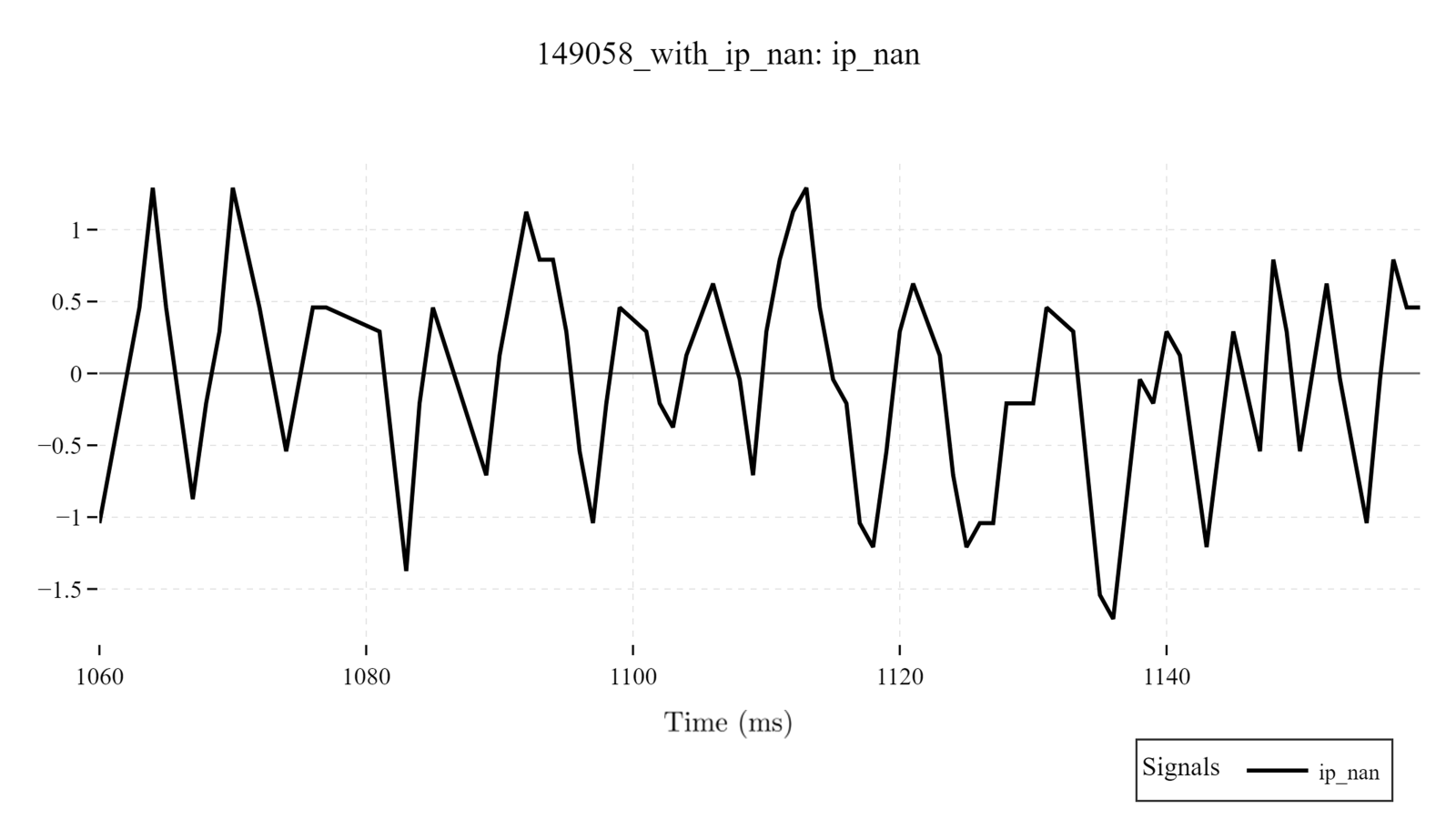} \caption{The same 100 point data segment shown in figures \ref{fig:fill}-\ref{fig:smooth}, but here normalized after fill using Median-IQR (units are dimensionless).} \label{fig:norm} 
\vspace{-10pt} 
\end{wrapfigure} The role of normalization extends beyond mere unit adjustment: it is a deliberate transformation that places diverse features on a comparable scale while preserving their intrinsic relationships and physical interpretability. For example, a model ingesting raw temperature and density measurements without normalization may incorrectly prioritize the variable with the largest numerical range, masking subtler but physically significant variations in other channels. This is particularly detrimental in fusion research, where meaningful cross-diagnostic correlations, such as between density gradients and magnetic topology, may occur at vastly different scales. 

Different normalization strategies (e.g., Min-Max scaling, $z$-score standardization, robust scaling) impose different statistical and physical constraints on the transformed data. The appropriate choice depends on the intended downstream use, the noise characteristics of each diagnostic, and whether the preservation of relative amplitudes or the enforcement of statistical standardization is more critical. In the context of fusion workflows, normalization is not just a statistical convenience, it is an essential precondition for integrating diverse measurements into unified, physically coherent datasets suitable for reliable feature extraction, model training, and multi-source data fusion. 



\subsection{Integrating Custom Filters \& Feature Maps into dFL }

Any custom filter (be it a DSP filter, a custom normalization, resampling, smoothing, imputation, etc.), or, more generally, feature map may be integrated into dFL using either the \texttt{Fetch Data} functionality in the  dFL \texttt{data coordinator} API discussed in the \href{http://dfl.sophelio.io/documentation}{dFL documentation}, or by integrating custom filters into the smoothing or normalization GUIs (NOTE: these can be custom filters for any purpose, that will only show up in the smoothing or normalization dFL dropdowns, along with any provided parameters). It is worth noting that feature maps often serve as the inputs to feature-level data fusion (e.g., combining per-mode amplitudes from Modespyec with edge profile gradients); where dFL preserves their lineage and parameters so that fused conclusions remain auditable.

\begin{figure} \centering \includegraphics[width=0.90\linewidth]{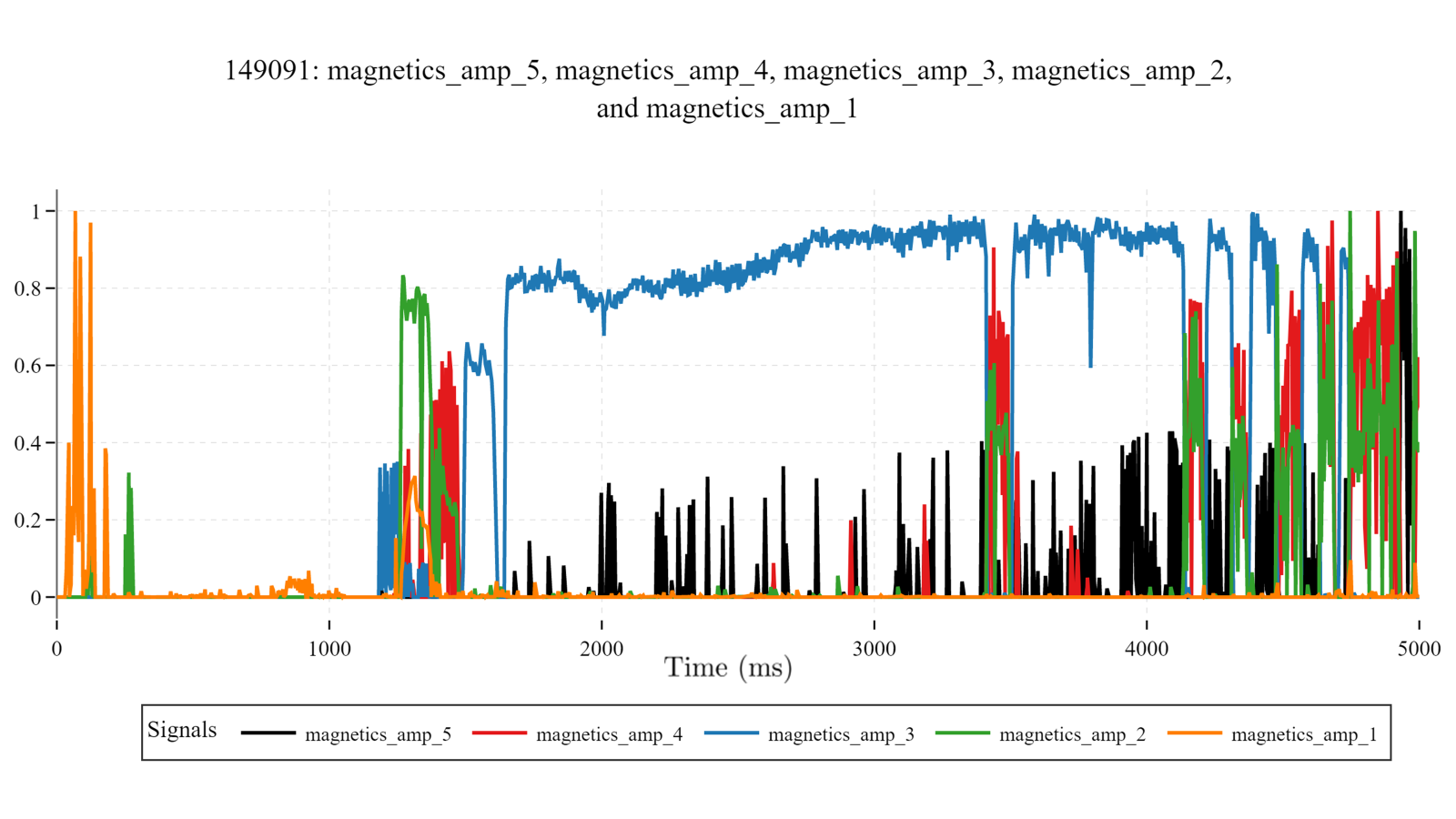} \caption{RMS amplitude vs.\ time of first 5 (positive) coherent toroidal modes \(n\) for shot 149091 (Dark theme), projected from the 2D mode spectrum data in Fig.~\ref{fig:modespec}. Estimates use two toroidally separated probes \((\Delta\theta)\), keeping only frequency bins with high cross-coherence and cross-phase \(\phi_{12}\!\approx\!-n\Delta\theta\); the mean autospectral power over selected bins yields the output. Traces are Min-Max normalized for visualization (units are dimensionless).} \label{fig:modespec_down_mag} \end{figure}

A simple example of using a custom feature map to extract relevant 1D time-series signals from 2D source signals is shown in Figure \ref{fig:modespec_down_mag}, where amplitudes associated with toroidal mode numbers are extracted from the Modespyec spectrum.  More details and full code is available in the \href{http://dfl.sophelio.io/documentation}{dFL documentation}.  

\subsection{Order of Operations \& Exporting to Databases}\label{OOO}

The preprocessing of fusion data---prior to export for simulation, analysis, machine learning (ML) and artificial intelligence (AI) tasks---involves a sequence of operations that act as functional transformations on noisy, sparse, and multimodal inputs. Each operation constitutes a potentially non-commutative map on the input space, and therefore, the order in which these transformations are applied is mathematically and empirically critical. This section introduces a mathematically principled view of the recommended pipeline implemented in the Data Fusion Labeler (dFL), emphasizing the critical importance of operator ordering, and illustrating the consequence through practical examples.

\subsubsection*{Operator Ordering in Data Harmonization}

Let $\mathcal{X}$ denote the space of input signals, such that $x \in \mathcal{X}$ is a univariate or multivariate time series. Each preprocessing step, Trim ($T$), Fill ($F$), Resample ($R$), Smooth ($S$), Normalize ($N$) acts as an operator on $\mathcal{X}$.  Generally any two of these operators $A$ and $B$ commute only if $A \circ B(x) = B \circ A(x)$ for all $x \in \mathcal{X}$. In practice, most preprocessing operators do \emph{not} commute, i.e., \[
N \circ S \neq S \circ N, \quad S \circ R \neq R \circ S, \quad R \circ F \neq F \circ R.\] This non-commutativity implies that arbitrary reordering of operations yield different outputs, even when the same raw signal is used as input. 


\subsubsection*{Recommended Preprocessing Pipeline}

To ensure statistical consistency, scientific reproducibility, and to minimize the risk of introducing artifacts that propagate into numerical models (e.g. classical simulations, ML/AI pipelines, data analysis, etc.) we recommend (by default) the following composite operation $O_3$ when dealing with large data sets on irregular or multiscale grids, given by: \[
O_3 := \textcolor{blue}{\text{Trim}} \rightarrow \textcolor{blue}{\text{Fill}} \rightarrow \textcolor{blue}{\text{Resample}} \rightarrow \textcolor{blue}{\text{Smooth}} \rightarrow \textcolor{blue}{\text{Normalize}} \rightarrow \textcolor{blue}{\text{Export}},\] where each step in $O_3$ is chosen to logically prepare the data for the next corresponding step.  This workflow often makes sense when each step is needed, though different orderings can be used to greater or lesser effectiveness given a specific context (examples below).  However, it should be noted that classical DSP theory strongly recommends smoothing first (before resampling) to avoid generating aliasing error when dealing with uniformly sampled data at a fixed sample rate that has real high-frequency content (i.e. edges, spikes, oscillations close to the Nyquist frequency), and is being downsampled by a large factor, and/or is intended to be analyzed in frequency space (e.g. Fourier transformed, etc.).  Another example of where $O_3$ may be non-optimal is smoothing before normalizing.  Since smoothing generally changes the signals variance, normalizing should generally be done second, but often numerically ill-conditioned physical units (e.g. 10$^{23}$) can make tuning smoothing parameters challenging and fraught, making the seemingly redundant workflow, $\text{Normalize} \rightarrow \text{Smooth} \rightarrow \text{Normalize}$, much more practical in order to preserve signal variances between channels.  As a consequence, depending on how many different types of data are being incorporated into your fusion pipeline, determining the correct order of operations can be subtle, and should be done in a principled way. 

\subsubsection*{Simple Example: Non-Commutativity of Operations}

Consider a raw signal sampled every 1 second, and given by the data vector, $ x = [2,\; \text{NaN},\; 6,\; 5]$. Let us define the preprocessing pipeline with the following configurations: (\emph{i}) Fill ($F$) via linear interpolation; (ii) Resample ($R$) to 0.5 s (upsampling by a factor of 2); (iii) Smooth ($S$) via a 3-point moving average; and (iv) Normalize ($N$) using z-score standardization. Performing these operations on a pipeline $A$, such that $A\colon F \rightarrow R \rightarrow S \rightarrow N$, directly yields the following: \begin{enumerate} \item Fill: $x_{\text{filled}} = [2,\; 4,\; 6,\; 5]$ \item Resample: $x_{\text{resampled}} = [2,\; 3,\; 4,\; 5,\; 6,\; 5.5]$ \item Smooth: $x_{\text{smoothed}} = [\text{NaN},\; 3,\; 4,\; 5,\; 5.5,\; \text{NaN}]$ \item Normalize: $\mu = 4.38,\; \sigma \approx 0.96 \Rightarrow x_{\text{norm}} \approx [\text{NaN},\; -1.43,\; -0.39,\; 0.65,\; 1.17,\; \text{NaN}]$ \end{enumerate} Now, performing these same operations in a different order, namely pipeline $B$, which simply changes the order of one operation, $B\colon F\rightarrow R \rightarrow N \rightarrow S$, yields: \begin{enumerate} \item Fill: $x_{\text{filled}} = [2,\; 4,\; 6,\; 5]$ \item Resample: $x_{\text{resampled}} = [2,\; 3,\; 4,\; 5,\; 6,\; 5.5]$ \item Normalize: $\mu = 4.25,\; \sigma \approx 1.41 \Rightarrow x_{\text{norm}} \approx [-1.6,\; -0.89,\; -0.18,\; 0.53,\; 1.24,\; 0.89]$ \item Smooth: $x_{\text{smoothed}} \approx [\text{NaN},\; -0.89,\; -0.18,\; 0.53,\; 0.89,\; \text{NaN}]$ \end{enumerate} Note that the final output differ in both magnitude and structure between the pipelines. Notably, applying smoothing after normalization distorts the statistical properties that the normalization was intended to standardize, showing that $N \circ S \neq S \circ N$ and highlighting the mathematical necessity of order-awareness in data processing workflows.

\subsubsection*{Advanced Example: Non-Commutative Smoothing \& Polyphase Kaiser Down-sampling}

\begin{figure}[ht] \centering \begin{subfigure}{0.75\linewidth} \centering \includegraphics[width=\linewidth]{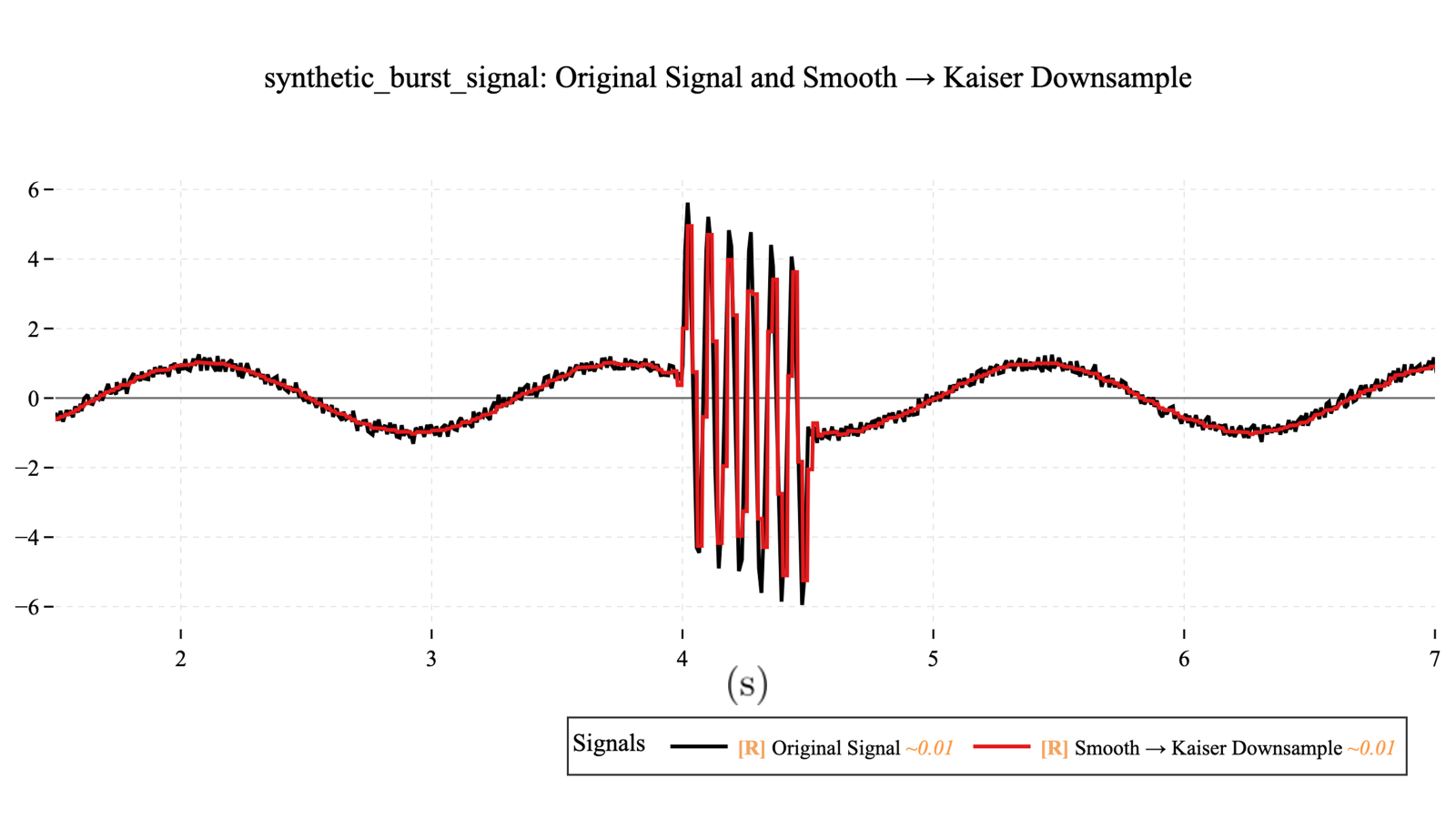} \caption{Pipeline A: Smooth \(\rightarrow\) Kaiser Downsample. In this case the burst survives, attenuated but unmistakably still present.} \end{subfigure} 
\begin{subfigure}{0.75\linewidth} \centering \includegraphics[width=\linewidth]{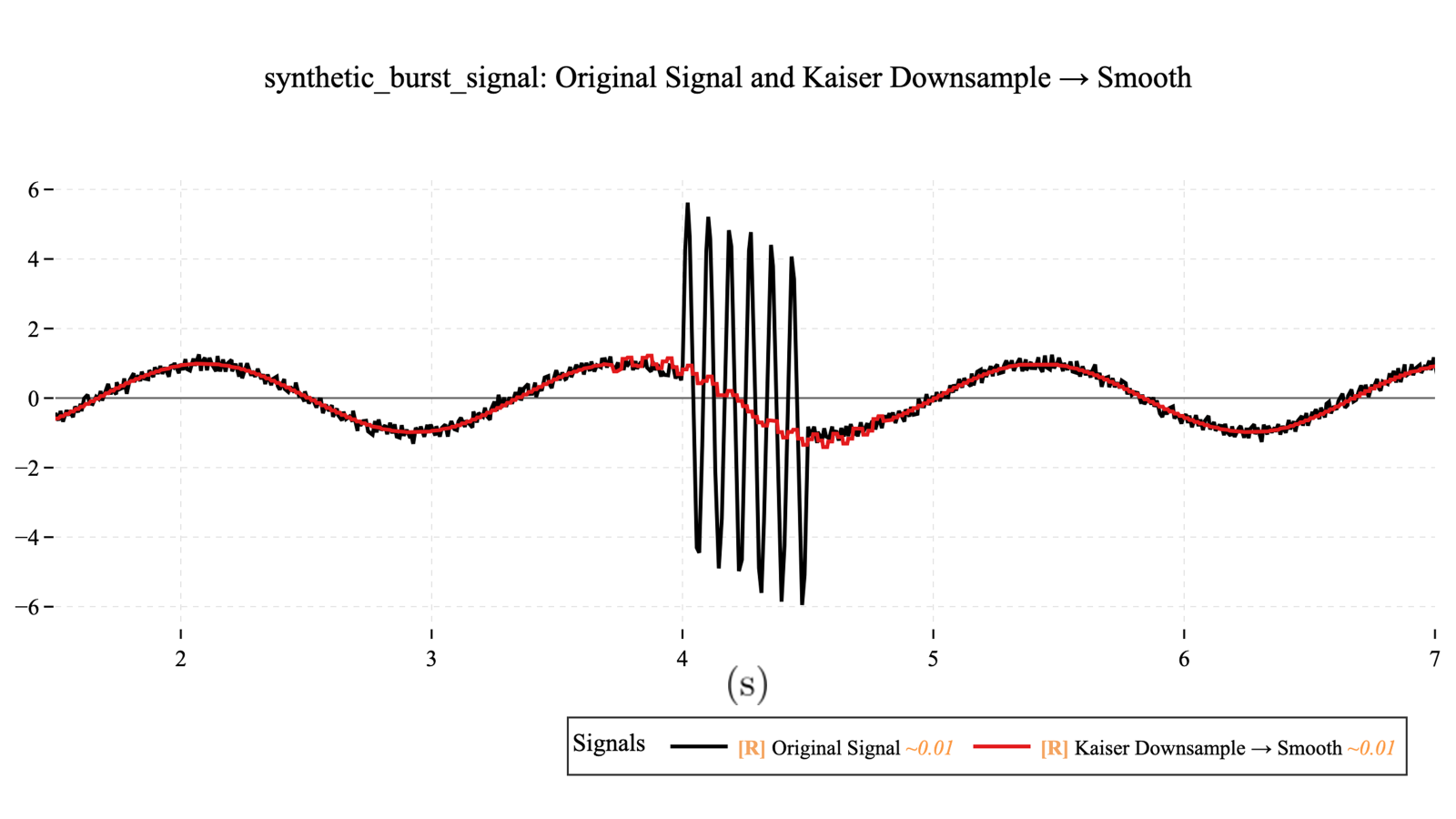} \caption{Pipeline B: Kaiser Downsample \(\rightarrow\) Smooth. Anti‑alias filtering eliminates most high-frequency energy before the smoother sees it; the burst is nearly erased.} \end{subfigure} \caption{Operator order matters even with an `ideal' polyphase FIR. A single transient critical for fault detection can vanish when down‑sampling precedes smoothing, underscoring the non‑commutativity of these preprocessing steps even when utilizing advanced/composite signal processing tools (Publication theme).} \label{fig:example2} \end{figure} To highlight how operator order alone can turn a dramatic transient into a barely perceptible wiggle (even when using advanced signal processing tools), we synthesize a \(10\) s record that contains a slow \(0.6\;\text{Hz}\) sine, mild Gaussian noise, and a brief, high-amplitude \(12\;\text{Hz}\) burst confined to \(t\in(4,4.5)\,\text{s}\) (see \ref{AppA} for details). The two preprocessing pipelines applied are: \[ \begin{aligned} \text{Pipeline \ A}:& \ \text{SG}(31,3)\;\longrightarrow\;\text{Kaiser}(d{=}10,\beta{=}2), \\[4pt] \text{Pipeline \ B}:& \ \text{Kaiser}(d{=}10,\beta{=}2)\;\longrightarrow\;\text{SG}(31,3), \end{aligned} \] where `SG'' is a Savitzky-Golay local cubic smoothing \citep{SciPy_SavGol} with a window of 31, and Kaiser'' is a Kaiser polyphase FIR (Finite Impulse Response) \citep{SciPy_ResamplePoly} with a downsampling factor $d=10$ and the Kaiser-Bessel parameter set to $\beta =2$. The results are stretched back to the original grid for direct comparison (working code included in \ref{AppA}). Figure \ref{fig:example2} shows that Pipeline A retains the burst’s \(\pm5\) swing, whereas Pipeline B suppresses it, leading to a dramatic numerical difference in the two simple processing pipelines, \(\max|\text{A}-\text{B}|=4.6\). It is worth noting that polyphase Kaiser down-sampling is popular in part because it applies a low-pass filter whose cutoff is matched to the new Nyquist rate, and that internal filter is the `smooth-before-downsample'' safeguard from classical DSP theory \citep{CrochiereRabiner1983}; especially when applied to regularly gridded data. Doing it here ensures alias-free rate conversion before you touch any cosmetic noise shaping, but at the cost of a loss in potentially meaningful physically relevant information (i.e. the burst in Fig.~\ref{fig:example2}).

\subsubsection*{Exporting Harmonized Datasets}

Once the preprocessing pipeline $O_3$ (or a custom mapping) has been applied, the processed dataset $\hat{x} = O_3(x)$ is ready for export. The dFL supports schema-aware export (e.g., CMF, IMAS, TokSearch), preserving both the harmonized data and associated metadata. While the default $O_3$ flow may be safest assuming a high dimensional irregular multimodal dataset, users may easily override it with a custom sequence tailored to their specific scientific goals, provided they take note of the consequences of non-commutativity and information loss, and the critical need to store operation order as metadata in the data provenance pipeline (discussed in more detail in Section \ref{sec:provenance}). Extensive options are available in dFL for custom exporting of datasets and details can be found in the \href{http://dfl.sophelio.io/documentation}{dFL documentation}. 

\section{Data and Metadata Standardization and Provenance}\label{sec:provenance}

Provenance in dFL is a critical and central design principle. 
Every action taken within the system---from data ingestion, to preprocessing, to manual or automated labeling, to export---is recorded and can be extended by the user to capture custom metadata. 
This is achieved through three extensibility points: the \texttt{manual\_labeling\_hook}, the \texttt{label\_export\_hook}, and the \texttt{data\_export\_hook}. 
Each hook exposes the internal state of the workflow at a critical transition, allowing users to inject additional context, perform side logging, or direct data and labels into external provenance frameworks such as CMF. 

The \texttt{manual\_labeling\_hook} is invoked each time a user creates a label. 
By default, dFL records dataset identifiers, shot numbers, and temporal bounds, but this hook allows the label row to be augmented with richer metadata---for example, the annotator identity, timestamp, or derived features computed on-the-fly. 
This ensures that manual operations are traceable and reproducible across campaigns. 
The \texttt{label\_export\_hook} intercepts the labeling dataset at the moment of export, enabling versioning, stamping, and redirection into institutional repositories or metadata services. 
Finally, the \texttt{data\_export\_hook} governs the export of harmonized signal data. 
It is invoked both for individual graph-level exports and for bulk operations, providing not only the processed signals but also a structured metadata document describing the applied pipeline (e.g., fill strategy, resampling parameters, smoothing operations, and order of execution). 

Together, these three hooks establish a flexible but rigorous provenance layer: all transformations are explicit, all exports are accompanied by machine-readable metadata, and users can extend the default tracking to meet local compliance or archival requirements. 
This architecture ensures that downstream analyses in fusion science are not only reproducible in principle but auditable in practice, with full transparency about what data were transformed, how, and by whom.

\subsection{Benefits of Standardized Data Schemas (e.g., IMAS, OMAS)}

The International Magnetics Fusion Research (IMAS) data standard is a cornerstone framework for organizing, storing, and exchanging data within the nuclear fusion community. It defines a unified schema capable of representing heterogeneous data sources (e.g., from diagnostic measurements and simulation outputs to control system telemetry) in a consistent and interoperable manner. By enforcing a common data model across institutions and experiments, IMAS enables seamless integration, comparison, and cross-validation of results, thereby accelerating collaborative research and reducing redundancy. This interoperability is particularly critical for multi-facility projects and for building machine learning (ML) models that require large, diverse, and standardized datasets.

OMAS (Ordered Multidimensional Array Structures) complements IMAS by providing a flexible yet schema-driven data structure optimized for the complex, high-dimensional datasets produced in fusion experiments and simulations. OMAS facilitates hierarchical organization of time-dependent and multidimensional data, ensuring compatibility with multiple software tools and analysis pipelines. Its efficient serialization formats and scalable architecture make it well-suited for both archival and real-time processing needs. Together, IMAS and OMAS form a robust ecosystem for ensuring that data are not only interoperable but also provenance-aware, enabling transparent and reproducible analysis in both research and operational contexts.

\subsection{dFL Integrated Tools for Data \& Metadata Management, Validation, and Tracking}

The Data Fusion Labeler (dFL) also integrates with a variety of backend systems and frameworks to ensure that data ingestion, labeling, curation, and export occur in a standardized, reproducible, and provenance-rich manner. These integrations allow users to seamlessly access heterogeneous datasets, maintain metadata integrity, and comply with international data schema standards.

\subsubsection{TokSearch}\label{toksearch}

TokSearch is a high-performance query and retrieval engine developed for fusion energy research. Integrated into dFL as a backend data provider, TokSearch enables rapid, targeted queries across massive, multimodal experimental datasets (i.e. often spanning decades of operation) without requiring manual file handling. Its query system supports flexible searches by shot number, diagnostic type, temporal range, operational regime, or derived physical parameters. In practice, TokSearch can, for example, return all bolometry and magnetic fluctuation signals from shots exhibiting specific confinement transitions, or retrieve synchronized diagnostics and actuator traces for disruption analysis.

When used within dFL, TokSearch ensures that labeled datasets are drawn from reliable sources and are consistently aligned across diagnostics. Queries can be embedded directly in dFL ingestion scripts, so labeling workflows always operate on well-prepared subsets of data. Importantly, TokSearch has been demonstrated to scale efficiently on HPC systems, with parallel backends (Ray, Spark, and SLURM integration) enabling throughput improvements of several orders of magnitude compared to traditional access methods. This scalability makes it practical to apply dFL to petabyte-scale archives typical of modern fusion facilities, and its ongoing extension to multiple devices further broadens the scope of labeling and analysis that dFL can support.

\subsubsection{CMF}
\label{CMF}

The HPE Common Metadata Framework (CMF) \citep{koomthanam2024integrated,hewlettpackard_cmf} provides a modular, service-oriented architecture for managing the complex, heterogeneous datasets inherent to fusion energy research. CMF addresses challenges such as handling multimodal, imbalanced, noisy, and harmonized datasets, as well as the sparse labeling characteristic of fusion diagnostics. Its composable design allows seamless integration of preprocessing tasks, normalization, resampling, smoothing, etc., while providing robust infrastructure for manual and automated labeling. CMF manages code, data and metadata (parameters, metrics) in Git-like fashion in a single framework, enabling reproducibility and facilitating reuse of workflows that preprocess multi-diagnostic data streams into a harmonized format, ensuring consistency before they enter the dFL labeling interface.

Crucially, CMF offers advanced metadata management capabilities aligned with IMAS and OMAS standards, supporting dynamic updates as new labels or harmonized datasets are generated. This allows provenance metadata to be recorded at fine granularity whether during batch exports, individual label creation, or automated labeling runs, capturing timestamps, personnel identifiers, data versions, and processing history. With comprehensive data, metadata and lineage query capabilities, CMF helps development of data discovery tools based on analyzing correlations between quality of results, workflow parameters, and input data. CMF supports management of labels as metadata, enabling rapid analysis of model sensitivities to preprocessing tasks and label sets without the need to access the data stores. Integration with dFL is achieved through Python ingestion scripts that invoke CMF commands at key stages of the workflow, enabling transparent, version-controlled curation.

It should be noted that CMF can be readily integrated at the dFL GUI level, where dFL will handle metadata handling for all harmonization and labeling tasks natively.  However, if the user wants to incorporate custom preprocessing using the \texttt{Fetch Data} approach shown in \ref{AppE}, then the metadata handling must be dealt with at the CMF level.

\subsubsection{MGKDB}

The Multiscale Gyrokinetic Database (MGKDB) is a specialized data infrastructure for storing, managing, and analyzing the large, high-dimensional outputs of gyrokinetic simulations---essential for advancing the understanding of plasma turbulence and transport phenomena. Uniquely, MGKDB natively stores its contents in an \textit{IMAS-compliant schema}, ensuring that simulation results share a common structure with experimental data repositories. This schema conformity enables seamless interoperability with other IMAS-aware tools and workflows, and allows simulation data to be cross-referenced, merged, and jointly analyzed alongside diagnostic measurements from operational devices.

MGKDB’s hierarchical organization accommodates complex, multi-parameter simulation outputs, capturing both scalar and field quantities across space, time, and configuration parameters. When integrated with the Data Fusion Labeler (dFL), synthetic diagnostics generated from simulations can be labeled using the same conventions and interfaces as experimental diagnostics. This alignment not only streamlines data curation but also improves the fidelity of machine learning models that bridge theory and experiment, by ensuring that training datasets reflect a harmonized representation of both sources. Furthermore, MGKDB’s standardized access patterns and version-controlled metadata allow simulation-derived labels and processed datasets to be reproduced, audited, and compared across research institutions, fostering collaborative validation and model benchmarking at scale.

\subsubsection{OMFIT}

The One Modeling Framework for Integrated Tasks (OMFIT) is a modular environment for managing, analyzing, and visualizing fusion energy data. Its built-in modules for profile analysis, MHD stability, transport modeling, and equilibrium reconstruction make it highly relevant to labeling workflows where physical interpretation and consistency checks are essential. OMFIT’s scripting capabilities allow for the automation of preprocessing, labeling, and export tasks within dFL pipelines. Furthermore, OMFIT’s native support for IMAS/OMAS ensures schema compliance, while its ability to handle dynamic, nonstationary, and multiscale datasets makes it well-suited for labeling tasks that require time-dependent context. In addition, TokSearch is integrated with OMFIT to support large-scale data ingestion, providing a seamless path from high-throughput retrieval to analysis workflows. The integration of OMFIT into dFL enables a unified workflow in which advanced modeling and manual or automated labeling reinforce one another, improving both label quality and downstream scientific utility.

\section{Data Labeling}

The value of any machine learning (ML) or artificial intelligence (AI) framework in fusion energy science hinges critically on the availability of accurate, context-rich, and physically meaningful labels. In the absence of such labels, supervised learning pipelines cannot reliably distinguish between stochastic fluctuations, diagnostic artefacts, and the true signatures of underlying plasma phenomena. Data labeling is thus not an ancillary step, but a central act of \emph{knowledge encoding}: it embeds domain expertise into the dataset, transforming raw multimodal streams into structured corpora that can train models, validate theories, and guide operational decisions.

\begin{wrapfigure}{l}{0.21\textwidth}  
    \centering
    \vspace{-10pt} 
    \includegraphics[width=0.2\textwidth]{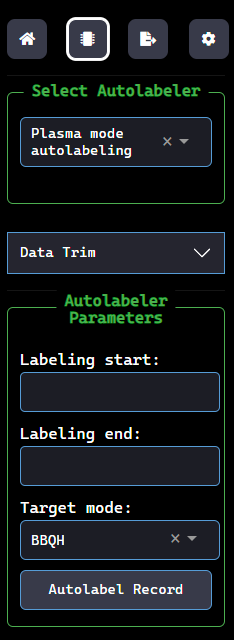}
    \caption{The classifier-based plasma mode autolabeler in the dFL GUI (Sophelio dark theme).}
    \label{fig:pmclass-al}
    \vspace{-10pt} 
\end{wrapfigure}Fusion energy datasets pose unique labeling challenges compared to other scientific domains. First, their complexity is inherently \emph{multimodal}, combining signals from diverse systems, such as magnetic diagnostics, bolometry, reflectometry, Thomson scattering, neutron detection, imaging systems, synthetic diagnostics from multiscale simulations, etc. Each modality captures different projections of the same evolving plasma state, with diverse units, sampling rates, noise characteristics, and temporal alignments. This heterogeneity demands a harmonization step before labeling can even begin, or else labels refer to inconsistent or misaligned events. Second, fusion data exhibit extreme \emph{class imbalance}: events of high scientific and operational significance---such as major disruptions, edge-localized modes (ELMs), or neoclassical tearing modes---occur far less frequently than stable operating intervals, yet their correct identification is crucial for both understanding plasma physics and ensuring machine safety.

Labeling in this context is further complicated by the \emph{nonstationary} nature of the plasma and its surrounding environment. Operational conditions, control schemes, and even diagnostic calibrations evolve across shots and campaigns, so that the same physical event may present differently depending on machine configuration, plasma scenario, or sensor drift. In this sense, temporal context becomes essential: a label attached to a transient must be interpretable relative to preceding and subsequent plasma conditions. Moreover, many phenomena of interest often unfold across disparate timescales, from microsecond turbulence bursts to confinement regime changes over seconds, necessitating multi-resolution labeling strategies that preserve hierarchical temporal relationships.

From a methodological perspective, the sparsity of human-labeled data in fusion research cannot be overstated. Expert labeling is expensive in both time and cognitive load, and is often constrained to a handful of well-characterized discharges. This scarcity has motivated the adoption of semi-supervised, weakly supervised, and active learning paradigms \citep{pavone2023machine,wang2023scientific}, in which algorithms leverage limited labeled data to guide the selection of new samples for expert annotation or to infer labels across unlabeled regions. In these frameworks, label uncertainty must be explicitly represented and propagated to downstream models, particularly when labels are derived from heuristic thresholds, statistical classifiers, or surrogate models trained on simulation outputs \citep{churchill2020machine,hatch2022reduced}.

Finally, the integration of labeling into the broader \emph{data fusion} workflow (cf. Section~\ref{sec:harmonization}) has deep implications for reproducibility and scientific validity. Labels that are not tied to harmonized, provenance-aware data streams risk becoming ambiguous or non-transferable between facilities, campaigns, or code bases. Conversely, when labeling is embedded in a harmonized pipeline, complete with temporal alignment, unit standardization, and metadata preservation, it becomes a reusable asset, enabling cross-experiment benchmarking, multi-device model training, and transparent physics discovery. In this way, robust labeling not only serves immediate ML tasks but also contributes to the long-term goal of building a federated, interoperable corpus of fusion knowledge.

Below we explore the manual, automated, and hybrid approaches to labeling in fusion energy science, detailing the algorithmic strategies, human-machine interaction models, and integration points with dFL that make high-quality labeling tractable at scale.

\subsection{Manually Labeling Multimodal Data Sets}

Manual labeling remains one of the most reliable and context-aware methods for annotating complex fusion energy datasets. Its primary strength lies in the ability of expert human judgment to identify subtle, context-dependent phenomena, such as precursor oscillations to disruptions, nuanced changes in edge fluctuations, or diagnostic artefacts, that might evade purely automated algorithms. When performed carefully, manual labeling can capture the richness and ambiguity inherent in real-world experimental data, ensuring that event definitions are physically meaningful, aligned with research objectives, and robust against spurious detections. This process is particularly valuable in early-stage campaigns, during exploratory analyses of new diagnostics, or when curating benchmark datasets for algorithm validation.

However, manual labeling also comes with inherent limitations. It is time-consuming, often requiring subject-matter experts to inspect large volumes of data point-by-point or shot-by-shot, which can become prohibitively labor-intensive in the petabyte-scale data regimes typical of modern tokamaks and stellarators. Furthermore, the process can be susceptible to human variability, e.g., different annotators may interpret borderline cases differently, leading to inconsistencies across large datasets. In such contexts, manual labeling is best employed in a targeted manner, either as a gold-standard reference set for validating automated pipelines, or as an expert-in-the-loop refinement step where algorithmic pre-labeling provides candidate regions that are then confirmed or adjusted by humans.

The Labeler (dFL) provides a robust, intuitive, and scientifically oriented interface purpose-built to simplify the manual labeling of complex fusion datasets, be they multidimensional, multimodal, heterogeneous, imbalanced, or contaminated by noise. Its design enables researchers to work with both precision and efficiency, while preserving the full informational richness and physical context of the data.

Within dFL, data are ingested as discrete \emph{records}, selectable from the \texttt{Select Record} drop-down menu. A ``record'' can correspond to a specific \emph{shot} or \emph{discharge} number, a simulation \emph{model} type, a diagnostic \emph{component}, or any other uniquely identifiable data grouping defined by the user. Once a record is selected, all available \emph{signals} associated with that record are loaded into dFL and can be visualized by selecting \texttt{Add Graph}. The Labeler natively supports multiple graph types, as outlined in Section~\ref{vis}. The default view---a high-resolution time-series plot---often serves as the optimal choice for labeling, but alternative visualizations such as spectral plots or user-defined custom plots can be seamlessly integrated and used for manual labeling.

Label creation begins by defining a label name in the \texttt{Settings} tab, after which it becomes available in the \texttt{Select Labels} drop-down of the main control panel. Additional customization of labels, such as pre-defining categories, structures, or metadata, can be implemented directly within the \texttt{data provider} ingestion script.  Additional full details are provided in the \href{http://dfl.sophelio.io/documentation}{dFL documentation}.

\begin{figure}
    \centering
    \includegraphics[width=0.9\linewidth]{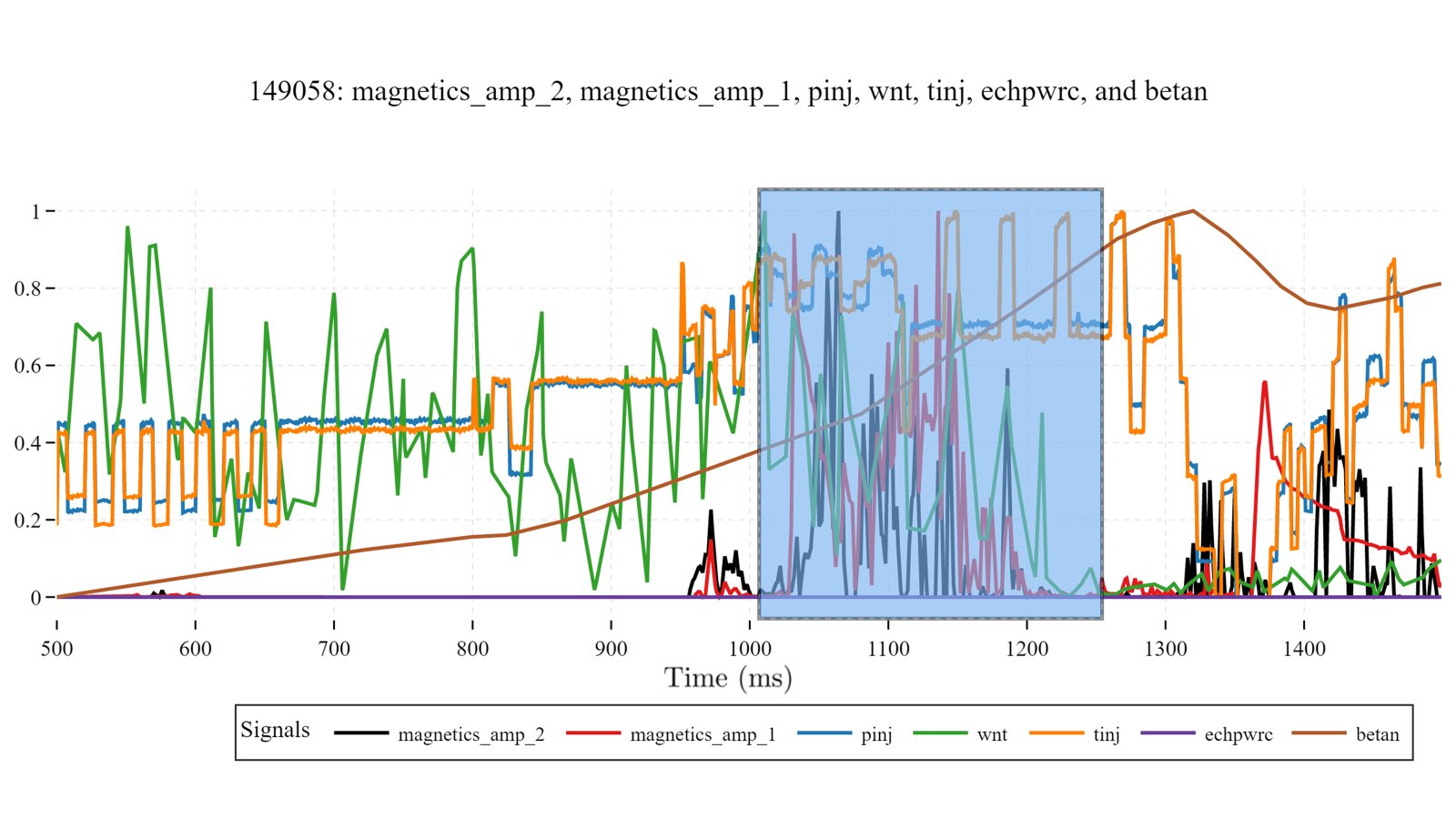}
    \caption{Here we manually label a generic `magnetic burst corresponding to an average pedestal width ``wnt'' event' where seven different signals (by TokSearch point names) of a single discharge/shot from DIII-D are loaded on a single plot, and the signals are Min-Max normalized to reveal synchronized onset behavior.}
    \label{fig:manual_label}
\end{figure}

Once a label name is chosen, the labeling process itself can proceed in one of two ways. The user may (\emph{i}) manually enter the start and end times for the label directly into the main control panel, or (\emph{ii}) enable \texttt{Label Selection Mode} from the display options. In the latter case, the plotted graphs become fully interactive, allowing the user to highlight the region of interest directly with the mouse. Upon confirming the selection via \texttt{Confirm Label}, the label is committed to the session and becomes visible in the \texttt{Current Labels} tab of the control panel. This tab not only lists all active labels for the session but also supports direct export of the labeling results to file, ensuring they are immediately ready for downstream analysis, archiving, or integration into machine-learning workflows. A simple manual labeling example is shown in Figure \ref{fig:manual_label}, and video demos and extensive user support may be found in the \href{http://dfl.sophelio.io/documentation}{dFL documentation}.

\subsection{Automated Labeling}
\label{sec:autolabeling}

The complexity, scale, and heterogeneity of data in fusion research---whether from experimental diagnostics, synthetic diagnostics embedded in simulations, or large-scale multiscale modeling pipelines---renders purely manual labeling infeasible at scale. Modern fusion experiments routinely produce petabyte-scale, multimodal time series and imaging datasets per campaign \citep{humphreys2020advancing,anirudh20232022}, encompassing diagnostics with widely varying sampling rates, signal-to-noise ratios, and spatial/temporal resolutions. Automated labeling frameworks are therefore indispensable for efficiently and consistently annotating these datasets with scientifically and operationally meaningful features. Once deployed, such frameworks not only enable rapid downstream analysis and machine learning (ML) model training but also support real-time or near-real-time operational decision-making, particularly in disruption prediction, regime identification, and advanced control scenarios.

Automated labeling in fusion can be systematically organized into five complementary categories:

\begin{enumerate}[label=\itshape\roman*)]

    \item \emph{Deterministic, physics-informed methods:}  
    closed-form, algorithmic approaches provide interpretable and computationally efficient labeling strategies for signals with reproducible, physically well-defined patterns. Examples include:
    \begin{itemize}
        \item Peak-finding algorithms to detect transient events such as Edge-Localized Modes (ELMs), sawtooth crashes, or pellet ablation signatures in high-cadence magnetic or bolometric diagnostics \citep{oshea2022edge}.
        \item Derivative thresholding to flag abrupt changes indicative of MHD instabilities, confinement transitions, or actuator faults.
        \item Turning point and zero-crossing detection to mark extrema in simulation outputs, such as minima in magnetic energy during relaxation events or maxima in temperature gradients near transport barriers.
    \end{itemize}
    Such methods are particularly effective when coupled to uncertainty-aware preprocessing pipelines (cf. Section~\ref{sec:harmonization}), but their reliance on fixed thresholds or prescribed signal morphology can make them sensitive to noise and less adaptable to multivariate dependencies.

    \item \emph{Statistics-based methods:}  
    Statistical inference and change detection techniques identify anomalous or regime-shift behavior by quantifying deviations from nominal baselines in a probabilistic framework. Approaches include classical hypothesis testing, control charts, parametric and non-parametric density estimation, and time-series change-point detection \citep{truong2020selective,gelman2013bayesian,qin2003statistical}. For example, statistical process control can flag gradual drifts in diagnostic calibration, while kernel density estimation can detect distributional shifts in reflectometry or bolometry signals. These methods balance interpretability, computational efficiency, and formal uncertainty quantification, qualities that are highly valued in high-consequence scientific workflows. However, they often assume stationarity or specific distributional forms, and performance may degrade in highly non-linear, multi-regime operational settings.

    \item \emph{Data-driven and adaptive methods:}  Machine learning-based approaches, ranging from supervised classifiers to fully unsupervised representation learning, offer greater flexibility in handling noisy, high-dimensional, and partially labeled datasets. These methods often rely on learning low-dimensional \emph{embeddings}---compact, structured representations that preserve salient physical or statistical relationships while suppressing noise---to enable robust downstream analysis. Examples include:
    \begin{itemize}
        \item Pre-trained classifiers fine-tuned on fusion-specific data, such as generic convolutional neural networks for identifying plasma boundary features or classifying MHD modes from diagnostic imagery \citep{bustos2021automatic}.
        \item Transfer learning from related tasks, enabling rapid adaptation of general-purpose models (e.g., video anomaly detectors, time-series transformers) to specialized fusion contexts without requiring prohibitively large labeled corpora.
        \item Unsupervised clustering (e.g., $k$-means, Gaussian Mixture Models, spectral clustering) to reveal latent operational regimes, confinement states, or instability precursors without pre-existing labels \citep{meng2020survey}.
        \item Representation learning via autoencoders or contrastive methods to generate feature embeddings that are amenable to downstream classification or event detection.
    \end{itemize}

    \item \emph{Simulation and model driven autolabelers:}  
    These methods leverage outputs from high-fidelity physics simulations, synthetic diagnostics, or reduced-order models to generate labels in both simulated and experimental datasets. In fusion research, first-principles simulations such as extended-MHD (e.g., NIMROD, M3D-C1) \citep{sovinec2004nimrod,ferraro2023m3d} and gyrokinetic turbulence codes (e.g., GENE, GS2) \citep{jenko2000gene,Kotschenreuther1995} can be post-processed to extract physically meaningful event markers---such as instability onset times, fluctuation amplitudes, or mode structures---that are then mapped onto experimental data streams via synthetic diagnostic forward models \citep{shi2016synthetic}. Reduced-order transport solvers (e.g., TGLF, QuaLiKiz) \citep{staebler2007tglf,bourdelle2015core} can similarly provide regime classification boundaries (e.g., L- to H-mode transitions, internal transport barrier formation) that serve as consistent labeling criteria across large archives. 

    This simulation-informed labeling approach is particularly valuable in data-sparse or rare-event regimes, where experimental coverage is limited but predictive models are available. It enables pre-labeling of operational scenarios that have yet to be experimentally realized, accelerates the training of machine learning models in unexplored parameter spaces, and supports physics-based consistency checks in hybrid labeling pipelines. The primary limitation lies in the fidelity and validation status of the underlying models; any bias, missing physics, or diagnostic misalignment in the simulation may propagate directly into the labels. Consequently, best practice involves cross-validating model-generated labels with statistically independent experimental datasets and quantifying label uncertainty via ensemble or Bayesian simulation frameworks \citep{michoski2024gaussian}.

    \item \emph{Hybrid and expert-in-the-loop systems:}  
    The most powerful automated labeling pipelines often blend deterministic physics-informed algorithms, statistical detectors, and adaptive ML-based components, while incorporating expert oversight. For instance, an ELM detector may first use derivative-based thresholds to produce candidate events, which are then filtered and refined by a neural network trained on expert-labeled exemplars. This hybrid approach enhances robustness to noise, addresses rare-event scarcity, and retains interpretability---factors that are essential for operational trust and scientific reproducibility \citep{wang2023scientific,pavone2023machine}. When integrated into platforms such as the Data Fusion Labeler (dFL), hybrid pipelines can generate, validate, and deploy labels at scale across archival and real-time data streams.

\end{enumerate}

In the broader context of fusion informatics, automated labeling is not merely a convenience, it is an enabling technology for the next generation of data-driven discovery, model validation, and real-time plasma control. When combined with standardized data models (e.g., IMAS/OMAS) and provenance-aware harmonization workflows, these techniques enable researchers to rapidly extract structured knowledge from complex, multimodal datasets, bridge experimental and simulation domains, and feed high-fidelity, physics-consistent labels into predictive models. As ML and AI methods continue to mature, automated labeling will become a cornerstone of integrated modeling, cross-facility data sharing, and autonomy in the fusion energy ecosystem.  Sections \ref{sec:stats_autolabeling}--\ref{sec:phys-inf} below discuss autolaber types already integrated into dFL.

\subsubsection{Statistics-based Autolabeling}
\label{sec:stats_autolabeling}

Statistics-based autolabeling constitutes a critical component of automated annotation pipelines in fusion research, providing interpretable, uncertainty-aware mechanisms for detecting, characterizing, and segmenting features of interest in large-scale, heterogeneous datasets. These approaches derive their strength from formal statistical inference, allowing event detection to be framed in terms of hypothesis testing, threshold exceedance, or confidence interval breaches. They have been applied to a wide range of fusion data analysis tasks, from identifying abrupt confinement transitions and fluctuation-level changes preceding disruptions, to detecting anomalous sensor behavior in magnetic, bolometric, and reflectometric diagnostics. 

\begin{wrapfigure}{r}{0.7\textwidth} 
    \centering
    \vspace{-10pt} 
    \includegraphics[width=0.69\textwidth]{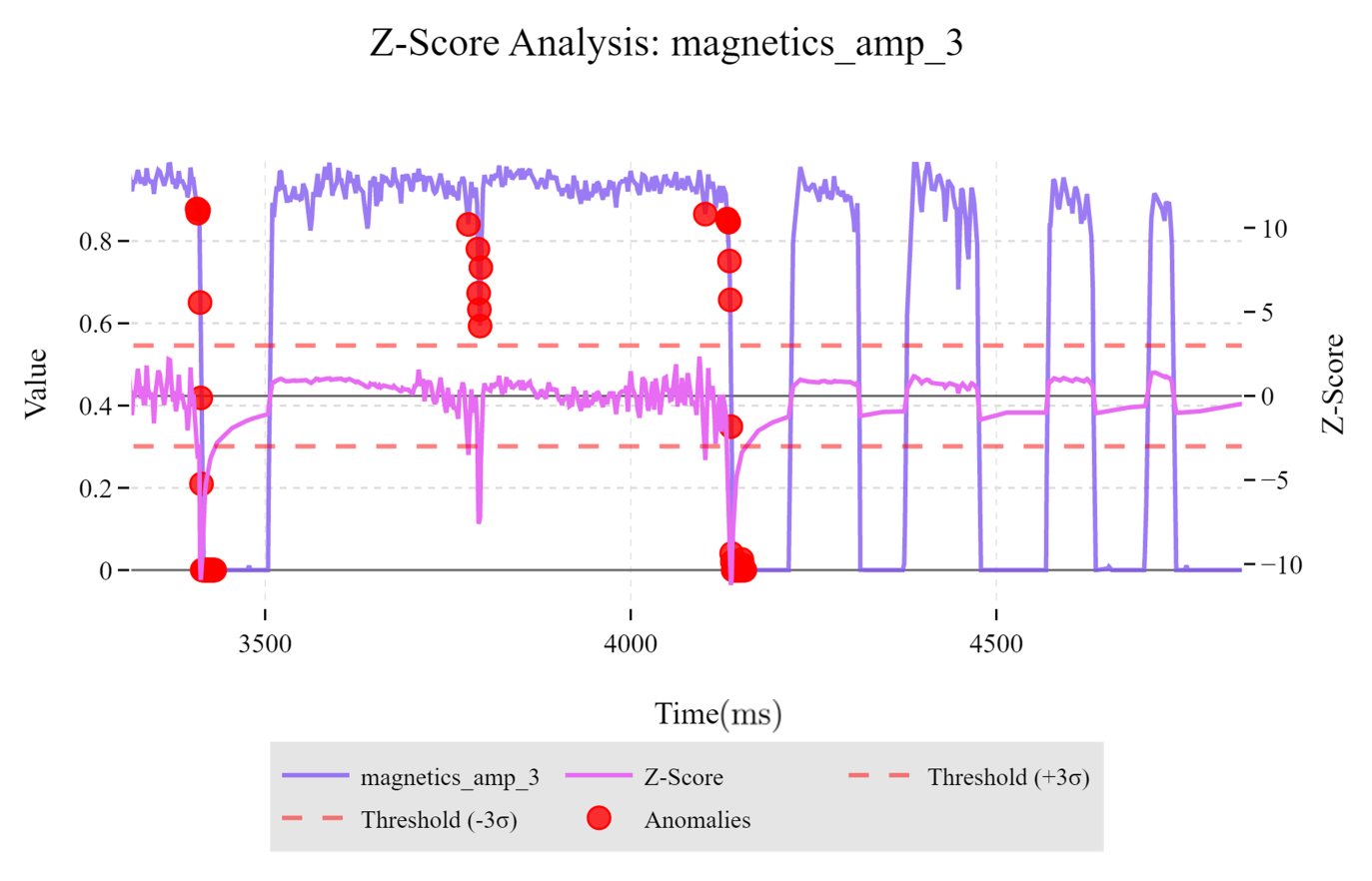}
    \caption{A moving-window $z$-score deviation anomile detector, setting the window size to 200 ms and the threshold to 3$\sigma$ on a Min-Max normalized Mirnov coil magnetics signal (units are dimensionless).}
    \label{fig:stats}
    \vspace{-10pt} 
\end{wrapfigure} In the \emph{Data Fusion Labeler} (dFL), three widely used statistical anomaly detection methods are implemented within the \texttt{Stats} graph type (e.g., see Figure~\ref{fig:stats}). Once the user selects \texttt{Stats} under the \emph{Graph Type} menu, they may choose from:  
(\emph{i}) a moving-window $z$-score deviation detector,  
(\emph{ii}) a moving-window cumulative sum (CUSUM) chart, and  
(\emph{iii}) a moving-average with confidence intervals (CI).  Each method operates over a user-defined window size $W$ and includes an additional sensitivity parameter (e.g., threshold or confidence level) that governs detection strictness.  Viewing options include whether or not to highlight the location of anomalies in addition to viewing the selected base signal, and selected threshold cutoff lines.  In all three cases the \texttt{Anomalies} may be autolabeled by selecting \texttt{Capture Anomalies into Proposed Labels}; at which point labels will appear in ``Current Proposed Labels" under the \texttt{Auto Labeler} tab, where clicking \texttt{Confirm Autolabels for ALL Records} will save them.  To export the resulting labels click the \texttt{Current Labels} button in the main control panel tab.

\begin{enumerate}
\item Moving-Window $z$-Score Deviation:  
this method computes, for each time index $t$, the standardized deviation of the current sample $x_t$ from the mean $\mu_t$ of the preceding $W$ samples:
\begin{equation}
z_t = \frac{x_t - \mu_t}{\sigma_t}, \quad 
\mu_t = \frac{1}{W} \sum_{i=t-W+1}^t x_i, \quad 
\sigma_t = \sqrt{\frac{1}{W-1} \sum_{i=t-W+1}^t \left(x_i - \mu_t\right)^2}.
\end{equation}
An anomaly is flagged whenever $|z_t| > \tau_\sigma$, where $\tau_\sigma$ is the user-specified $\sigma$-threshold.  
\emph{Pros:} Interpretable, computationally efficient, and effective for stationary signals with approximately Gaussian noise.  
\emph{Cons:} Sensitive to nonstationarity and outliers; performance degrades when the baseline distribution changes rapidly.  
For a review of $Z$-score-based detection in time series, see \citep{chandola2009anomaly}.

\item Moving-Window CUSUM Chart:
the cumulative sum (CUSUM) method detects small, persistent shifts in the mean of a process by recursively computing:
\begin{align}
C_t^+ &= \max\left[0, C_{t-1}^+ + (x_t - \mu_0 - k)\right], \\
C_t^- &= \max\left[0, C_{t-1}^- + (\mu_0 - x_t - k)\right],
\end{align}
where $\mu_0$ is the nominal process mean (estimated from the window), $k$ is the reference value controlling sensitivity, and $C_t^+$/$C_t^-$ track positive and negative mean shifts, respectively.  
An alarm is raised when $C_t^+ > h$ or $C_t^- > h$, where $h$ is the user-specified threshold.  
\emph{Pros:} Highly sensitive to small, sustained changes; robust for quality-control-type monitoring.  
\emph{Cons:} Requires a good estimate of the nominal mean; less effective for highly transient or oscillatory events.  
For applications and theory, see \citep{page1954continuous,montgomery2009statistical}.
\item Moving Average with Confidence Intervals:  
this method computes a rolling mean $\mu_t$ over a window $W$ and constructs a $(1-\alpha)$ confidence interval assuming approximate normality:
\begin{equation}
\text{CI}_t = \mu_t \pm z_{1-\alpha/2} \cdot \frac{\sigma_t}{\sqrt{W}},
\end{equation}
where $z_{1-\alpha/2}$ is the quantile of the standard normal distribution corresponding to the desired CI level (user-specified), and $\sigma_t$ is the rolling standard deviation. Points outside $\text{CI}_t$ are flagged as anomalies.  
\emph{Pros:} Statistically rigorous, interpretable bounds; adjustable false positive rate through $\alpha$.  
\emph{Cons:} Assumes approximate normality and independent samples within the window; performance may degrade for autocorrelated or heavy-tailed noise.  
A statistical foundation for CI-based monitoring can be found in \citep{james2013introduction}.
\end{enumerate}

\noindent\emph{Role in Fusion Data Pipelines:}  
while each of these methods has distinct strengths, their statistical assumptions (e.g., stationarity, Gaussian noise) may be violated in fusion datasets characterized by multi-regime behavior, non-linearities, and mixed-mode fluctuations. In practice, they can be made more robust by combining them with physics-informed preprocessing (Section~\ref{sec:phys-inf}) or by embedding them in hybrid expert-in-the-loop pipelines where human operators verify algorithmic detections. As components of dFL, these detectors allow rapid, first-pass identification of candidate events for subsequent expert refinement or ML-driven classification.

\subsubsection{Classifier-based Autolabeling}
\label{sec:classifier_autolabeling}

Classifier-based autolabeling leverages supervised machine learning (ML) models to automatically assign labels to incoming data streams, making it particularly powerful when high-quality, expert-labeled datasets already exist or when pre-trained classifiers have been developed for the signals of interest. In dFL such models can be seamlessly integrated through the data ingestion script, enabling automated label generation on both archival datasets and real-time diagnostic feeds.

In dFL the \texttt{Plasma Mode Autolabeler} is included under the \texttt{Auto Labeler} tab, in the \texttt{Select Autolabeler} dropdown menu, as shown in Figure \ref{fig:pmclass-al}.  The record may then be autolabeled by simply clicking \texttt{Autolable Record}.  The resulting labels are stored in the \texttt{Current Proposed Labels} section, where confirming the autolabels will save them to the \texttt{Main Control Panel} tab for analysis and/or export.  This autolabeler is specifically designed to identify plasma modes in DIII-D data, and will be discussed separately in Section \ref{sec:case2}. More broadly, any classification-based autolabeler may be integrated into dFL, which will be discussed in more detail in Section \ref{al-custom}. Below is a brief overview of types of classification models that can be readily incorporated into dFL's autolabeler options. \\

\noindent\emph{Mathematical Formulation}: given an input signal segment $\mathbf{x} \in \mathbb{R}^d$ (where $d$ represents the dimensionality of the feature space, which may include time-domain, frequency-domain, or multimodal fusion features), a classifier implements a mapping:
\begin{equation}
f_\theta: \mathbb{R}^d \to \{1, 2, \dots, K\},
\end{equation}
where $K$ is the number of possible classes (labels), and $\theta$ denotes the model parameters learned from training data $\{(\mathbf{x}^{(i)}, y^{(i)})\}_{i=1}^N$. The predicted class is:
\begin{equation}
\hat{y} = \arg \max_{k} \; p_\theta(y = k \mid \mathbf{x}),
\end{equation}
where $p_\theta(y = k \mid \mathbf{x})$ is the model’s estimated posterior probability for class $k$. These probabilities can also be used to define confidence scores for downstream filtering or expert review. \\

\noindent\emph{Supported Model Types:} a wide range of classifiers may be deployed in dFL, including:
\begin{itemize}
    \item \textit{Linear Models:} Logistic regression, linear discriminant analysis (LDA) — interpretable and efficient, suitable for low-dimensional or linearly separable feature spaces \citep{bishop2006pattern}.
    \item \textit{Tree-based Models:} Random forests, gradient boosted trees — robust to heterogeneous features and noise, often requiring minimal feature scaling \citep{friedman2001elements}.
    \item \textit{Kernel Methods:} Support vector machines (SVM) with non-linear kernels — effective for complex decision boundaries in moderate-dimensional spaces \citep{cortes1995support}.
    \item \textit{Neural Networks:} Fully connected feedforward networks, convolutional neural networks (CNNs) for spatial/temporal patterns, and recurrent neural networks (RNNs) or transformers for sequential dependencies \citep{lecun2015deep,vaswani2017attention}.
\end{itemize}

\noindent\emph{Pros and Cons:} classifier-based autolabeling offers several important advantages. Once trained, such models enable rapid, large-scale labeling, making it feasible to process vast quantities of fusion data with minimal human intervention. They can capture complex, multi-dimensional dependencies among features---relationships that are often inaccessible to simple threshold-based or rule-driven methods. Furthermore, the probabilistic outputs of many classifiers provide confidence scores that can be used for probabilistic label validation, enabling human-in-the-loop workflows where experts focus their attention on lower-confidence or ambiguous cases. These strengths, however, are balanced by notable limitations. Achieving high accuracy typically requires substantial quantities of high-quality labeled training data, unless techniques such as transfer learning are employed to adapt models from related domains. Classifiers can also be vulnerable to dataset shift, where performance degrades if diagnostic conditions evolve or the system operates in previously unseen regimes. Finally, compared to physics-informed or purely statistical detectors, many classifiers (particularly deep neural networks) may lack interpretability, making it more challenging to directly link classification decisions to underlying plasma physics. \\

\noindent\emph{Role in Fusion Data Pipelines:} classifier-based autolabeling is particularly valuable for tasks such as identifying instability types from magnetic coil arrays, classifying plasma shapes from real-time EFIT reconstructions, or tagging operational modes from multiple diagnostics. In hybrid workflows, classifier predictions can serve as initial candidate labels, which are then confirmed or corrected by domain experts or further refined by statistical post-processing (cf. Section~\ref{sec:stats_autolabeling}). This synergy allows fusion researchers to combine the scalability of ML with the rigor and interpretability of statistical and physics-informed methods. \\
 
\subsubsection{Physics-informed Autolabeling}
\label{sec:phys-inf}

Physics-informed autolabeling leverages explicit knowledge of the governing equations, conservation laws, and phenomenological models underlying plasma behavior to identify events, regimes, or structural features of interest directly from raw or preprocessed diagnostics. By embedding first-principles physics into the detection algorithms, these methods provide both interpretability and strong generalization within the operational space of interest \citep{raissi2019physics,pavone2023machine}. Such approaches can operate on scalar time series, spatially resolved measurements, or volumetric simulation data, using algorithms tuned to detect physically meaningful features.

Common examples include:
\begin{itemize}
    \item \textbf{ELM detection via magnetic fluctuation envelopes:} using physics-motivated thresholds on the amplitude of edge magnetic fluctuations to detect the onset of Edge Localized Modes (ELMs) \citep{snyder2009edge}.
    \item \textbf{Sawtooth crash identification:} tracking core temperature profiles or $q$-profile evolution to flag abrupt changes associated with sawtooth collapses \citep{porcelli1996model}.
    \item \textbf{Transport barrier detection:} locating sharp gradients in temperature or density profiles, often associated with H-mode or internal transport barriers, via derivative-based criteria \citep{conway2004plasma}.
\end{itemize}

\begin{table}[!ht]
\centering
\caption{Zero-crossing orders $\mathcal{Z}_n$ available in the Archaieus Autolabeler, their mathematical definitions, and representative use-cases.}
\label{tab:zero_crossings}
\renewcommand{\arraystretch}{1.2}
\begin{tabular}{||p{1.cm} | p{2.0cm} | p{9.2cm}||}
\hline
\textit{Order} & \textit{Definition} & \textit{Representative Use-Case} \\
\hline
$\mathcal{Z}_0$ & $f(t) = 0$ & Crossing of reference level (e.g., density perturbations crossing baseline, mode amplitude sign changes). \\ \hline
$\mathcal{Z}_1$ & $f'(t) = 0$ & Detection of local extrema in diagnostic signals (e.g., peak stored energy before disruption). \\ \hline
$\mathcal{Z}_2$ & $f''(t) = 0$ & Inflection point detection (e.g., onset of rapid confinement degradation). \\ \hline
$\mathcal{Z}_3$ & $f^{(3)}(t) = 0$ & Curvature extrema, can signal transition between growth/decay phases of instabilities. \\ \hline
$\mathcal{Z}_4$ & $f^{(4)}(t) = 0$ & Higher-order changes, useful for detecting subtle pre-cursors in mode chirping or nonlinear saturation. \\
\hline
\end{tabular}
\end{table}

The main advantage of physics-informed labeling is that it encodes invariances and scaling laws intrinsic to the plasma, making the output robust to diagnostic noise, moderate parameter drifts, and changes in operating regime. However, such methods can be brittle when underlying physics assumptions break down or when phenomena are driven by unexpected mechanisms not covered by the detection model. \\

\noindent\emph{The Archaieus Autolabeler:} among the physics-informed plugins available in dFL, the \emph{Archaieus Autolabeler} occupies a unique niche by enabling the user to smooth any scalar time-series signal to $C^5$ continuity before performing feature extraction. This degree of differentiability ensures stable and accurate estimation of up to fourth-order temporal derivatives, even in the presence of moderate diagnostic noise \citep{deboor2001practical,holoborodko2008smooth}. Once smoothed, the algorithm computes the zero-crossing sets $\mathcal{Z}_n$ for $n=0,1,2,3,4$, where
\[
\mathcal{Z}_n = \{\, t_i \;|\; f^{(n)}(t_i) = 0 \,\}
\]
denotes the set of times at which the $n$-th derivative vanishes.

\begin{figure}
    \centering
        \includegraphics[width=0.99\linewidth]{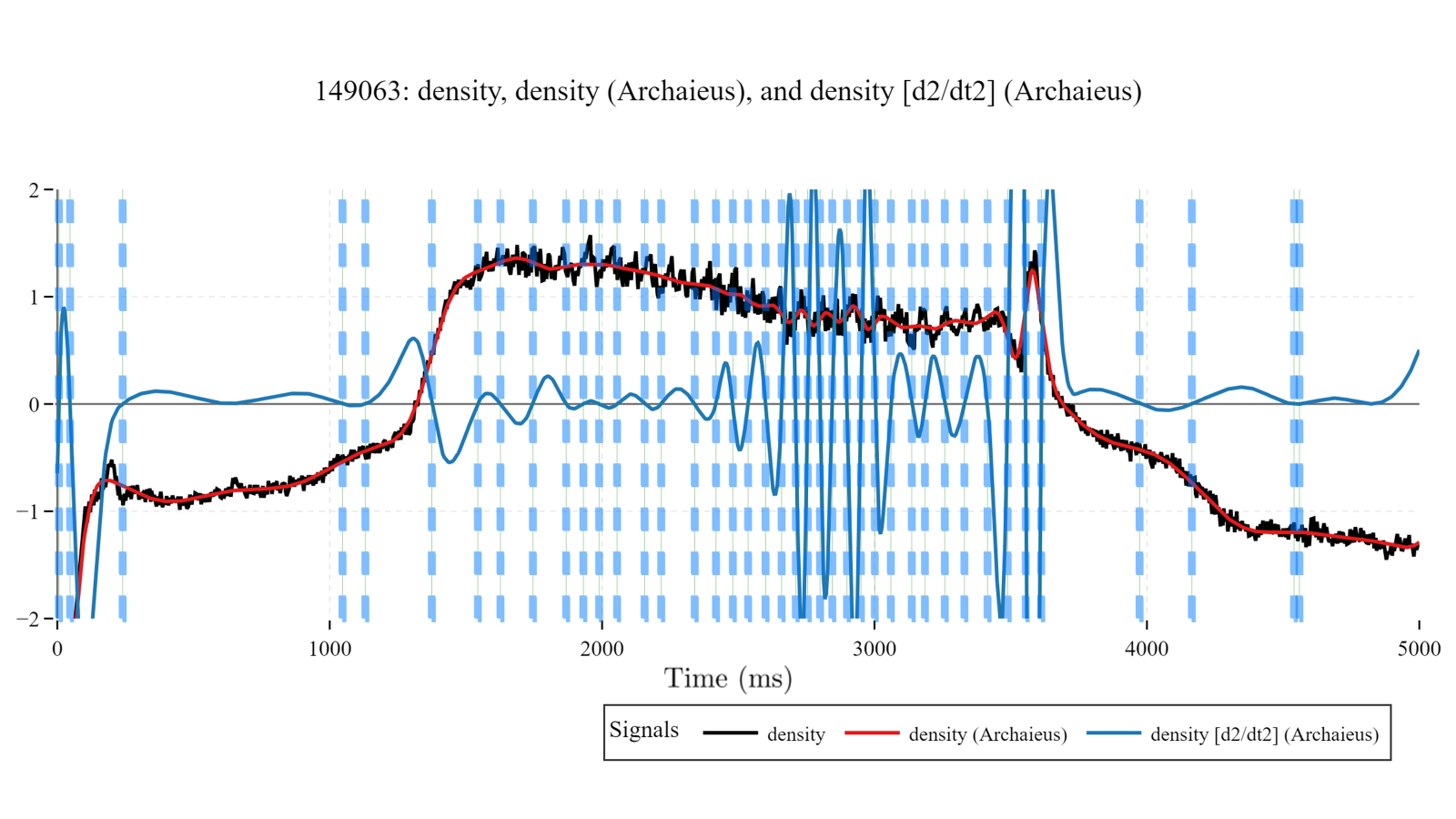}
    \caption{Inflection points automatically detected and selected (e.g., for labeling) with blue-dashed lines using the zero point crossing autolabeler derived from the second derivatives of the plasma density (i.e. $\partial^{2} n/\partial t^2=0$), after $z$-score normalizing all  then denoising the density using the Archaieus smoother (dimensionless units).}
    \label{fig:inflection}
\end{figure}

These zero crossings provide a structured hierarchy of dynamical markers that can reveal subtle precursor phenomena, oscillatory regime changes, or shifts in waveform morphology that might otherwise remain undetected. Table~\ref{tab:zero_crossings} summarizes their definitions, physical interpretations, and representative fusion-related use-cases.

Because higher-order derivatives amplify noise, the $C^5$ smoothing stage is essential: it suppresses spurious oscillations without attenuating physically relevant features at the chosen derivative order. Lower-order zero crossings are generally robust and suited to coarse event detection, while higher-order crossings are best reserved for well-conditioned signals or post-processed simulation outputs where derivative stability is assured.

Physics-informed methods, including tools like the \emph{Archaieus Autolabeler}, thus combine domain knowledge with algorithmic precision, offering interpretable, reproducible, and operationally useful event detection capabilities that remain aligned with the underlying plasma physics.

\subsection{Integrating Custom Autolabelers into dFL}\label{al-custom}

The autolabeling system provides a flexible framework for automatically detecting and labeling temporal events in time-series, or sequential, data. The architecture follows a modular design where domain-specific autolabeling functions are registered with a central coordinator that manages data access, parameter handling, and result aggregation. The system supports both single-shot and bulk processing modes, with the ability to apply custom algorithms to multiple data records simultaneously.

The core autolabeling workflow consists of three main components: (1) a \texttt{data coordinator} API that abstracts data access patterns, (2) a registry of autolabeling functions that implement domain-specific detection algorithms, and (3) a callback system that integrates with the web interface for user interaction. Each autolabeling function follows a standardized signature that accepts dataset identifiers, shot IDs, a data coordinator reference, and additional parameters. The system automatically handles parameter validation, data fetching, and result formatting, allowing researchers to focus on implementing their detection logic rather than infrastructure concerns.

The autolabeling functions are defined by the user and can implement various detection strategies, from simple signal processing methods like threshold crossing and peak finding, to more sophisticated algorithms that can use machine learning to classify the data. As an example we show a threshold autolabeling function that detects events by monitoring when signal values cross user-defined thresholds.  A code example of how to write such a custom Python autolabeler is presented in \ref{AppD}.  Extensive examples and demos are available in the  \href{http://dfl.sophelio.io/documentation}{dFL documentation}.

\section{Fusion Energy Case Studies}

Below are a set of example use-cases from DIII-D data provided through the fully integrated TokSearch backend integration. The following Table shows the case studies examined in this section, with a high level overview of how dFL impacts the relevant workflows.  Beyond single-shot demonstrations, each workflow can be executed in dFL's bulk mode over \(\mathcal{O}(10^6)\) discharges in the DIII-D archive via TokSearch-backed iterators, producing machine-actionable labels and summary statistics that enable systematic indexing, search, and meta-analysis across campaigns. These labels can also be exported in IMAS/OMAS schema for cross-device training and benchmarking, thereby supporting multi-facility generalization.

\begin{table}[!ht]
\centering
\footnotesize
\renewcommand{\arraystretch}{1.15}
\setlength{\tabcolsep}{4.5pt}
\textbf{Summary of Fusion Energy Case Studies} \\[6pt]
\label{tab:case_studies}

\begin{adjustbox}{max width=\linewidth}
\begin{tabular}{|p{3.2cm}|p{3.2cm}|p{3.2cm}|p{3.2cm}|}
\hline
\rowcolor[gray]{0.9}
\textbf{Case Study} & \textbf{Methodology} & \textbf{Key Challenges} & \textbf{Efficiency Gain} \\
\hline
\emph{Automated ELM Identification} &
Moving-window $z$-score and slope-based outlier detection applied to filterscope \(D_{\alpha}\) signals. &
Variability in amplitude/shape, diagnostic saturation, baseline drift, high-frequency noise. &
Reduced inspection from hours per shot to minutes in DIII-D ELM detection pilot project; $>$50X faster than manual labeling with dFL. \\
\hline
\emph{Manual Plasma Mode Labeling} &
Expert-in-the-loop classification using magnetics, filterscopes, BES, and pedestal widths. &
Requires expert judgment; labor-intensive inspection of multimodal diagnostics; subjectivity in borderline cases. &
Baseline for comparison; typically days of expert time to label $\mathcal{O}(100)$ discharges. dFL improves by 5--8X. \\
\hline
\emph{Autolabeling of Plasma Modes} &
XGBoost multiclass classifier trained on $\sim$792k samples from 360 shots. &
Differentiating QH, BBQH, WPQH regimes; class imbalance; subtle mode signatures; reproducibility. &
Classifies hundreds of discharges in minutes; $\sim50$X faster than expert manual classification with dFL. \\
\hline
\end{tabular}
\end{adjustbox}
\end{table}

\subsection{Case Study: Automating the Labeling of Edge-Localized Modes (ELMs)}

The dataset for this case study is sourced from the DIII-D National Fusion Facility and retrieved via the TokSearch interface \citep{toksearch_docs} and Section \ref{toksearch}, which provides structured, programmatic access to large volumes of experimental signals.  
For post-discharge ELM identification, the primary diagnostic is the filterscope system, which measures spectrally filtered \(D_{\alpha}\) emission from the plasma edge \citep{burrell2001}.  
These measurements offer high temporal resolution and strong sensitivity to the rapid light bursts associated with ELM events, making them well-suited for automated detection.  
Key challenges include shot-to-shot variability in signal amplitude and waveform shape due to changes in plasma configuration, fueling, and control schemes, as well as occasional diagnostic saturation, baseline drift, and data dropouts.  
Manual labeling can be subjective and inconsistent across operators, motivating the development of automated threshold-based and machine learning-assisted labeling workflows to provide consistent, reproducible identification across large datasets. 

Several recent studies provide robust, peer-reviewed benchmarks for automated ELM identification.  O’Shea et al.\ demonstrated a parameter-free statistical algorithm for detecting ELMs in DIII-D filterscope data, achieving over 97\% precision and recall across hundreds of discharges \citep{o_shea2023automatic}.  In parallel, Song et al.\ developed a convolutional neural network–based detector for KSTAR that achieved comparable accuracy without hyperparameter tuning \citep{song2023automatic}.  

The present dFL implementation does not aim to replace these established algorithms but rather to integrate them within a unified, provenance-aware framework that also supports statistical and slope-based detectors.  This enables consistent execution and comparative benchmarking across devices and diagnostics.  A formal benchmark of dFL’s built-in ELM detectors against these state-of-the-art approaches is planned as future work.

At scale, the same detectors can be swept over the full DIII-D archive to produce per-shot ELM corpora (onset/offset times, waiting-time distributions, burst statistics) stored with provenance. These labels support archive-wide queries (e.g., ``find shots with high ELM frequency at fixed \(q_{95}\)’’), enable data-balanced training sets for disruption/ELM predictors, and systematically organize under-studied discharges rather than focusing on a few canonical shots. Beyond local usage, the same corpora can be exported in IMAS/OMAS schema to seed cross-device generalization (e.g., training predictors simultaneously on DIII–D and NSTX–U), and to align experimental ELM intervals with synthetic diagnostics from nonlinear MHD simulations for model validation.

\begin{figure}
    \centering
    \includegraphics[width=0.99\linewidth]{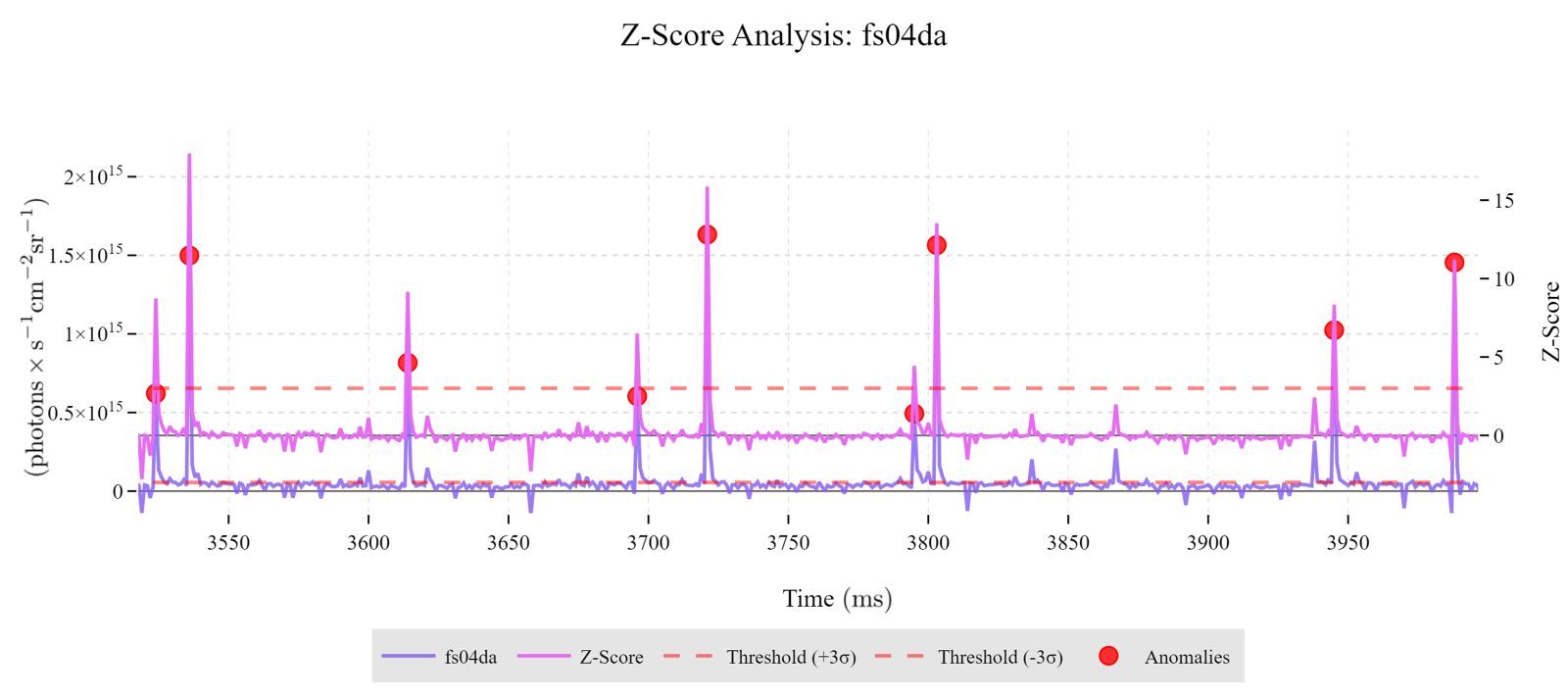}
    \caption{Automatically Identifying ELMs from filterscope data using the simple built-in $z$-score deviation detector from Section \ref{sec:stats_autolabeling}, setting \texttt{Window Size}=1000 and \texttt{Threshold}=3 standard deviations, shot \#149092.}
    \label{fig:elms}
\end{figure}

To autolabel ELMs using filterscope data we have a number of options natively supported in dFL. The simplest (and surprisingly effective approach) is to utilize a statistics-based autolabeler. More specifically, using the moving-window $z$-score deviation detector from Section \ref{sec:stats_autolabeling}, we can easily identify ELMs as simple statistical anomalies of the filterscope signal. Figure \ref{fig:elms} shows that all ELMs are identified as statistical outliers of the signal. The remaining challenge, in order to automate the labeling over all ELMy shots then becomes how to set the statistical autolabeler’s free parameters \texttt{Window Size} and standard deviation \texttt{Threshold} over all shots of interest \emph{a priori}. If this cannot be easily accomplished, then an \emph{expert-in-the-loop} system can be performed, allowing the user to set these parameters independently on each shot (i.e. still much faster than manually labeling each ELM). In bulk execution, dFL supports parameter grids (or adaptive parameterization via robust statistics of each shot) and writes out per-shot quality metrics so that archive-wide sensitivity analyses are possible. For real-time application, the same $z$-score detector can be implemented causally (one-sided window) with bounded latency and deployed as a lightweight module running alongside the Plasma Control System (PCS). This module would stream real-time event trigger signals (e.g., imminent ELM onset) that the PCS can subscribe to, enabling automated responses such as pellet pacing or RMP coil actuation, or alternatively provide a feature stream for model predictive control (MPC) that incorporates predicted ELM likelihood into constraint handling. Additional downstream applications include mining rare-event statistics (e.g., compound bursts or compound ELM-RMP interactions), generating benchmark datasets for ML training, and feeding uncertainty-aware ELM probability traces into Bayesian fusion frameworks that propagate label uncertainty through to disruption forecasts.

\subsubsection{Case Study: Automating ELM Labeling with Slope-Based Outlier Detection}
\label{sec:slope_elm}

An alternative approach to automated ELM identification leverages slope-based outlier analysis of the filterscope \(D_{\alpha}(t)\) signal, and is available in dFL natively or as a script provided as input into the dFL data provider found in the \href{http://dfl.sophelio.io/documentation}{dFL documentation}. Previous work by Eldon and Xing \citep{xing2020cake} demonstrated that rapid variations in filterscope \(D_{\alpha}\) signals could be used to detect and filter ELMs for equilibrium reconstruction. Building on this idea, the method presented here detects events by examining the time derivative of the filtered signal. Large positive spikes in \(D_{\alpha}(t)\) correspond to the onset of an ELM, while sharp negative slopes denote the termination of a burst. These features are statistically evaluated using $z$-scores of the positive and negative derivative distributions computed over the entire shot, allowing the algorithm to robustly identify start and end points. In this way, slope-based detection complements amplitude-based methods by focusing on the transient growth and crash dynamics that characterize ELM activity, providing an alternative approach that emphasizes slope dynamics rather than amplitude.

Slope-based methods are inherently more sensitive to high-frequency noise, so preprocessing tools are provided to help users improve detection reliability. Low-pass filtering and signal smoothing (mean, median, exponential moving average, or Gaussian) allow users to reduce high-frequency fluctuations before computing slopes, with the low-pass cutoff frequency typically selected by an expert in the loop to balance noise suppression and temporal resolution. Once ELMs are identified, an ELM merging window can be applied to combine closely spaced bursts into a single event. This reflects the physical behavior observed in strongly ELMy plasmas, where regions often produce frequent, small-amplitude events that are best interpreted as part of a single extended ELMing episode rather than independent bursts. Together, these options give users the flexibility to tailor detection to the characteristics of each shot, producing reproducible, interpretable ELM labels while preserving the underlying dynamics captured by slope-based analysis. Full details of the algorithm and code are available in the \href{http://dfl.sophelio.io/documentation}{dFL documentation}. 

For archive-scale processing, the derivative-based detector can be run in batch with per-shot adaptive smoothing (e.g., bandwidth set by noise floors estimated from non-ELMy intervals), yielding consistent onset/offset annotations across hundreds of thousands of discharges and enabling population-level statistics (e.g., \(D_{\alpha}\) slope distributions conditioned on fueling or \(q_{95}\)).  Such systematic slope-based catalogs can also be used to construct archive-level scaling laws for ELM crash dynamics and to cross-validate synthetic slope signals from nonlinear MHD simulations, etc.

\subsection{Case Study: Automated Plasma Mode Identification} \label{sec:case2}

The dFL provides the fusion data community with an automated labeling (autolabeler) capability built on an Extreme Gradient Boosting (XGBoost) classifier \citep{chen2016xgboost}, trained on a curated dataset comprising approximately 792{,}000 samples drawn from 360 plasma shots. The classifier is designed to identify three distinct advanced tokamak confinement regimes of quiescent H-mode (QH-mode) operation.
As widely acknowledged in the magnetically confined fusion community, bursty intermittent disruptive events, known as edge-localized modes (ELMs), eject uncontrollable particle and heat flux, leading to severe damage to plasma-facing components. One of the most promising candidates for ELM-free operation is the quiescent H-mode (QH-mode), which preserves good overall confinement performance while maintaining ELM suppression.  

In DIII-D, three types of QH-mode have been identified: (\emph{i}) \textbf{QH-mode}---the canonical quiescent H-mode, characterized by the absence of type-I ELMs and sustained edge harmonic oscillations (EHOs) \citep{Burrell2002,Burrell2009}; (\emph{ii}) \textbf{BBQH-mode}---broadband QH-mode, exhibiting a broad spectral signature of edge fluctuations rather than a narrow EHO peak, yet retaining the ELM-free condition \citep{Chen2023}; and (\emph{iii}) \textbf{WPQH-mode}---wide-pedestal QH-mode, obtained by reducing external injected torque and marked by unusually broad pedestal widths in edge temperature and density profiles while maintaining quiescence \citep{Burrell2016,Chen2017}. These regimes are of particular interest due to their potential for enabling high-performance, ELM-free operation in future reactors. Understanding the entry conditions and key parameters for accessing these regimes is therefore of great importance.  

It is also worth noting that while QH-mode remains a leading candidate for sustained ELM-free operation, the expert-in-the-loop methodology developed here is not restricted to this specific regime.  A similar workflow in dFL can be extended to other confinement and exhaust-handling states that are of growing interest for reactor design and control optimization.  For example, I-mode operation (first identified on Alcator C-Mod) achieves improved energy confinement without triggering ELMs, characterized by a strong edge temperature pedestal but weak particle confinement \citep{Whyte2010}.  Automated and reproducible identification of I-mode intervals using dFL would enable large-scale, cross-device studies of pedestal formation and access conditions, advancing predictive models for next-generation devices.  Likewise, systematic labeling of divertor detachment states, which play a central role in managing heat and particle fluxes to plasma-facing components \citep{Leonard2018}, could facilitate data-driven training of control algorithms for advanced exhaust scenarios.  Together, these extensions emphasize that dFL provides a generalizable infrastructure for supervised regime identification and expert-in-the-loop training across confinement and detachment states—not solely for QH-mode classification.

Manual labeling of QH-mode discharges in DIII-D has been performed using the extensive suite of diagnostics, and forms the basis of the experimental QH database covering 2012--2018. The labeling procedure is as follows:  
\begin{enumerate}
    \item Verify plasma density, current, and magnetic field to confirm a valid plasma shot in H-mode;  
    \item Inspect \(D_{\alpha}\) signals from filterscopes to identify periods with ELM activity;  
    \item Check the pedestal width and compute the ratio between the measured pedestal width and the conventional pedestal width scaling \citep{Snyder2011}. A ratio above 1.25 is labeled as WPQH;
    \item  If the pedestal width ratio is lower than 1.25, analyze magnetics data to determine the presence of coherent harmonic modes and confirm edge localization with other diagnostics, such as Beam Emission Spectroscopy (BES). If an edge harmonic oscillation (EHO) is confirmed, the discharge is labeled as QH-mode; otherwise it will be labelled as BBQH; 
    \item Perform cross-checks with other key signals (e.g., total injected torque, edge rotation) to confirm the consistency of the classification.  
\end{enumerate}

\begin{figure}[!ht]
    \centering
    \includegraphics[width=0.80\linewidth]{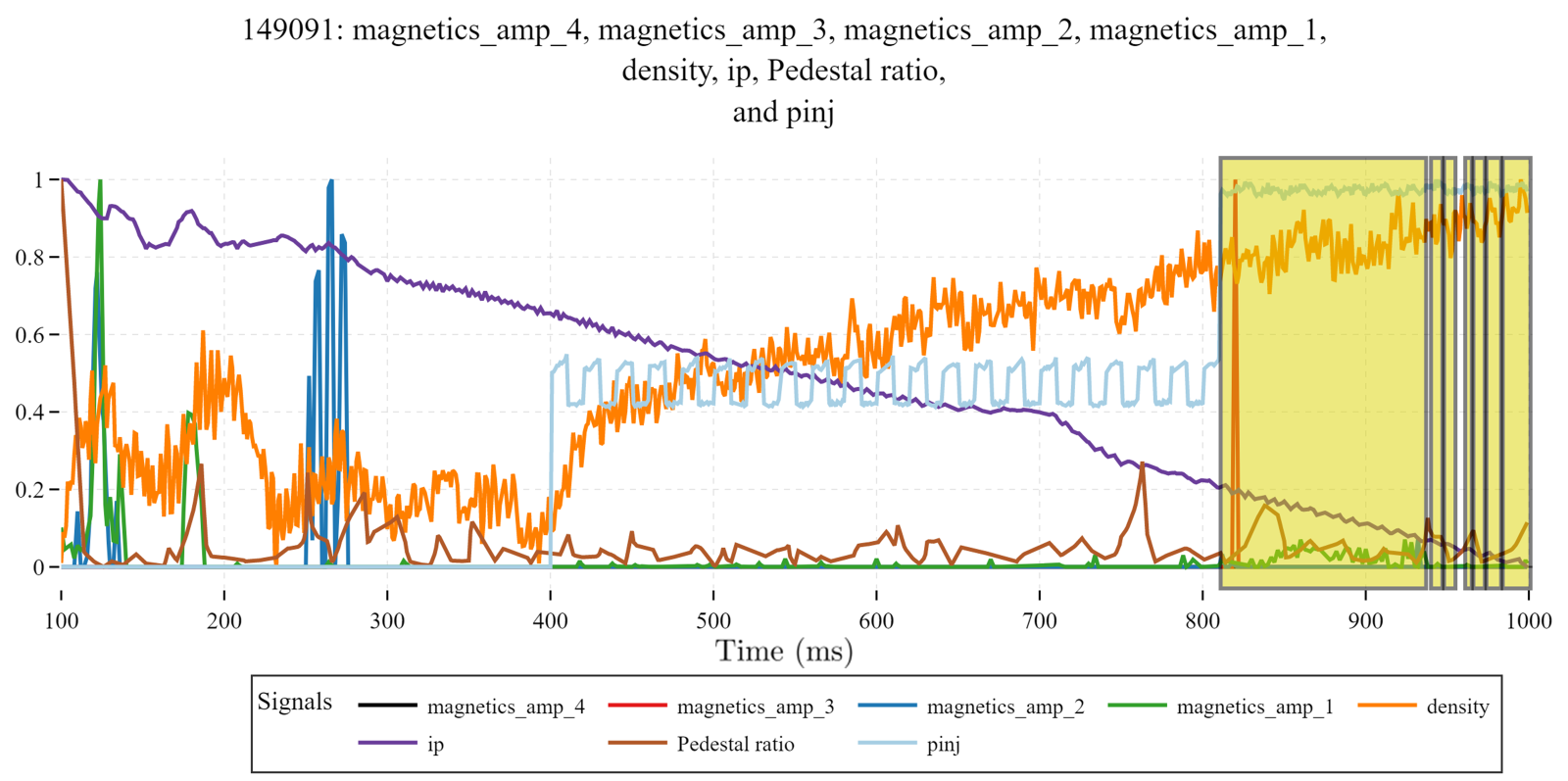}\\
        \includegraphics[width=0.80\linewidth]{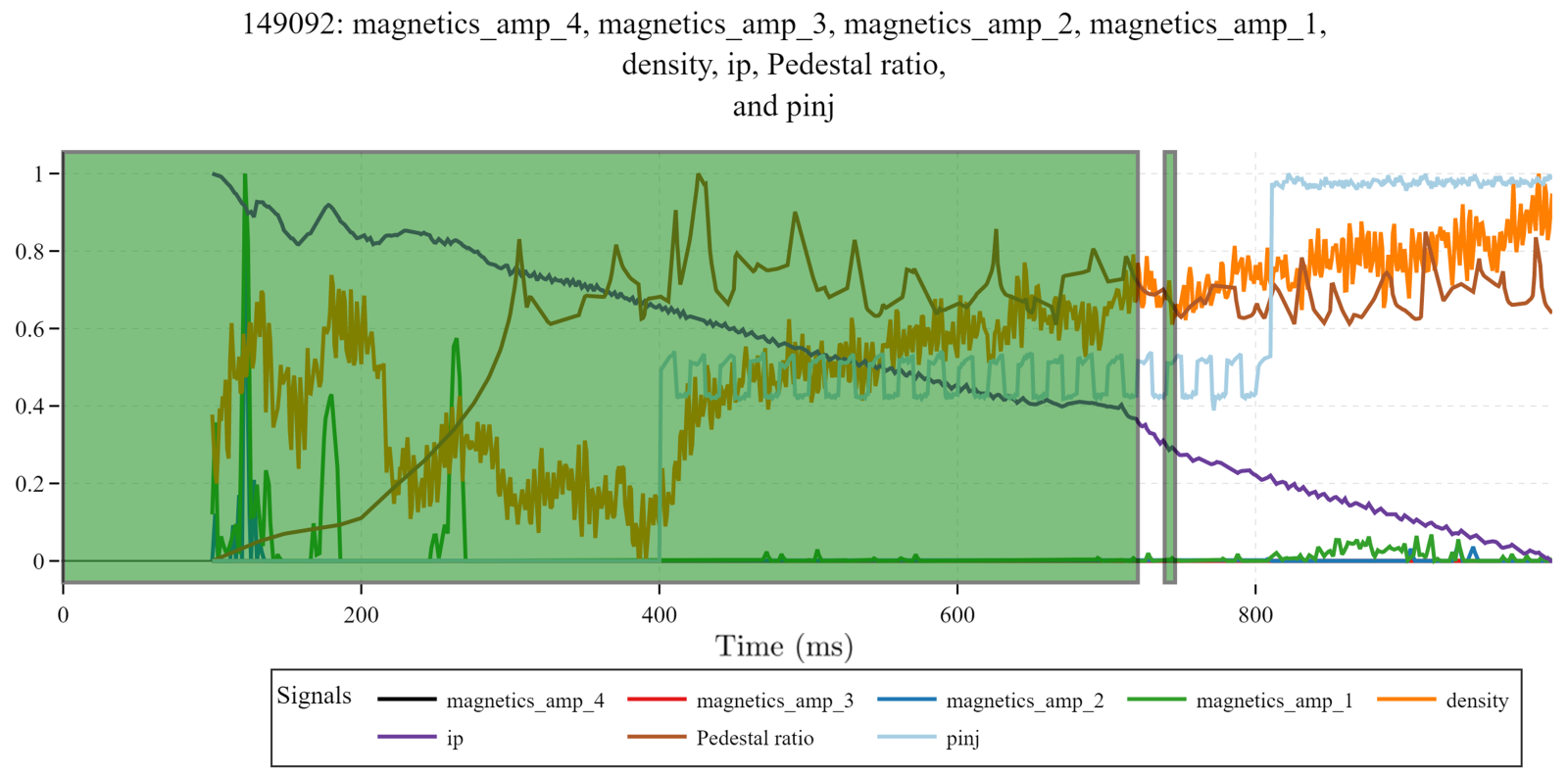}\\
            \includegraphics[width=0.80\linewidth]{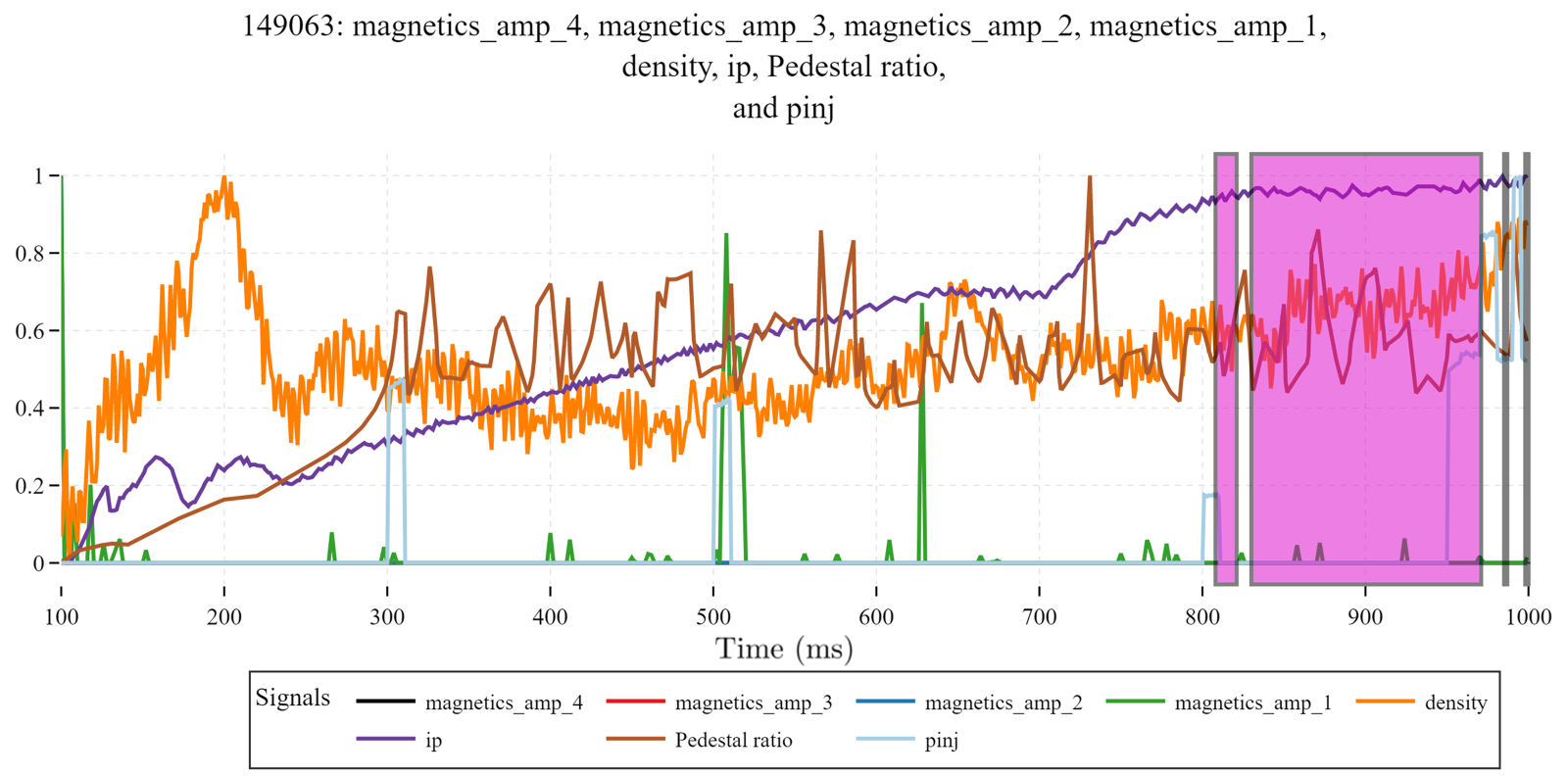}
    \caption{Automated Plasma Mode Identification on three separate shots/discharges (zoomed in) from DIII-D, all signals feature scaled (Min-Max normalized), where the top (yellow labeled section) is QH mode, the middle (green labeled section) BBQH mode, and the bottom (magenta labeled section) WPQH mode (units dimensionless).}
    \label{fig:plasmamode}
\end{figure}

The labeled dataset was constructed using an expert-in-the-loop approach, in which experienced plasma physicists manually classified operating regimes from multiple DIII-D diagnostics (including magnetic fluctuations, reflectometry, and pedestal profile data) following the established mode-identification criteria above. This process ensured that the training set reflected accurate, domain-validated mode boundaries, including cases with subtle or borderline signatures.

The machine learning pipeline employs a multiclass classifier using the `multi:softprob' objective in XGBoost \citep{chen2016xgboost}, which outputs a probability distribution over the three classes for each sample, with model evaluation based on the multiclass logarithmic loss (mlogloss). Training was performed using a 5-fold GroupKFold cross-validation strategy, grouping by shot identifier to prevent data leakage between training and validation folds. To address moderate class imbalance, balanced class weights were computed with the \texttt{compute\_class\_weight} utility from scikit-learn \citep{pedregosa2011scikit}. Hyperparameters were tuned to promote generalization, including L2 regularization (\texttt{reg\_lambda}=0.5), L1 regularization (\texttt{reg\_alpha}=0.1), a maximum tree depth of \texttt{max\_depth}=5, a minimum split loss (\texttt{gamma}) of 0.1, and a \texttt{min\_child\_weight} of 3. Early stopping was applied to prevent overfitting.

Feature preprocessing included renaming columns into a $\{f_0, f_1, \dots, f_n\}$ format to ensure compatibility with the Open Neural Network Exchange (ONNX) serialization standard \citep{onnx}, while retaining a mapping to the original physical feature names for interpretability. Final evaluation across the complete dataset, using the best-performing cross-validation model, achieved an overall classification accuracy exceeding 90\%, indicating robust performance across diverse plasma conditions. Deployed across the full DIII-D archive, the autolabeler yields a time-indexed catalog of QH/BBQH/WPQH intervals with calibrated class posteriors, enabling large-scale mining of entry/exit conditions, actuator settings, and robustness margins; this moves the community beyond a handful of well-studied shots to statistically grounded conclusions over \(\gg 10^6\) discharges. For operationalization, the ONNX-exported classifier can run as a low-latency PCS sidecar that streams the posterior probability of the mode given the most recent diagnostic features $\mathbf{x}_t$, i.e., \(p(\text{mode}| \mathbf{x}_t)\), over each control cycle; MPC can then incorporate these posteriors to enforce mode-aware constraints (e.g., avoiding ELM-prone regimes) or to track toward QH targets by adjusting torque, fueling, or RMP settings in closed loop. In addition, probabilistic mode posteriors enable uncertainty-aware control (risk-weighted constraints), support cross-device transfer learning when exported in IMAS schema, and populate archive-level regime maps (e.g., pedestal-width vs.\ torque scans) that reveal physics trends across campaigns. These datasets can further serve as benchmark corpora for the community, accelerating the development of standardized ML pipelines in fusion.

Taken together, these case studies illustrate the dual role of dFL autolabeling pipelines: (i) enabling archive-scale mining that systematically organizes hundreds of thousands of discharges across DIII-D (and beyond, when exported in schema-compliant formats), and (ii) providing causal, low-latency outputs that integrate directly into PCS and MPC workflows. Beyond immediate classification, they open downstream opportunities for simulation-experiment fusion, rare-event discovery, cross-device benchmarking, uncertainty-aware control, and the creation of standardized benchmark datasets that catalyze community-wide ML development.

\section{Discussion}


The results presented here demonstrate that data harmonization, fusion, and provenance in fusion energy science arise not only from software design choices but from the underlying physics and engineering constraints of confined plasmas. 
The diversity of temporal and spatial scales, the heterogeneity of diagnostic and simulation modalities, and the nonstationary nature of operating conditions impose constraints that shape every stage of the information pipeline. In this environment, harmonization is not an optional preprocessing step; it is the means by which raw measurements are transformed into a physically coherent basis suitable for inference, control, and reproducibility.

From a systems perspective, the Data Fusion Labeler (dFL) occupies a strategic position in this pipeline. Its capacity to ingest multimodal, asynchronous data streams, align them in both temporal and spatial coordinates, and normalize them into schema-compliant structures ensures that downstream fusion and labeling operations are physically and statistically consistent. This capability is particularly impactful in scenarios where data imbalance and sparsity of high-value events---such as disruptions, ELMs, or internal transport barrier formation---would otherwise hinder supervised learning or bias automated detection. By embedding uncertainty-aware harmonization early, dFL preserves the subtle signal features that are often diagnostic of rare events, enabling their recovery in subsequent fusion and classification stages.

The integration of automated and physics-informed labeling modules within a provenance-aware framework further addresses two persistent challenges in fusion data science: scalability and reproducibility. Automated labeling pipelines, whether statistical, classifier-based, or simulation-driven, accelerate the curation of large event corpora, while data provenance capture ensures that every label can be traced to its originating signals, processing steps, and algorithmic parameters. This linkage is crucial for scientific auditability; without it, labeled datasets risk becoming ``orphaned'' from their context, undermining cross-device benchmarking and model transferability. Conversely, when labels are embedded in a harmonized, versioned data context, they can be federated across facilities and campaigns, contributing to a shared, interoperable knowledge base.

The implications extend beyond the immediate case studies. The architectural principles embodied in dFL---operator-order awareness, multimodal synchronization, schema conformity, and integrated provenance---are directly portable to other high-consequence, sensor-rich scientific domains, from climate modeling to high-energy physics. Within fusion, these capabilities form the connective tissue between advanced diagnostics, real-time control, and predictive modeling, enabling a closed-loop \emph{measure $\rightarrow$ harmonize $\rightarrow$ fuse $\rightarrow$ label $\rightarrow$ learn/control $\rightarrow$ reproduce} cycle. As devices approach reactor-relevant performance, and as campaigns become increasingly multi-institutional, such integrated, provenance-rich data systems will be indispensable for both operational safety and accelerated physics discovery.

In summary, the present work demonstrates that the rigorous, principled treatment of harmonization, fusion, and labeling---supported by a robust provenance model---is not simply good data hygiene, but a scientific necessity in fusion energy research. The dFL provides a concrete, operational realization of these principles---reducing time-to-analysis, improving label quality, and enabling reproducible--- producing physics-aligned workflows that will be critical for the data-driven control and optimization of future burning plasma devices.

\section{Conclusion}

The Data Fusion Labeler (dFL) addresses a fundamental gap in fusion energy research workflows by unifying data harmonization, data fusion, and data provenance into a coherent, operationally deployable system. Modern fusion devices and their multiscale simulations generate petabyte-scale, heterogeneous, and temporally asynchronous datasets whose scientific value cannot be realized without rigorous preprocessing and alignment. By implementing uncertainty-aware harmonization at ingestion, embedding both manual and automated labeling within a schema-compliant architecture, and capturing complete provenance metadata, dFL ensures that downstream analysis, machine learning, and control tasks operate on physically consistent, reproducible data streams.

The case studies presented demonstrate that such integration is not merely a convenience, but a scientific enabler. Automated detection of ELMs, classification of advanced confinement regimes, and simulation-informed event identification all benefit from a pipeline in which alignment, normalization, and label generation are intrinsically linked and version-controlled. This tight coupling reduces time-to-insight from weeks to hours, improves the statistical robustness of predictive models, and facilitates cross-device and cross-campaign comparability.

Looking ahead, the principles embodied in dFL---data harmonization, operator-order awareness, multimodal synchronization, and embedded provenance---are extensible to future fusion facilities and multi-facility collaborations. As burning plasma experiments and reactor prototypes emerge, the demands on data systems will intensify: higher diagnostic densities, greater operational variability, and tighter control loops will require harmonization and labeling at unprecedented scale and speed. The architecture presented here positions dFL to meet these challenges, offering a path toward fully integrated, reproducible, and scientifically transparent fusion informatics pipelines. Beyond its immediate utility, dFL also has the potential to serve as a benchmarking platform for comparing alternative data harmonization and labeling methods, while hosting effectiveness-proven algorithms contributed by the broader fusion community.

In essence, dFL operationalizes the aspirational cycle of \emph{measure $\rightarrow$ harmonize $\rightarrow$ fuse $\rightarrow$ label $\rightarrow$ learn/control $\rightarrow$ reproduce} for the fusion energy community. By doing so, it not only accelerates data-to-insight conversion, but also establishes a durable framework for knowledge retention, model transferability, and collaborative physics discovery in the era of large-scale, data-driven fusion research.

\section*{Acknowledgments}

This work was partially supported by the U.S. Department of Energy, Office of Science, Office of Fusion Energy Sciences, using the DIII-D National Fusion Facility, a DOE Office of Science user facility, under Award No. DE-FC02-04ER54698, along with Office of Fusion Energy Sciences Awards No. DE-SC0024426 and No. DE-SC0024399.

\section*{Disclaimer}

Disclaimer: This report was prepared as an account of work sponsored by an agency of the United States Government. Neither the United States Government nor any agency thereof, nor any of their employees, makes any warranty, express or implied, or assumes any legal liability or responsibility for the accuracy, completeness, or usefulness of any information, apparatus, product, or process disclosed, or represents that its use would not infringe privately owned rights. Reference herein to any specific commercial product, process, or service by trade name, trademark, manufacturer, or otherwise does not necessarily constitute or imply its endorsement, recommendation, or favoring by the United States Government or any agency thereof. The views and opinions of authors expressed herein do not necessarily state or reflect those of the United States Government or any agency thereof.

During the preparation of this work the author(s) used Large Language Model (LLM) tools as collaborative aids (similar to google search) to refine language, enhance clarity, verify consistency, and explore alternative technical framings across multiple disciplinary contexts in fusion energy science and data harmonization in order to edit, improve, discover, and deepen understanding of the presented topics through an iterative process of human–machine co-working. After using this tool/service, the author(s) reviewed and edited the content as needed and take(s) full responsibility for the content of the published article.

\pagebreak
\appendix
\section{Incorporating Custom Graphs into dFL}\label{AppC}

The custom graphing system within dFL (introduced in Section \ref{vis}) provides a flexible framework for creating domain-specific visualizations that integrate seamlessly with the main labeling application. The architecture follows a plugin-based design where custom graphing functions are registered with a central coordinator that handles parameter management, data access, and user interface generation. The system supports both simple text-based outputs and complex interactive visualizations using Plotly, with automatic parameter validation and dynamic UI generation.

The custom graphing workflow consists of three main components: a registry of graphing functions that implement domain-specific visualization logic, an automatic parameter generation system that creates user interface components based on function specifications, and a callback system that manages the interaction between user inputs and graph updates. Each custom graphing function follows a standardized signature that accepts application control parameters (containing global state like current shot ID, trim values, and theme settings) and function-specific parameters (user-configurable options like multipliers, signal selections, and visualization options).

The system automatically generates user interface components for parameter configuration based on declarative specifications in the function registry. Parameters can be defined with various types (numeric inputs, dropdowns, text fields) and constraints (minimum/maximum values, default values, option lists). The framework supports nested parameter structures, allowing complex configurations with conditional parameter visibility based on user selections. This declarative approach enables rapid development of new visualization types without requiring custom UI code.
Here is a Plotly-based visualization function that imitates Seaborn styling.  For more information on the \texttt{data provider} see \ref{AppA} and the \href{http://dfl.sophelio.io/documentation}{dFL documentation}.

\begin{python}
def seaborn_plotly_example(app_control_parameters, parameters):
    """
    Generate a Plotly line chart with Seaborn-like styling.
    Args:
        app_control_parameters (dict): Global parameters from the application controller
        parameters (dict): Parameters specific to this graphing function
    Returns:
        plotly.graph_objects.Figure: A Plotly figure object
    """
    # Extract user-configurable parameters with defaults
    multiplier = parameters.get("seaborn_plotly_example_multiplier", 1.0)
    signal_option = parameters.get("seaborn_plotly_example_signal_option", "Temperature")
    # Fetch time-series data using the data coordinator
    shot_data = app_control_parameters["data_coordinator"].fetch_data_async(
        app_control_parameters["data_coordinator"].data_folder,
        dataset_id=None,
        shot_id=app_control_parameters["shot_id"],
        signals=[signal_option],
        global_data_params={},
        trim_1=app_control_parameters["trim_t1"],
        trim_2=app_control_parameters["trim_t2"],
        timeout=10,
    )
    # Extract and process signal data
    signal = shot_data["signals"][0]["data"]
    times = shot_data["signals"][0]["times"]
    signal = np.array(signal).flatten() * multiplier
    times = np.array(times).flatten()
    # Create Plotly figure with custom styling
    fig = go.Figure()
    fig.add_trace(go.Scatter(
        x=times, y=signal, mode='lines',
        line=dict(width=2.5), name=signal_option
    ))
    # Apply theme-aware styling
    fig.update_layout(
        title=dict(text=f"{signal_option} Over Time (Multiplier: {multiplier})"),
        xaxis=dict(title="Time", showgrid=True, gridcolor='rgba(0,0,0,0.1)'),
        yaxis=dict(title=signal_option, showgrid=True, gridcolor='rgba(0,0,0,0.1)'),
        template="plotly_white" if not app_control_parameters["theme_value"] else "plotly_dark",
        plot_bgcolor='rgba(0,0,0,0)', paper_bgcolor='rgba(0,0,0,0)'
    )
    return fig
# Function registration in the data provider
custom_grapher_dictionary = {
    "seaborn_plotly_example": {
        "display_name": "Seaborn Plotly Example",
        "parameters": {
            "multiplier": {
                "default": 1.0,
                "min": 0.0,
                "max": None,
                "display_name": "Multiplier"
            },
            "signal_option": {
                "default": "Temperature",
                "options": {
                    "Temperature": "Temperature",
                    "Pressure": "Pressure",
                    "Precipitation": "Precipitation"
                },
                "display_name": "Signal Option"
            }
        },
        "function": seaborn_plotly_example,
    }
}
\end{python}

\section{Incorporating Custom Filters into dFL \& Advanced Operator Ordering}\label{AppA}
Here we show the entire pipeline of generating a \texttt{data provider} and \texttt{utilities} script for a custom python filter showing the non-commutativity of smoothing and downsampling from Section \ref{OOO}. Note that custom filters may be performed in sequence in the \texttt{data provider}, thus customizing the operator ordering as much as needed, e.g., performing multiple normalizations sequentially.

\begin{python}
"""
Data Provider Module

This module provides a data provider interface for paper research signals demonstrating
the non-commutative nature of smoothing and downsampling operations.

The provider generates synthetic signals with high-frequency bursts and provides
custom processing functions for:
- Savitzky-Golay smoothing
- Kaiser windowed downsampling
- Pipeline comparison visualization

Key Features:
- Generates synthetic signals with controllable parameters
- Provides custom graphing functions for pipeline comparison
- Supports data processing options (smoothing, downsampling)
- Includes robust error handling and logging
"""

from datetime import timezone
from pathlib import Path
import numpy as np
import pandas as pd
import logging
import traceback
import matplotlib
matplotlib.use('Agg')  # Use non-interactive backend to avoid GUI warnings
import matplotlib.pyplot as plt
import plotly.graph_objects as go
from paper_utilities import (
    matplotlib_to_plotly_base64, 
    savitzky_golay_smoothing, 
    kaiser_downsample_processing,
    parse_value, 
    trim_data,
    generate_synthetic_signal
)
from scipy.signal import resample_poly, savgol_filter

def pipeline_comparison_grapher(app_control_parameters, parameters):
    """
    Generate a comparison plot showing both processing pipelines side by side.
    
    This function demonstrates the non-commutative nature of smoothing and downsampling
    by showing:
    - Pipeline A: Smooth → Kaiser Downsample
    - Pipeline B: Kaiser Downsample → Smooth
    
    Args:
        app_control_parameters (dict): Global parameters from the application controller
        parameters (dict): Parameters specific to this graphing function
        
    Parameters Keys:
        - "downsample_factor": int (default: 10)
        - "beta": float (default: 2.0) 
        - "window_length": int (default: 31)
        - "polyorder": int (default: 3)
        - "show_both_plots": bool (default: True)
        
    Returns:
        plotly.graph_objects.Figure: A Plotly figure object
    """
    print(f"Pipeline comparison parameters: {parameters}")
    
    # Extract parameters with defaults
    downsample_factor = int(parameters.get("pipeline_comparison_grapher_downsample_factor", 
                                         parameters.get("downsample_factor", 10)))
    beta = float(parameters.get("pipeline_comparison_grapher_beta", 
                              parameters.get("beta", 2.0)))
    window_length = int(parameters.get("pipeline_comparison_grapher_window_length", 
                                     parameters.get("window_length", 31)))
    polyorder = int(parameters.get("pipeline_comparison_grapher_polyorder", 
                                 parameters.get("polyorder", 3)))
    show_both_plots = parameters.get("pipeline_comparison_grapher_show_both_plots", 
                                   parameters.get("show_both_plots", True))
    
    # Get the signal data
    print(f"Pipeline Comparison: Fetching Original Signal for shot {app_control_parameters['shot_id']}")
    
    try:
        shot_data = app_control_parameters["data_coordinator"].fetch_data_async(
            app_control_parameters["data_coordinator"].data_folder,
            dataset_id=None,
            shot_id=app_control_parameters["shot_id"],
            signals=["Original Signal"],
            global_data_params={},
            trim_1=app_control_parameters.get("trim_t1"),
            trim_2=app_control_parameters.get("trim_t2"),
            timeout=10,
        )
    except Exception as e:
        print(f"ERROR: Pipeline comparison data fetch failed: {e}")
        raise
    
    if len(shot_data["signals"]) == 0:
        raise ValueError(f"No signals could be fetched for pipeline comparison. Available signals: {app_control_parameters['data_coordinator'].all_possible_signals}")

    signal = np.array(shot_data["signals"][0]["data"]).flatten()
    times = np.array(shot_data["signals"][0]["times"]).flatten()
    N = len(signal)

    # Define operators (from the original code)
    def smooth(sig):
        """Savitzky-Golay local cubic smoothing (window = window_length samples)."""
        # Ensure window_length is odd and valid
        wl = window_length
        if wl 
            wl += 1
        if wl > len(sig):
            wl = len(sig) if len(sig) 
        if wl < 1:
            wl = 1
        po = min(polyorder, wl - 1)
        return savgol_filter(sig, wl, po)

    def kaiser_downsample(sig, factor=downsample_factor, b=beta):
        """Downsample by 'factor' using a polyphase FIR whose taps are shaped by a Kaiser window."""
        return resample_poly(sig, up=1, down=factor, window=('kaiser', b))

    # Two non-commuting pipelines
    pipe_A = kaiser_downsample(smooth(signal), downsample_factor)  # Smooth → Downsample
    pipe_B = smooth(kaiser_downsample(signal, downsample_factor))  # Downsample → Smooth

    # Stretch results back to original grid for overlay
    def stretch(arr, factor, target_len):
        return np.repeat(arr, factor)[:target_len]

    pipe_A_stretched = stretch(pipe_A, downsample_factor, N)
    pipe_B_stretched = stretch(pipe_B, downsample_factor, N)

    # Create subplots if showing both, otherwise just one
    if show_both_plots:
        fig = go.Figure()
        
        # Plot 1: Pipeline A (Smooth → Kaiser Downsample)
        fig.add_trace(go.Scatter(
            x=times,
            y=signal,
            mode='lines',
            line=dict(width=1, color='lightblue'),
            name='Original',
            opacity=0.4,
            yaxis='y1'
        ))
        
        fig.add_trace(go.Scatter(
            x=times,
            y=pipe_A_stretched,
            mode='lines',
            line=dict(width=2, color='orange'),
            name='Smooth → Kaiser Downsample',
            yaxis='y1'
        ))

        # Plot 2: Pipeline B (Kaiser Downsample → Smooth)  
        fig.add_trace(go.Scatter(
            x=times,
            y=signal,
            mode='lines',
            line=dict(width=1, color='lightblue'),
            name='Original (B)',
            opacity=0.4,
            showlegend=False,
            yaxis='y2'
        ))
        
        fig.add_trace(go.Scatter(
            x=times,
            y=pipe_B_stretched,
            mode='lines',
            line=dict(width=2, color='lightcoral'),
            name='Kaiser Downsample → Smooth',
            yaxis='y2'
        ))

        # Update layout for subplots
        fig.update_layout(
            title=dict(
                text="Pipeline Comparison: Order Matters in Signal Processing",
                font=dict(size=16),
                y=0.95
            ),
            xaxis=dict(
                title="Time [s]",
                domain=[0, 1]
            ),
            yaxis=dict(
                title="Pipeline A: Smooth → Kaiser Downsample",
                domain=[0.52, 1],
                anchor='x'
            ),
            yaxis2=dict(
                title="Pipeline B: Kaiser Downsample → Smooth", 
                domain=[0, 0.48],
                anchor='x'
            ),
            plot_bgcolor='rgba(0,0,0,0)',
            paper_bgcolor='rgba(0,0,0,0)',
            template="plotly_white" if not app_control_parameters["theme_value"] else "plotly_dark",
            margin=dict(l=60, r=20, t=80, b=60),
            height=600
        )
    else:
        # Show only Pipeline A
        fig = go.Figure()
        
        fig.add_trace(go.Scatter(
            x=times,
            y=signal,
            mode='lines',
            line=dict(width=1, color='lightblue'),
            name='Original',
            opacity=0.4
        ))
        
        fig.add_trace(go.Scatter(
            x=times,
            y=pipe_A_stretched,
            mode='lines',
            line=dict(width=2, color='orange'),
            name='Smooth → Kaiser Downsample'
        ))

        fig.update_layout(
            title=dict(
                text="Pipeline A: Smooth then Kaiser Downsample",
                font=dict(size=16),
                y=0.95
            ),
            xaxis=dict(title="Time [s]"),
            yaxis=dict(title="Amplitude"),
            plot_bgcolor='rgba(0,0,0,0)',
            paper_bgcolor='rgba(0,0,0,0)',
            template="plotly_white" if not app_control_parameters["theme_value"] else "plotly_dark",
            margin=dict(l=50, r=50, t=80, b=50)
        )

    return fig

def simple_signal_plot(app_control_parameters, parameters):
    """
    A simple signal plot showing the selected signal.
    
    Args:
        app_control_parameters (dict): Global parameters from the app
        parameters (dict): Parameters specific to this function
        
    Returns:
        plotly.graph_objects.Figure: A Plotly figure object
    """
    # Try different possible parameter keys
    signal_name = (
        parameters.get("simple_signal_plot_signal_name") or 
        parameters.get("signal_name") or 
        "Original Signal"
    )
    
    print(f"Simple Signal Plot: Requesting '{signal_name}' for shot {app_control_parameters['shot_id']}")
    
    try:
        shot_data = app_control_parameters["data_coordinator"].fetch_data_async(
            app_control_parameters["data_coordinator"].data_folder,
            dataset_id=None,
            shot_id=app_control_parameters["shot_id"],
            signals=[signal_name],
            global_data_params={},
            trim_1=app_control_parameters.get("trim_t1"),
            trim_2=app_control_parameters.get("trim_t2"),
            timeout=10,
        )
    except Exception as e:
        print(f"ERROR: Failed to fetch signal '{signal_name}': {e}")
        raise

    if len(shot_data["signals"]) == 0:
        # Fallback to the first available signal if the requested one wasn't found
        print(f"Signal '{signal_name}' not found, using fallback: '{app_control_parameters['data_coordinator'].all_possible_signals[0]}'")
        fallback_signal = app_control_parameters["data_coordinator"].all_possible_signals[0]
        
        try:
            shot_data = app_control_parameters["data_coordinator"].fetch_data_async(
                app_control_parameters["data_coordinator"].data_folder,
                dataset_id=None,
                shot_id=app_control_parameters["shot_id"],
                signals=[fallback_signal],
                global_data_params={},
                trim_1=app_control_parameters.get("trim_t1"),
                trim_2=app_control_parameters.get("trim_t2"),
                timeout=10,
            )
        except Exception as e:
            print(f"ERROR: Fallback signal fetch failed: {e}")
            raise
            
        signal_name = fallback_signal

    if len(shot_data["signals"]) == 0:
        raise ValueError(f"No signals could be fetched. Available signals: {app_control_parameters['data_coordinator'].all_possible_signals}")

    signal = np.array(shot_data["signals"][0]["data"]).flatten()
    times = np.array(shot_data["signals"][0]["times"]).flatten()

    # Create Plotly figure
    fig = go.Figure()
    
    fig.add_trace(go.Scatter(
        x=times,
        y=signal,
        mode='lines',
        line=dict(width=2),
        name=signal_name
    ))
    
    fig.update_layout(
        title=dict(
            text=f"{signal_name} Over Time",
            font=dict(size=16),
            y=0.95
        ),
        xaxis=dict(title="Time [s]"),
        yaxis=dict(title="Amplitude"),
        plot_bgcolor='rgba(0,0,0,0)',
        paper_bgcolor='rgba(0,0,0,0)',
        template="plotly_white" if not app_control_parameters["theme_value"] else "plotly_dark",
        margin=dict(l=50, r=50, t=80, b=50)
    )

    return fig

def get_provider(_):
    """
    Get the complete data provider configuration for the Paper App.
    
    This function sets up and returns a comprehensive configuration dictionary
    that defines the entire data provider interface for demonstrating signal
    processing pipeline comparisons.
    
    Args:
        _ (Any): Unused parameter for compatibility with the interface.
        
    Returns:
        dict: A dictionary containing the full provider configuration.
    """
    # --- Logging Setup ---
    log_file_path = Path(__file__).parent / "paper_data_provider_errors.log"
    logger = logging.getLogger(__name__)
    logger.setLevel(logging.ERROR)
    
    # Prevent duplicate handlers if get_provider is called multiple times
    if not logger.handlers:
        file_handler = logging.FileHandler(log_file_path)
        formatter = logging.Formatter('
        file_handler.setFormatter(formatter)
        logger.addHandler(file_handler)
    # --- End Logging Setup ---

    try:
        def fetch_shot_ids_for_dataset_id(data_folder, _=None):
            """
            Fetch all available shot IDs from the data folder.
            
            For the paper app, we provide synthetic shot IDs.
            """
            # Return some synthetic shot IDs
            return ["synthetic_burst_signal", "synthetic_clean_signal", "synthetic_noisy_signal"]

        def fetch_data(data_folder, _dataset_id, shot_id, signals, _global_data_params, data_trim_1=None, data_trim_2=None):
            """
            Fetch and process data for a specific shot ID and signals.
            
            Generates synthetic signals based on the shot_id parameters.
            """
            if shot_id is None:
                return []
            if signals is None:
                signals = ["original_signal"]
            
            # Generate different synthetic signals based on shot_id
            if shot_id == "synthetic_burst_signal":
                signal, times = generate_synthetic_signal(N=5000, seed=1)
            elif shot_id == "synthetic_clean_signal":
                # Clean signal without noise
                np.random.seed(1)
                times = np.linspace(0, 10, 5000)
                signal = np.sin(2 * np.pi * 0.6 * times)
                burst_mask = (times > 4.0) & (times < 4.5)
                signal += burst_mask * 5 * np.sin(2 * np.pi * 12 * times)
            elif shot_id == "synthetic_noisy_signal":
                # Very noisy signal
                np.random.seed(42)
                times = np.linspace(0, 10, 5000)
                signal = np.sin(2 * np.pi * 0.6 * times)
                burst_mask = (times > 4.0) & (times < 4.5)
                signal += burst_mask * 5 * np.sin(2 * np.pi * 12 * times)
                signal += 0.5 * np.random.randn(5000)  # More noise
            else:
                # Default to burst signal
                signal, times = generate_synthetic_signal(N=5000, seed=1)

            # Use raw numerical time values (since is_date=False)
            times_array = times

            data_array = []
            errored_signals = []

            for signal_name in signals:
                if signal_name == "Original Signal":
                    record = {
                        "data": signal, 
                        "data_name": signal_name, 
                        "times": times_array, 
                        "units": {"Original Signal": "amplitude"}, 
                        "dims": ("times",)
                    }
                    data_array.append(record)
                elif signal_name == "Smooth → Kaiser Downsample":
                    # Pre-compute Pipeline A result for display
                    from scipy.signal import savgol_filter, resample_poly
                    smoothed = savgol_filter(signal, 31, 3)
                    downsampled = resample_poly(smoothed, up=1, down=10, window=('kaiser', 2.0))
                    stretched = np.repeat(downsampled, 10)[:len(signal)]
                    record = {
                        "data": stretched, 
                        "data_name": signal_name, 
                        "times": times_array, 
                        "units": {"Smooth → Kaiser Downsample": "amplitude"}, 
                        "dims": ("times",)
                    }
                    data_array.append(record)
                elif signal_name == "Kaiser Downsample → Smooth":
                    # Pre-compute Pipeline B result for display
                    from scipy.signal import savgol_filter, resample_poly
                    downsampled = resample_poly(signal, up=1, down=10, window=('kaiser', 2.0))
                    smoothed = savgol_filter(downsampled, 31, 3)
                    stretched = np.repeat(smoothed, 10)[:len(signal)]
                    record = {
                        "data": stretched, 
                        "data_name": signal_name, 
                        "times": times_array, 
                        "units": {"Kaiser Downsample → Smooth": "amplitude"}, 
                        "dims": ("times",)
                    }
                    data_array.append(record)
                else:
                    errored_signals.append(signal_name)

            return {"id": shot_id, "signals": data_array, "errored_signals": errored_signals}

        # --- Module Initialization ---
        data_folder = Path(__file__).parent / "paper_data"
        
        # Available signals
        all_possible_signals = ["Original Signal", "Smooth → Kaiser Downsample", "Kaiser Downsample → Smooth"]

        # --- Configuration Dictionaries ---
        custom_smoothing_dictionary = {
            "savitzky_golay_smoothing": {
                "display_name": "Savitzky-Golay Smoothing",
                "parameters": {
                    "window_length": {"default": 31, "min": 3, "max": 101, "display_name": "Window Length"},
                    "polyorder": {"default": 3, "min": 1, "max": 10, "display_name": "Polynomial Order"}
                },
                "function": savitzky_golay_smoothing,
            },
            "kaiser_downsample_processing": {
                "display_name": "Kaiser Downsample Processing",
                "parameters": {
                    "downsample_factor": {"default": 10, "min": 1, "max": 50, "display_name": "Downsample Factor"},
                    "beta": {"default": 2.0, "min": 0.1, "max": 10.0, "display_name": "Kaiser Beta"}
                },
                "function": kaiser_downsample_processing,
            },
        }

        custom_grapher_dictionary = {
            "pipeline_comparison_grapher": {
                "display_name": "Pipeline Comparison",
                "parameters": {
                    "downsample_factor": {"default": 10, "min": 1, "max": 50, "display_name": "Downsample Factor"},
                    "beta": {"default": 2.0, "min": 0.1, "max": 10.0, "display_name": "Kaiser Beta"},
                    "window_length": {"default": 31, "min": 3, "max": 101, "display_name": "Window Length"},
                    "polyorder": {"default": 3, "min": 1, "max": 10, "display_name": "Polynomial Order"},
                    "show_both_plots": {"default": True, "display_name": "Show Both Pipelines"}
                },
                "function": pipeline_comparison_grapher,
            },
            "simple_signal_plot": {
                "display_name": "Simple Signal Plot",
                "parameters": {
                    "signal_name": {
                        "default": "Original Signal", 
                        "options": {
                            "Original Signal": "Original Signal",
                            "Smooth → Kaiser Downsample": "Pipeline A Result",
                            "Kaiser Downsample → Smooth": "Pipeline B Result"
                        }, 
                        "display_name": "Signal to Plot"
                    }
                },
                "function": simple_signal_plot,
            },
        }

        custom_normalizing_dictionary = {}
        
        # --- Final Provider Assembly ---
        data_coordinator_info = {
            "fetch_data": fetch_data,
            "dataset_id": "paper_signals",
            "fetch_shot_ids_for_dataset_id": fetch_shot_ids_for_dataset_id,
            "all_possible_signals": all_possible_signals,
            "custom_smoothing_options": custom_smoothing_dictionary,
            "custom_normalizing_options": custom_normalizing_dictionary,
            "spline_path": "spline_parameters.csv",
            "auto_label_function_dictionary": {},
            "all_labels": ["Burst", "Clean", "Noise"],
            "custom_grapher_dictionary": custom_grapher_dictionary,
            "is_date": False,
            "trim_data": trim_data,
            "data_folder": data_folder,
            "layout_options": {}
        }
        return data_coordinator_info
    except Exception as e:
        logger.error("Exception occurred in get_provider:")
        logger.error(traceback.format_exc())
        raise  # Re-raise the exception to be caught by the caller

\end{python}

\begin{python}
"""
Paper App Utilities

This module contains utility functions for signal processing used in the paper app,
including the signal processing pipelines from the research paper:
- Savitzky-Golay smoothing
- Kaiser windowed downsampling
- Pipeline comparison functions

The app demonstrates the non-commutative nature of smoothing and downsampling operations.
"""

import numpy as np
import pandas as pd
from scipy.signal import resample_poly, savgol_filter
from datetime import timezone
import matplotlib
matplotlib.use('Agg')  # Use non-interactive backend to avoid GUI warnings
import matplotlib.pyplot as plt
import plotly.graph_objects as go
import io
import base64
from pathlib import Path

def parse_value(value):
    """
    Parse a value to its most appropriate data type (datetime, float, int, or original).
    
    This function attempts to convert input values in the following priority order:
    1. Datetime (with automatic unit detection for timestamps)
    2. Float
    3. Integer
    4. Original value (if all conversions fail)
    
    Args:
        value: Input value to parse (can be string, number, or datetime)
        
    Returns:
        Parsed value in most appropriate type, or None for empty strings
    """
    # First, try to parse as datetime
    try:
        # Determine unit based on timestamp length
        if isinstance(value, (int, float)):
            timestamp_str = str(value)
            length = len(timestamp_str)

            if length == 10:
                unit = "s"  # Seconds
            elif length == 13:
                unit = "ms"  # Milliseconds
            elif length == 16:
                unit = "us"  # Microseconds
            elif length == 19:
                unit = "ns"  # Nanoseconds
            else:
                # Default to milliseconds if unsure
                unit = "ms"

            # Use UTC timezone to avoid timezone conversion issues
            dt = pd.to_datetime(value, unit=unit, utc=True)
        else:
            dt = pd.to_datetime(value, utc=True)  # For strings or already datetime objects

        if pd.isna(dt):  # Check if the result is NaT
            raise ValueError
        return dt.to_pydatetime()
    except (ValueError, TypeError):
        pass

    # If not int, try to convert to float
    try:
        return float(value)
    except (ValueError, TypeError):
        pass

    # If not datetime, try to convert to int
    try:
        return int(value)
    except (ValueError, TypeError):
        pass

    # If all conversions fail, return the original value
    if isinstance(value, str) and len(value) == 0:
        return None
    return value

def find_nearest_datetime_index(arr, target_datetime):
    """
    Find the index of the nearest datetime in an array to a target datetime.
    
    Args:
        arr (array-like): Array of datetime values to search
        target_datetime (datetime): Target datetime to find nearest match for
        
    Returns:
        int or None: Index of nearest datetime, or None if array is empty or error occurs
    """
    if target_datetime is None:
        return None

    # Convert array to numpy array if it's not already
    arr = np.array(arr)

    if len(arr) == 0:
        return None

    # Check if the array contains timezone-aware datetimes
    try:
        sample_time = pd.Timestamp(arr[0])
        arr_has_timezone = sample_time.tzinfo is not None
    except:
        arr_has_timezone = False

    # Check if target datetime has timezone
    target_has_timezone = hasattr(target_datetime, "tzinfo") and target_datetime.tzinfo is not None

    # Make timezone consistent between array and target
    if target_has_timezone and not arr_has_timezone:
        # Remove timezone from target to match array
        target_datetime = target_datetime.replace(tzinfo=None)
    elif not target_has_timezone and arr_has_timezone:
        # Add UTC timezone to target to match array
        target_datetime = target_datetime.replace(tzinfo=timezone.utc)

    # Convert target_datetime to nanoseconds since epoch
    try:
        target_ns = int(target_datetime.timestamp() * 1e9)
    except (AttributeError, TypeError, ValueError) as e:
        print(f"Error converting target datetime to timestamp: {e}")
        return None

    try:
        # Convert datetime64 array to nanoseconds since epoch
        arr_ns = arr.astype("datetime64[ns]").astype(np.int64)

        # Find the index of the nearest value
        nearest_index = np.abs(arr_ns - target_ns).argmin()
        return nearest_index
    except Exception as e:
        print(f"Error finding nearest datetime index: {e}")
        return None

def savitzky_golay_smoothing(_shot_name, _signal_name, raw_signal, _times, parameters):
    """
    Apply Savitzky-Golay smoothing to a 1D array.
    
    This function smooths data using a Savitzky-Golay filter which fits local polynomials
    to preserve features better than simple moving averages.
    
    Args:
        _shot_name (str): The name of the shot being analyzed (unused)
        _signal_name (str): The name of the signal (unused)
        raw_signal (array-like): Input data to smooth
        _times (array-like): Time array of the shot (unused)
        parameters (dict): Dictionary containing smoothing parameters
        
    Parameters Dictionary Keys:
        - "window_length": int, Window size for Savitzky-Golay filter (default: 31, must be odd)
        - "polyorder": int, Polynomial order for fitting (default: 3)
    
    Returns:
        numpy.array: Smoothed data with same length as input
    """
    # Convert input to numpy array
    data = np.array(raw_signal)
    
    # Extract parameters with defaults
    window_length = int(parameters.get("savitzky_golay_smoothing_window_length", 
                                     parameters.get("window_length", 31)))
    polyorder = int(parameters.get("savitzky_golay_smoothing_polyorder", 
                                 parameters.get("polyorder", 3)))

    # Ensure window_length is odd and valid
    if window_length 
        window_length += 1
        
    if window_length < 1:
        window_length = 1
        
    if window_length > data.shape[0]:
        window_length = data.shape[0] if data.shape[0] 
        if window_length < 1:
            window_length = 1
        
    # Ensure polyorder is valid
    if polyorder >= window_length:
        polyorder = window_length - 1
        
    if polyorder < 0:
        polyorder = 0

    # Apply Savitzky-Golay filter
    try:
        smoothed = savgol_filter(data, window_length, polyorder)
    except ValueError as e:
        print(f"Savitzky-Golay filter error: {e}")
        # Fallback to original data if filtering fails
        smoothed = data

    return smoothed

def kaiser_downsample_processing(_shot_name, _signal_name, raw_signal, times, parameters):
    """
    Apply Kaiser windowed downsampling to a 1D array.
    
    This function downsamples data using a polyphase FIR filter with Kaiser windowing.
    The low beta value deliberately relaxes stop-band attenuation to make aliasing visible.
    
    Args:
        _shot_name (str): The name of the shot being analyzed (unused)
        _signal_name (str): The name of the signal (unused)
        raw_signal (array-like): Input data to downsample
        times (array-like): Time array of the shot
        parameters (dict): Dictionary containing downsampling parameters
        
    Parameters Dictionary Keys:
        - "downsample_factor": int, Downsampling factor (default: 10)
        - "beta": float, Kaiser window parameter (default: 2.0)
    
    Returns:
        tuple: (downsampled_data, downsampled_times)
    """
    # Convert input to numpy arrays
    data = np.array(raw_signal)
    time_array = np.array(times)
    
    # Extract parameters with defaults
    factor = int(parameters.get("kaiser_downsample_processing_downsample_factor", 
                              parameters.get("downsample_factor", 10)))
    beta = float(parameters.get("kaiser_downsample_processing_beta", 
                              parameters.get("beta", 2.0)))

    # Ensure factor is valid
    if factor < 1:
        factor = 1
        
    if factor > data.shape[0]:
        factor = data.shape[0]

    # Apply polyphase FIR downsampling with Kaiser window
    try:
        downsampled_data = resample_poly(data, up=1, down=factor, window=('kaiser', beta))
        
        # Downsample times accordingly
        downsampled_times = time_array[::factor][:len(downsampled_data)]
        
        # If lengths don't match, truncate the longer one
        min_len = min(len(downsampled_data), len(downsampled_times))
        downsampled_data = downsampled_data[:min_len]
        downsampled_times = downsampled_times[:min_len]
        
    except Exception as e:
        print(f"Kaiser downsampling error: {e}")
        # Fallback to simple decimation
        downsampled_data = data[::factor]
        downsampled_times = time_array[::factor]

    return downsampled_data, downsampled_times

def trim_data(shot_data, trim_1=None, trim_2=None):
    """
    Trim shot data based on time range specifications.
    
    Args:
        shot_data (dict): Shot data dictionary containing signals
        trim_1 (str, optional): Start time for trimming (various formats accepted)
        trim_2 (str, optional): End time for trimming (various formats accepted)
        
    Returns:
        dict: Modified shot_data with trimmed signals
    """
    signals = []
    for signal in shot_data["signals"]:
        data = signal["data"]
        times = signal["times"]
        index1 = None
        index2 = None

        # Check if times array has timezone info
        has_timezone = False
        if len(times) > 0:
            try:
                # Convert back to pandas datetime temporarily to check timezone
                sample_time = pd.Timestamp(times[0])
                has_timezone = sample_time.tzinfo is not None
            except:
                has_timezone = False

        if trim_1 is not None and len(trim_1) != 0:
            # Use parse_value to handle different timestamp formats
            trim1 = parse_value(trim_1)
            # Make timezone consistent with times array
            if hasattr(trim1, "tzinfo") and trim1.tzinfo is not None:
                if not has_timezone:
                    trim1 = trim1.replace(tzinfo=None)
            else:
                if has_timezone:
                    trim1 = trim1.replace(tzinfo=timezone.utc)

            index1 = find_nearest_datetime_index(times, trim1)

        if trim_2 is not None and len(trim_2) != 0:
            # Use parse_value to handle different timestamp formats
            trim2 = parse_value(trim_2)
            # Make timezone consistent with times array
            if hasattr(trim2, "tzinfo") and trim2.tzinfo is not None:
                if not has_timezone:
                    trim2 = trim2.replace(tzinfo=None)
            else:
                if has_timezone:
                    trim2 = trim2.replace(tzinfo=timezone.utc)

            index2 = find_nearest_datetime_index(times, trim2)

        if index1 is not None and index2 is not None:
            data = data[index1:index2]
            times = times[index1:index2]
        if index1 is not None and index2 is None:
            data = data[index1:]
            times = times[index1:]
        if index1 is None and index2 is not None:
            data = data[:index2]
            times = times[:index2]

        signal["data"] = data
        signal["times"] = times
        signals.append(signal)

    shot_data["signals"] = signals
    return shot_data

def matplotlib_to_plotly_base64(fig_mpl):
    """
    Convert matplotlib figure to base64-encoded PNG for embedding in Plotly.
    
    Args:
        fig_mpl (matplotlib.figure.Figure): Matplotlib figure to convert
        
    Returns:
        str: Base64-encoded PNG data URL ready for use in Plotly layout images
    """
    buf = io.BytesIO()
    fig_mpl.savefig(buf, format='png', bbox_inches='tight', dpi=150)
    buf.seek(0)
    img_base64 = base64.b64encode(buf.read()).decode('utf-8')
    buf.close()
    plt.close(fig_mpl)  # Important: close the matplotlib figure to free memory
    return f"data:image/png;base64,{img_base64}"

def generate_synthetic_signal(N=5000, seed=1):
    """
    Generate the synthetic signal with ultrahigh frequency burst as described in the paper.
    
    Args:
        N (int): Number of samples (default: 5000)
        seed (int): Random seed for reproducibility (default: 1)
        
    Returns:
        tuple: (signal, time_array) where signal is the synthetic data and time_array is the time values
    """
    np.random.seed(seed)
    x = np.linspace(0, 10, N)
    
    # Slow sine + short, high freq, high amplitude burst + mild noise
    signal = np.sin(2 * np.pi * 0.6 * x)
    burst_mask = (x > 4.0) & (x < 4.5)
    signal += burst_mask * 5 * np.sin(2 * np.pi * 12 * x)  # 12Hz inside 4-4.5s
    signal += 0.1 * np.random.randn(N)
    
    return signal, x

\end{python}

\section{Incorporating Custom Autolabelers into dFL}\label{AppD}

Simiarly to \ref{AppA}, this autolabeling script can be integrated into the \texttt{data provider}.  This autolabeler detects and marks events in a signal when its value exceeds a user-specified threshold.  It scans the selected time-series, identifies contiguous regions above the threshold, and generates labeled intervals that extend slightly before and after the excursion to capture context.  
These automatically generated labels can then be used dFL for anomaly tagging, event detection, or downstream model training.

\begin{python}
def perform_threshold_autolabeling(
    dataset_id, shot_id, data_coordinator, additional_parameters, trim_1=None, trim_2=None
):
    """Detect events when signal values exceed a threshold."""
    if shot_id is None:
        return []
    
    # Extract parameters from the configuration
    signals_to_autolabel = additional_parameters["threshold_signals"]
    threshold = additional_parameters["threshold_autolabel_threshold"]
    
    # Fetch time-series data for the specified signals
    shot_data = data_coordinator.fetch_data_async(
        data_coordinator.data_folder, dataset_id, shot_id, 
        signals_to_autolabel, {}, trim_1=trim_1, trim_2=trim_2
    )
    
    if len(shot_data["signals"]) == 0:
        return []
    
    signal = shot_data["signals"][0]
    signal_data = signal["data"]
    times = signal["times"]
    
    # Implement threshold crossing detection
    labels = []
    start = None
    finish = None
    
    for value_index, value in enumerate(signal_data):
        if value > threshold and start is None:
            start = times[value_index]
        if value < threshold and start is not None:
            finish = times[value_index]
        if start is not None and finish is not None:
            # Create label with temporal bounds and classification
            label = {}
            for possible_label in data_coordinator.all_labels:
                label[possible_label] = None
            label["Anomaly"] = True
            label["T1"] = pd.Timestamp(start) - pd.Timedelta(seconds=10)
            label["T2"] = pd.Timestamp(finish) + pd.Timedelta(seconds=10)
            labels.append(label)
            start = None
            finish = None
    
    return labels
\end{python}

\section{Preprocessing with Fetch Data}\label{AppE}

dFL allows for a user to supply their own script to fetch the data given parameters from the dFL GUI, such as the data trim, and the signals requested. The user can additionally make any adjustments they want to in the data, in addition to any custom filters, such as normalizations, smoothings, resamplings, etc. defined independently in the script (e.g. see \ref{AppA}).

The following is an example of taking data from a saved file, in this case a \texttt{parquet} file, and fetching additional magnetics data in case the user has requested a signal relevant to a spectrogram.

\begin{python}
def fetch_data(data_folder, dataset_id, shot_id, signals, global_data_params, data_trim_1=None, data_trim_2=None):
        """
        Fetch and process data for a specific shot ID and signals.
        
        This function is the core data loader for the TokSearch offline provider. It handles:
        1. Loading standard signals from a main parquet file.
        2. Loading specialized 'modespec' probe signals from a separate pickle file.
        3. Applying numerical trimming based on time values (not indices).
        4. Consolidating data into a standardized shot dictionary.
        
        Args:
            data_folder (Path): Directory containing data files
            dataset_id (str): Dataset ID
            shot_id (str): Unique identifier for the data shot
            signals (list[str]): List of signal names to extract
            global_data_params (dict): Global data parameters for additional project-wide customizations
            data_trim_1 (float, optional): Start time for data trimming
            data_trim_2 (float, optional): End time for data trimming
            
        Returns:
            dict: Shot data dictionary with signals and errors.
        """
        if shot_id is None:
            return []
        if signals is None:
            signals = [all_possible_signals[0]] if all_possible_signals else []
        
        df = pd.read_parquet(Path(data_folder) / f"{shot_id}.parquet")
        main_times = df["times"].to_numpy()
        data_array = []
        errored_signals = []

        for signal in signals:
            # Handle special 'modespec' signals from pickle file
            modespec_signals = ["mpi66m307d", "mpi66m340d", "mpi66m307e", "mpi66m340e"]
            if signal in modespec_signals:
                file_path_probe = Path(data_folder) / "probe_data" / f"spectrum_{shot_id}.parquet"

                if os.path.exists(file_path_probe):
                    probe_df = pd.read_parquet(file_path_probe)
                    if signal in probe_df.columns:
                        probe_signal = probe_df[signal]
                        probe_times = np.array(probe_df["times"])
                    else:
                        continue

                    index_start, index_end = find_trim_indices(
                        arr=probe_times, trim_beginning=data_trim_1, trim_end=data_trim_2
                    )
                    
                    times = probe_times[index_start:index_end]
                    signal_data = np.array(probe_signal)[index_start:index_end]
                        
                    record = {"data": signal_data, "data_name": signal, "times": times, "units": {}, "dims": ("times",)}
                    data_array.append(record)
                else:
                    errored_signals.append(signal)
            
            # Handle standard signals from main parquet file
            elif signal in df.columns.to_list():
                index_start, index_end = find_trim_indices(
                    arr=main_times, trim_beginning=data_trim_1, trim_end=data_trim_2
                )
                signal_data = df[signal].to_numpy()[index_start:index_end]
                times = main_times[index_start:index_end]
                
                record = {"data": signal_data, "data_name": signal, "times": times, "units": {}, "dims": ("times",)}
                data_array.append(record)
            else:
                errored_signals.append(signal)

        shot_dictionary = {"id": shot_id, "signals": data_array, "errored_signals": errored_signals}
        return shot_dictionary
\end{python}

The data could additionally come from a remote SQL table or database source such as TokSearch.  The shots that are available for selection from another user-supplied function, in this case, simply the filenames of the files of shot data:

\begin{python}
def fetch_shot_ids_for_dataset_id(data_folder, dataset_id=None):
        """
        Fetch all available shot IDs from the data folder.
        
        Shot IDs are derived from parquet filenames (without extension).
        """
        return [file.stem for file in data_folder.glob("*.parquet")]
\end{python}

The signals available come from a list supplied to the dictionary that is returned in the \texttt{get\_provider} method, such that the \texttt{get\_provider} method returns the following:

\begin{python}
# --- Final Provider Assembly ---
    data_coordinator_info = {
        "fetch_data": fetch_data,
        "dataset_id": "plasma_shots",
        "fetch_record_ids_for_dataset_id": fetch_shot_ids_for_dataset_id,
        "all_possible_signals": all_possible_signals,
        "all_labels": ["BBQH", "QH", "WPQH", "ELMy H"],
        "spline_path": "spline_parameters.csv",
        "custom_normalizing_options": custom_normalizing_options,
        "custom_smoothing_options": custom_smoothing_options,
        "auto_label_function_dictionary": plasma_mode_autolabeling_dictionary,
        "custom_grapher_dictionary": custom_grapher_dictionary,
        "is_date": False, # Important: This provider uses numerical time, not datetime
        "trim_data": trim_data,
        "data_folder": data_folder,
        "layout_options": layout_options,
        "label_export_hook": label_export_hook,
        "data_export_hook": data_export_hook,
        "manual_label_hook": manual_label_hook,
    }

    return data_coordinator_info    
\end{python}

\section{Upsampling in dFL}\label{AppF}

Upsampling involves inserting additional data points between existing input values using interpolation techniques.  Given a discrete set of data points \(\{(x_i, y_i)\}_{i=1}^{N}\), interpolation constructs a continuous function \( f(x) \) such that \( f(x_i) = y_i \) for known points and estimates intermediate values for unknown points. dFL supports the following interpolation-based methods for \emph{upsampling} and gap-filling. These techniques estimate values at intermediate time points between existing samples, increasing the effective sampling rate or regularizing irregularly sampled data for downstream processing. Each method differs in smoothness, monotonicity guarantees, computational complexity, and available free parameters.

\begin{enumerate}
    \item \textbf{Linear Interpolation} \citep{SciPyInterp} :  
    connects adjacent samples with straight-line segments, producing a continuous but non-differentiable function at the knots.
    \begin{itemize}
        \item \emph{Advantages}: computationally inexpensive; memory-efficient; never overshoots.
        \item \emph{Limitations}: corners at sample points; not suitable for high-curvature data.
    \end{itemize}

    \item \textbf{Quadratic Interpolation} \citep{StoerBulirsch} :  
    fits piecewise quadratic polynomials through consecutive triplets of points, ensuring $C^{1}$ continuity.
    \begin{itemize}
        \item \emph{Advantages}: smoother than linear interpolation with modest computational overhead.
        \item \emph{Limitations}: can produce mild overshoot and oscillations in regions of high curvature.
    \end{itemize}

    \item \textbf{Cubic Spline Interpolation} \citep{DeBoor2001,SciPyCubic}:  
    constructs $C^{2}$-continuous cubic polynomials between points, minimizing global curvature under specified boundary conditions.
    \begin{itemize}
        \item \emph{Advantages}: very smooth; widely used in scientific computing and graphics.
        \item \emph{Limitations}: no monotonicity guarantee; may exhibit ringing near steep gradients.
    \end{itemize}

    \item \textbf{Piecewise Cubic Hermite Interpolation (PCHIP, Mono-PCHIP)} \citep{FritschCarlson1980,SciPyPCHIP}: 
    uses cubic Hermite polynomials with slope adjustments to preserve monotonicity where present.
    \begin{itemize}
        \item \emph{Advantages}: shape-preserving; no overshoot; ideal for cumulative or strictly monotonic data.
        \item \emph{Limitations}: only $C^{1}$-continuous.
    \end{itemize}

    \item \textbf{Akima Interpolation (Mono-Akima)} \citep{Akima1970,SciPyAkima}: 
    employs locally weighted cubic polynomials with slopes estimated from neighboring intervals; can be made monotonic.
    \begin{itemize}
        \item \emph{Advantages}: robust to outliers; minimal ringing; suitable for nonuniformly spaced data.
        \item \emph{Limitations}: higher computational cost than PCHIP.
    \end{itemize}
\end{enumerate}

\noindent\textit{Upsampling Method Selection Considerations:} the optimal choice of interpolation method used for upsampling depends critically on the smoothness requirements, noise characteristics, and application context. Linear interpolation offers maximum computational efficiency and minimal implementation overhead but may introduce abrupt slope changes that degrade the quality of highly smooth or curvature-sensitive signals. Quadratic and cubic spline methods provide progressively smoother approximations, with cubic splines achieving $C^{2}$ continuity at the expense of increased computational cost and potential ringing near steep gradients. Shape-preserving methods such as PCHIP and Akima are preferable when monotonicity is essential or when the data exhibit sharp transitions or outliers. In practice, selecting an appropriate method involves balancing computational constraints against the fidelity requirements of the downstream analysis or control task.

\section{Downsampling in dFL}\label{AppG}
Downsampling in dFL involves dividing the signal into contiguous \emph{blocks} and replacing each block with a single representative value. dFL provides five methods for downsampling time-series data, described below. In all methods, the first step is to specify the desired number of regularly spaced samples \(L_{\mathrm{target}}\) in the downsampled signal. This determines the number of output blocks \(M = L_{\mathrm{target}}\).  

Let the original signal be
\[
\mathbf{x} = \{(t_1, x_1), (t_2, x_2), \dots, (t_N, x_N)\}, \quad t_1 < t_2 < \dots < t_N.
\]
The nominal block size is
\[
b_{\mathrm{nom}} = \left\lfloor \frac{N}{M} \right\rfloor,
\] for floor funciton $\lfloor \cdot\rfloor$.
To ensure \(M\) complete blocks, any remainder samples \((N \bmod M)\) are discarded from the \emph{start} of the signal. The remaining samples are split contiguously into
\[
B_m = \{(t_{i_{m,1}}, x_{i_{m,1}}), \dots, (t_{i_{m,b_m}}, x_{i_{m,b_m}}) \}, \quad m = 1, \dots, M,
\]
where \(b_m\) is the actual number of samples in block \(m\).  
When the sampling grid is irregular, \(b_m\) may vary across blocks even though \(M\) is fixed.

All block-based downsampling methods in dFL operate by mapping each block \(B_m\) to a single representative value \(y_m\) according to a rule determined by the chosen method. All methods use the entire block (\(b_m\)) to compute statistics within that block.

dFL supports the following block-based and statistical downsampling techniques:

\begin{enumerate}
    \item \textbf{Block Average} \citep{SmithDSP,SciPyUniform1D}: for each block \(B_m\), compute the arithmetic mean over all \(b_m\) samples.
    \begin{itemize}
        \item \emph{Free parameters}: target length \(L_{\mathrm{target}}\) of downsampled signal.
        \item \emph{Advantages}: simple anti-aliasing for decimation; preserves average energy for white-noise signals.
        \item \emph{Limitations}: blurs sharp peaks and transients; sensitive to block boundaries.
    \end{itemize}

    \item \textbf{Block Max} \citep{SciPyMax1D}:  for each block \(B_m\), compute the maximum value over a sliding window of length \(b_m\). 
    \begin{itemize}
        \item \emph{Free parameters}: target length \(L_{\mathrm{target}}\) of downsampled signal.
        \item \emph{Advantages}: captures extreme peaks; useful for overload or fault detection.
        \item \emph{Limitations}: ignores average behavior; sensitive to outliers.
    \end{itemize}

    \item \textbf{Block Min} \citep{SciPyMin1D}:  
    for each block \(B_m\), compute the minimum value over the window of length \(b_m\). 
    \begin{itemize}
        \item \emph{Free parameters}: target length \(L_{\mathrm{target}}\) of downsampled signal.
        \item \emph{Advantages}: captures extreme valleys; complements Block Max for envelope estimation.
        \item \emph{Limitations}: ignores peak behavior; sensitive to negative spikes.
    \end{itemize}

    \item \textbf{Maximum-Likelihood of Gaussian Mixture} \citep{McLachlanPeel,MurphyML,SKLGMM}:  
    for each block \(B_m\), fit Gaussian mixture models (GMMs) to all \(b_m\) samples (\(k = b_m\)), testing component counts \(i \in \{1, \dots, n_{\mathrm{max}}\}\). Select the optimal model via the Bayesian Information Criterion (BIC) and return the maximum-likelihood value as the representative.  

    In more detail, MLE-GMM is a probabilistic downsampling method that estimates a representative value for each block \(B_m\) by fitting a Gaussian Mixture Model (GMM) to the block’s data and selecting the number of mixture components that best explains the local distribution.

    Given a block  
    \[
    B_m = \{x_{i_{m,1}}, \dots, x_{i_{m,b_m}}\},
    \]  
    the procedure is as follows:

\begin{enumerate}
    \item Fit a Gaussian Mixture Model (GMM) with \(i\) components to the \(b_m\) samples in the block, with \(i\) selected from a predefined set \(\{1, 2, \dots, n_{\mathrm{max}}\}\).
    \item Determine the optimal number of mixture components \(i_0\) by minimizing the Bayesian Information Criterion (BIC):
    \[
    \mathrm{BIC} = -2 \log \mathcal{L} + p \log b_m,
    \]
    where \(\mathcal{L}\) is the maximized likelihood of the model given the data in \(B_m\), and \(p\) is the number of free parameters in the GMM with \(i\) components.
    \item Using the selected model with \(i_0\) components, compute the \emph{maximum-likelihood estimate} (MLE) of the distribution:
    \[
    \hat{x}_m = \arg \max_{x} P(x \mid \theta_{i_0}),
    \]
    where \(P(x \mid \theta_{i_0})\) is the probability density function of the fitted GMM with parameters \(\theta_{i_0}\).
    \item Assign \(\hat{x}_m\) as the representative value for block \(B_m\). Repeating this process for \(m = 1, \dots, M\) produces the downsampled signal \(\{\hat{x}_1, \dots, \hat{x}_M\}\).
\end{enumerate}

The MLE-GMM method is particularly effective for blocks whose local distributions are multimodal or strongly non-normal. By fitting a mixture model and selecting the number of components via BIC, this approach preserves complex statistical characteristics, such as variance, skewness, and clustering structure, that would be lost under simple averaging. Empirical results indicate that reliable fitting requires at least \(b_m \geq 30\) samples; for smaller blocks, the method reverts to a block mean.  

Although MLE-GMM can deliver a more faithful representation of local structure, it is computationally expensive because it requires running the Expectation-Maximization (EM) algorithm for multiple candidate models per block and is sensitive to initialization. When the local distribution is far from a Gaussian mixture, or when block sizes are small, the method may yield suboptimal results. Nonetheless, for domains where accurate local distribution modeling is essential and sufficient data per block is available, MLE-GMM is a powerful probabilistic downsampling tool.

    \begin{itemize}
        \item \emph{Free parameters}: target length \(L_{\mathrm{target}}\); maximum mixture components \(n_{\mathrm{max}}\).
        \item \emph{Advantages}: preserves multimodal structure; adapts complexity to local distribution via BIC.
        \item \emph{Limitations}: computationally expensive (EM per block); sensitive to initialization; unreliable for \(b_m < 30\) (falls back to block mean).
    \end{itemize}

    \item \textbf{Importance Downsampling} \citep{faghihi2020moment,Paananen2021IWMM}:  
    for each block \(B_m\), the \(b_m\) sample points are used to selects values in each block so that chosen statistical central moments (zero, variance, skewness, etc.) match those of the original data and the samples remain close to their original time stamps.
    \begin{itemize}
        \item \emph{Free parameters}: target length \(L_{\mathrm{target}}\); set of central moments to preserve, e.g., $$\{\mu_1,\mu_2,\ldots,\mu_r\}=\{1,2,\ldots,r\}.$$
        \item \emph{Advantages}: explicitly preserves user-selected moments; temporal proximity constraint; adapts to local variability.
        \item \emph{Limitations}: requires at least \(k+1\) samples per block; computationally heavy/expensive (i.e. a nonlinear optimization per block), sensitive to initial guesses.
    \end{itemize}
\end{enumerate}

\noindent\textit{Downsampling Method Selection Considerations:} selecting an appropriate downsampling method in dFL depends on the signal’s characteristics, the analysis goals, and the downstream use case. For smooth, continuous signals where preserving the average trend is most important (e.g., low-frequency behavior for model fitting), the Block Average method offers a computationally inexpensive, anti-aliasing-friendly choice, but may obscure short-duration peaks. Conversely, Block Max and Block Min are well-suited for envelope extraction, overload detection, or fault monitoring, but will sacrifice statistical representativeness and can exaggerate the influence of single-sample outliers. The Maximum-Likelihood Gaussian Mixture method should be reserved for blocks with sufficient sample count (\(b_m \gtrsim 30\)) and suspected multimodal structure, as it is computationally expensive and sensitive to initialization, but excels at preserving distributional shape in complex data. Importance Downsampling provides a principled way to preserve selected statistical central moments while retaining temporal locality, making it valuable for maintaining statistical fidelity in irregularly spaced data; however, its nonlinear optimization cost can be prohibitive for large datasets or real-time pipelines.  

In all cases, the user should be mindful of aliasing risks when high-frequency content is present and large downsampling ratios are applied: while some dFL resampling methods implicitly smooth the data, block-based statistical methods without prior low-pass filtering may allow high-frequency components to fold into lower frequencies. In irregular-grid contexts, smoothing prior to downsampling may not be well-defined, so users may instead rely on the \texttt{Resample} stage of the pipeline to impose a regular grid before applying explicit smoothing. Care should also be taken when the data contain sharp transients or sparse events of interest---aggressive downsampling can irreversibly lose such features unless a peak-preserving method is chosen.  

\section{Smoothing in dFL}\label{AppH}

\subsection{Moving Average Approaches}

Moving averages are among the most widely used smoothing techniques for reducing high-frequency noise and stabilizing time-series signals. Two common variants supported in dFL are the \textit{simple moving average} (SMA) and the \textit{exponentially weighted moving average} (EMA), each with distinct weighting schemes and temporal sensitivities. \\

\noindent\emph{Simple Moving Average (SMA):}  
given a sequence $\{x_t\}$, the SMA of window length $N$ is
\[
\mathrm{SMA}_t = \frac{1}{N} \sum_{i=0}^{N-1} x_{t-i}.
\]
This true sliding window assigns equal weight to all samples within the window and ignores older points. Larger $N$ yields smoother results but increases lag by $(N-1)/2$ samples, while smaller $N$ preserves detail but is more susceptible to noise. Its mathematical transparency and deterministic behavior make it a useful baseline, though uniform weighting can blur sharp features and attenuate peaks. \\

\noindent\emph{Exponentially Weighted Moving Average (EMA):}  
the EMA replaces the uniform window with an exponentially decaying weight profile:
\[
\mathrm{EMA}_t = \alpha\, x_t + (1-\alpha)\, \mathrm{EMA}_{t-1}, 
\quad \alpha = \frac{2}{N+1},
\]
where $N$ is the \emph{span} parameter controlling the effective smoothing window.  
Although the update is a convex combination of the latest value and the previous EMA, unrolling the recursion reveals that past samples receive weights proportional to $(1-\alpha)^k$, which iteratively decay exponentially; i.e., older samples never vanish completely but their influence decays exponentially, imparting a strong recency bias. The EMA offers $O(1)$ memory and CPU cost, making it well suited for streaming or real-time systems. Its primary trade-off lies in tuning: too small an $N$ chases noise, too large an $N$ lags important events. \\

\noindent\emph{Comparative Considerations:} the SMA’s fixed-length, equal-weight window is preferable when a stable, interpretable filter is needed for exploratory analysis or when the noise level is relatively stationary. The EMA excels when responsiveness to recent changes is essential, such as in dashboards or real-time tokamak control loops, though its fixed decay rate limits adaptation to seasonal or cyclic structure. Both methods suppress high-frequency variance, but the EMA’s asymmetry in weighting recent points makes it better at tracking evolving plasma states, whereas the SMA provides a cleaner low-pass baseline.

\subsection{Savitzky-Golay Smoothing}

When the objective is to reduce noise without sacrificing fine-scale structure, the \textit{Savitzky-Golay} (SG) filter offers a principled alternative to simple moving averages. Rather than averaging, SG smoothing performs a local polynomial regression within a sliding window of odd length $m$, fitting a polynomial of degree $p < m$ to the samples and replacing the central point with the fitted value. Formally, for a window $\{x_{t-m'}, \dots, x_{t+m'}\}$ with $m = 2m' + 1$, the smoothed value is
\[
y_t = \sum_{i=-m'}^{m'} c_i\, x_{t+i},
\]
where the coefficients $\{c_i\}$ are precomputed from the least-squares fit of the polynomial basis to the window positions.  

The SG filter’s defining strength is \emph{shape preservation}: by approximating the signal locally with a low-order polynomial, it maintains the heights, widths, and positions of peaks, valleys, and inflection points, even while attenuating high-frequency noise. This makes it particularly effective for signals where higher-order features (e.g., slopes, curvatures, or spectral line shapes) carry critical physical meaning. Additionally, the same fitted polynomial can be analytically differentiated to yield smooth, low-noise first or second derivatives, a capability invaluable in spectroscopy, plasma diagnostics, and other derivative-sensitive analyses. \\

\noindent\emph{Key Parameters and Trade-offs}  
\begin{itemize}
    \item \textbf{Window length} ($m$): controls the degree of smoothing; larger $m$ yields stronger noise suppression but risks over-smoothing fine features.
    \item \textbf{Polynomial degree} ($p$): determines the local model’s flexibility; higher $p$ captures curvature but increases the risk of oscillatory artifacts in noisy data.
\end{itemize}
A poor choice of $(m,p)$ can lead to under-smoothing, artificial oscillations, or distortion near signal edges, where incomplete windows require padding or asymmetric fits. SG filters also tend to overshoot across step discontinuities, making them ill-suited for piecewise-constant or heavily quantized data.

The SG filter is well-suited for applications where both noise suppression and local feature fidelity are essential; such as magnetic probe traces, reflectometry profiles, or line-integrated density measurements in tokamaks, especially in the broader context of data harmonization, where retaining physically significant morphology is as important as removing stochastic fluctuations. Its efficiency (coefficients are reusable for a fixed $(m,p)$) makes it practical for high-throughput or streaming workflows, provided its parameters are tuned to the underlying physical and statistical structure of the data.

\subsection{Convolutional Smoothing}

\textit{Convolutional smoothing} generalizes the moving-average concept by allowing an arbitrary weighting kernel to be applied to the data. Given a discrete signal $\{x_t\}$ and a kernel $h[i]$ of width $2m+1$ satisfying $\sum_{i=-m}^{m} h[i] = 1$, the smoothed output is
\[
y_t = \sum_{i=-m}^{m} h[i]\, x_{t+i}.
\]
By choosing different kernel shapes, convolutional smoothing can be tuned to match the statistical and spectral characteristics of the data, making it highly adaptable for diverse signal environments.

A widely used special case is the \textit{Gaussian convolutional filter}, in which the weights follow a discrete Gaussian profile
\[
g(i) = \frac{1}{\sqrt{2\pi}\,\sigma} \exp\!\left(-\frac{i^{2}}{2\sigma^{2}}\right),
\]
where $\sigma$ controls the effective width of the smoothing window. The Gaussian kernel’s smooth, positive, and rapidly decaying form provides excellent time-domain fidelity with no oscillatory ringing, making it particularly well suited for attenuating high-frequency noise while preserving global signal shape. In practice, $\sigma$ serves as a single, interpretable scale parameter: larger values yield heavier smoothing, while smaller values retain more fine detail. \\

\noindent\emph{Gaussian Convolution Advantages and Trade-offs}  
\begin{itemize}
    \item \textbf{Minimal artifacts:} no Gibbs ringing, even in the presence of sharp transitions.
    \item \textbf{Edge handling:} reflective or symmetric padding reduces boundary bias.
    \item \textbf{Computational flexibility:} efficient implementations exist both in the direct time domain and via FFT for very long signals.
\end{itemize}
However, it is worth noting that excessive $\sigma$ blurs sharp edges and diminishes local detail. Single-scale kernels may also be inadequate for signals with noise spanning multiple characteristic scales, requiring either multi-pass filtering or wavelet-based alternatives.

Gaussian convolution is an excellent choice when the goal is to remove stochastic, high-frequency noise while preserving smooth, continuous structure, particularly in scenarios where physical interpretability of the signal’s overall morphology is critical. In the context of data harmonization, its non-parametric nature and minimal artifact profile make it an effective first-pass denoiser for heterogeneous signals prior to more specialized processing. Applications span both time-series domains (e.g., diagnostic traces in tokamaks) and spatial domains (e.g., imaging data from diagnostics or simulations), benefiting from its balance between robustness and computational efficiency.

\subsection{Butterworth Low-Pass Filter}

The \textit{Butterworth low-pass filter} is a classical digital filter design valued for its maximally flat pass-band response, eliminating ripple in the retained frequency range while providing rapid attenuation beyond the specified cutoff. For a given filter order $n$ and normalized cutoff frequency $\omega_c$, its magnitude response is
\[
\left| H(e^{j\omega}) \right| =
\frac{1}{\sqrt{1 + \left[ \frac{\tan(\omega / 2)}{\tan(\omega_c / 2)} \right]^{2n}}},
\]
where $\omega_c = \pi\,\text{normal\_cutoff}$ and
\[
\text{normal\_cutoff} = \frac{\texttt{low\_end\_cutoff}}{0.5\,\texttt{sampling\_freq}},
\quad 0 < \text{normal\_cutoff} < 1.
\]
Here, \texttt{low\_end\_cutoff} is the user-selected cutoff fraction relative to the Nyquist limit $f_{\mathrm{Nyq}} = \tfrac{1}{2}\,\texttt{sampling\_freq}$.

In \emph{non-causal} mode---achieved by forward-and-reverse-filtering the data (e.g., via filtfilt-style implementations \citep{Gustafsson1996})---the Butterworth filter becomes \textbf{zero-phase}, ensuring that all features remain temporally aligned without phase distortion. This property is essential in scientific contexts, such as tokamak diagnostics or experimental waveform analysis, where physical interpretability depends on accurate timing relationships between features. \\

\noindent\emph{Advantages and Limitations}  
\begin{itemize}
    \item \textbf{Flat pass-band:} minimal amplitude distortion for all frequencies below cutoff.
    \item \textbf{Steep attenuation:} rapid suppression of frequencies above cutoff.
    \item \textbf{Zero-phase option:} maintains temporal alignment when run forward and backward.
    \item \textbf{Broad applicability:} widely used in physics, audio engineering, and control systems.
\end{itemize} 

While effective at rejecting out-of-band noise, the Butterworth filter permanently removes spectral content above its cutoff, which may include legitimate high-frequency features if the cutoff is set too low. Edge artifacts may arise in short signals due to filter transients, necessitating windowing or padding prior to application. Moreover, because the cutoff is fixed, the method assumes stationarity in the underlying spectral structure; it is less effective for signals with time-varying noise profiles.

\subsection{Smoothing Method Selection Considerations}

Selecting an appropriate smoothing method is a critical step in the \emph{data harmonization} pipeline, as it directly impacts the preservation of physically meaningful structures and the suppression of noise in both experimental and simulated fusion datasets. The optimal choice depends on the statistical properties of the signal, the spectral separation between noise and features, the intended downstream analysis, and computational constraints. The four smoothing techniques discussed above span a spectrum of trade-offs in temporal responsiveness, shape preservation, and frequency selectivity.

Ultimately, smoothing method selection should be guided by:
\begin{itemize}
    \item \textbf{Signal morphology:} Are sharp transitions, peaks, or derivatives physically meaningful and thus in need of preservation?
    \item \textbf{Noise spectrum:} Is noise spectrally separated from the signal, or does it overlap with relevant frequencies?
    \item \textbf{Adaptivity needs:} Should the filter respond rapidly to evolving states (favoring EMA) or enforce stability (favoring SMA or Gaussian)?
    \item \textbf{Downstream requirements:} Will the smoothed data feed into derivative estimation, feature extraction, machine learning models, or cross-diagnostic fusion?
    \item \textbf{Computational constraints:} Is the filtering performed in real-time, on large archives, or within iterative analysis loops?
\end{itemize}

No single method is universally optimal. In practice, dFL’s design encourages an \emph{iterative and data-driven} selection process, leveraging exploratory visualization and quantitative diagnostics to ensure that smoothing serves the dual goals of noise suppression and preservation of the underlying plasma physics.

\section{Normalization in dFL}\label{AppI}

\subsubsection{Standard Score Normalization}

A widely used normalization strategy in both scientific data analysis and machine learning is the \textit{standard score} or $z$\textit{-score} standardization. Given a raw value $x$, the standard score is defined as  
\[
z \;=\; \frac{x - \mu}{\sigma},
\]
where $\mu$ is the mean and $\sigma$ is the standard deviation of the dataset (or of the specific feature in a multivariate setting). This transformation expresses each measurement in units of standard deviation relative to the feature’s mean, thereby producing a rescaled variable with zero mean and unit variance.

From a \emph{data harmonization} perspective, $z$-score normalization removes arbitrary offsets and scales, enabling direct comparison between heterogeneous variables, even those measured in entirely different units, without distorting their statistical structure. In multimodal fusion datasets, for example, plasma density, temperature, and magnetic field strength can differ by orders of magnitude in their raw numerical values, yet may exhibit comparable relative deviations from their respective means. Standardization equalizes these scales, allowing algorithms to operate on a balanced feature space where no single variable dominates due to its absolute magnitude.

The technique is especially valuable when the distribution of each variable is approximately Gaussian and when relative variability, rather than absolute magnitude, is of primary interest. In such cases, the $z$-score preserves the shape of the distribution, ensuring that the transformed feature retains its underlying statistical relationships. This is critical for distance-based methods (e.g., $k$-nearest neighbors), gradient-based optimizers in neural networks, and linear models (e.g., regression, PCA, SVMs), all of which benefit from features that are centered and scaled. In fact, standardization can reduce optimizer ill-conditioning, improve numerical stability, and shorten training convergence time in high-dimensional problems.

However, the $z$-score is not without limitations. Because $\mu$ and $\sigma$ are global statistics, the transformation is sensitive to outliers, which can skew both the center and the scale, reducing interpretability for the bulk of the data. Moreover, $z$-scores produce unbounded outputs, which may be suboptimal for models or activation functions expecting inputs in a fixed range. For streaming or time-evolving datasets, $\mu$ and $\sigma$ must be recomputed or incrementally updated to maintain consistency, adding computational overhead.

Despite these caveats, $z$-score normalization remains a robust, versatile default for a wide range of fusion data workflows, particularly when aiming to integrate variables from disparate diagnostics into a common, physically meaningful scale. In anomaly detection, the interpretability of $z$-scores---where values above, e.g., $|z|>3$ indicate strong deviations from nominal behavior---provides a direct statistical basis for identifying events of interest. Its balance of theoretical grounding, computational simplicity, and interpretive clarity makes it a core element of the \emph{data harmonization} pipeline.

\subsubsection{Feature Scaling (Min-Max Normalization)}

An alternative to standardization is \textit{feature scaling}, also known as \textit{Min-Max normalization}, which linearly maps the range of each variable to a fixed interval (in dFL between $[0,1]$) via:
\[
x' \;=\; \frac{x - x_{\min}}{x_{\max} - x_{\min}},
\]
where $x_{\min}$ and $x_{\max}$ denote, respectively, the minimum and maximum observed values of the feature. This transformation preserves the relative ordering of the data while compressing it into a uniform range, ensuring that all features have equal numerical bounds regardless of their original units or magnitudes.

In the context of data harmonization, Min-Max scaling is especially valuable when integrating multimodal datasets in which variables differ not only in units but also in numerical span. Without rescaling, features with large numerical ranges dominate distance-based measures such as Euclidean or cosine similarity, biasing clustering, regression, or classification models. By mapping all features to a common range, min–max scaling allows algorithms such as $k$-nearest neighbors, support vector machines, and neural networks to treat each variable with equal weight in geometric computations.

A key strength of Min-Max scaling is its simplicity; preserving the shape of the original distribution (apart from a linear stretch and shift), maintaining interpretability by bounding all values, and producing numerically stable input for optimization in models that are sensitive to input scale, particularly those employing bounded activation functions such as sigmoid or $\tanh$. This boundedness can accelerate convergence in deep networks and improve the conditioning of gradient-based learning.

However, because the transformation depends directly on $x_{\min}$ and $x_{\max}$, it is highly sensitive to outliers, which can cause the majority of the data to be compressed into a narrow subinterval of $[0,1]$, reducing effective resolution. It also lacks centering, meaning the mean of the transformed variable is not guaranteed to be zero, which may be suboptimal for algorithms that perform best with zero-centered features. In streaming or time-evolving datasets, min–max scaling requires ongoing tracking of extrema to avoid range drift, adding operational complexity. For features with intrinsically small ranges, rescaling may also reduce numerical precision.

When applied to fusion data, Min-Max scaling is well suited for harmonizing diagnostics whose absolute magnitudes differ but where bounded, order-preserving normalization is desired. This is common in preparing inputs for neural network surrogates of tokamak plasma states, where stable, bounded gradients improve training robustness. In such workflows, Min-Max normalization offers a low cost, model-friendly preprocessing step, provided that extreme values are handled appropriately, either through prior outlier mitigation or by adopting more robust scaling variants.

\subsubsection{Multiplicative Scalar Scaling}

\textit{Multiplicative scalar scaling} is the simplest normalization strategy, applying a uniform multiplicative factor to all samples in a dataset:
\[
x' \;=\; \lambda\,x,
\]
where $\lambda$ is a user-defined scaling constant. This transformation preserves all relative relationships within the data, ioncluding ratios, ordering, and distributional shape, while shifting the overall magnitude to a desired range or scale. When $\lambda$ is chosen such that the rescaled values fall within a fixed interval (e.g., $[-1,1]$ or $[0,1]$), scalar scaling serves as a lightweight alternative to more complex normalization procedures.

From a physical sciences perspective, multiplicative scaling is particularly relevant when adjusting data to match desired unit conventions or when harmonizing measurements from different sensors that are already on comparable scales but differ in absolute magnitude. In fusion research workflows, for example, $\lambda$ might be chosen to convert magnetic field measurements from millitesla to tesla, or to scale diagnostic traces into dimensionless form for nondimensional analysis. Crucially, because the transformation is linear and isotropic, it does not distort temporal or spatial patterns, making it ideal when preserving the full physical context of the waveform or field is essential.

Its computational triviality---requiring only a single multiplication per value---makes scalar scaling well suited for high-throughput, real-time, or streaming environments where processing latency is a constraint. Furthermore, it can be applied uniformly across large, homogeneous datasets without the need to compute statistics such as means, standard deviations, or extrema, avoiding potential biases introduced by nonstationary data.

However, scalar scaling does not address differences in variance between features, does not center the data, and leaves the influence of outliers unchanged. The choice of $\lambda$ is critical: a poor selection can either fail to address numerical conditioning issues or inadvertently amplify noise. Moreover, in heterogeneous, multimodal datasets—common in fusion experiments—assigning different $\lambda$ values to each feature can become logistically burdensome and risk introducing inconsistencies across experiments or analysis pipelines. 

When applied judiciously, multiplicative scalar scaling is an effective and low-cost normalization method, especially in physics-driven contexts where the preservation of physical units, relative proportions, and geometric structure is paramount. It is best viewed as a targeted tool for rapid magnitude adjustment rather than a comprehensive solution for feature standardization in complex, mixed-scale datasets.

\subsubsection{Box-Cox Transformation (Power Normalization)}

The \textit{Box-Cox transformation} is a parametric, power-based normalization technique designed to stabilize variance and reduce skew in strictly positive data. To handle datasets containing zeros or negative values, a constant \emph{shift parameter} $\delta$ is added before transformation, ensuring all shifted values are strictly positive. The general form is:
\[
x' =
\begin{cases}
\dfrac{(x + \delta)^{\lambda} - 1}{\lambda}, & \lambda \neq 0, \\[6pt]
\ln(x + \delta), & \lambda = 0,
\end{cases}
\]
where $x$ is the original data value, $\delta > -\min(x)$ is chosen so that $x + \delta > 0$ for all samples, and $\lambda$ is a continuous shape parameter controlling the degree and type of transformation.  Special cases include $\lambda = 1$ (identity transform), $\lambda = 0.5$ (square-root), and $\lambda = 0$ (logarithmic). In practice, $\lambda$ is often estimated via maximum likelihood to make the transformed distribution as close to Gaussian as possible, improving the performance of models that assume normality.

By applying a monotonic, non-linear mapping to $x + \delta$, the Box-Cox method can simultaneously correct for heteroscedasticity (non-constant variance) and reduce skew, producing more symmetric features. This is beneficial for algorithms sensitive to distributional shape, such as linear regression, ARIMA forecasting, and Gaussian-based classifiers. The transformation is invertible, allowing the data to be mapped back to the original physical units—an important property when interpretability or physical meaning must be preserved.

The shift parameter $\delta$ extends the applicability of Box-Cox to datasets that would otherwise violate its strict positivity requirement. However, its choice can influence the resulting transformation, particularly in the presence of extreme values. Additionally, $\lambda$ estimation adds computational overhead, and both $\lambda$ and $\delta$ can be biased by outliers. Because the transformation is non-linear, it may alter local ordering of values, affecting distance-based metrics.

In the context of fusion energy datasets, the Box-Cox transformation with shift is especially valuable for diagnostic signals that exhibit multiplicative noise or heavy-tailed distributions—for example, line-integrated density traces, fluctuation amplitudes, or power spectra with large dynamic ranges. By stabilizing variance and normalizing distribution shape, this method enhances the statistical robustness of downstream analyses, including principal component analysis, surrogate modeling, and anomaly detection.

\subsubsection{Robust Scaling (Median-IQR)}

\textit{Robust scaling} is a normalization technique specifically designed to mitigate the influence of extreme values and non-Gaussian distributions. It centers each feature by its median and scales by its inter-quartile range (IQR), defined as $Q_3 - Q_1$:
\[
x' = \frac{x - \mathrm{median}(x)}{Q_3 - Q_1}.
\]
Here, $Q_1$ and $Q_3$ denote the first and third quartiles, respectively, so that the transformation expresses each value in “IQR units” relative to the feature’s central tendency. Unlike mean-variance scaling, which can be heavily distorted by a small number of extreme observations, robust scaling relies entirely on rank-based statistics that are insensitive to outliers.

This property makes the method highly effective for datasets with heavy tailed distributions, skewness, or significant intermittent spikes; conditions  encountered in some experimental fusion diagnostics, where transient electromagnetic interference, sensor drift, or environmental noise can produce erratic high-amplitude readings. By removing the leverage of such extremes, robust scaling can produce stable, comparable feature magnitudes without requiring outlier removal or preprocessing heuristics.

Although robust scaling offers broad distributional adaptability, it does not bound the transformed data, and its units (IQRs) are less intuitively interpretable than standard deviations. In perfectly Gaussian data, $z$-score normalization achieves slightly lower variance and may yield marginally better statistical efficiency. Moreover, computing exact $Q_1$ and $Q_3$ quantiles on massive datasets can be computationally expensive, though streaming quantile approximation methods make it feasible for real-time or high-throughput workflows.

In \emph{data harmonization} pipelines for fusion data, robust scaling is best viewed as a cautious default for mixed-quality or partially vetted datasets---tempering occasional glitches, dropouts, and saturation without letting rare extremes dominate. For most well-calibrated diagnostics (magnetics, interferometry, Thomson, bolometry), which exhibit near-Gaussian, light-tailed noise after standard preprocessing, $z$-score (or Min-Max for bounded ranges) often yields more interpretable and statistically efficient features. Robust scaling remains useful for burst-prone channels in some contexts (e.g., filterscopes) or during early integration of heterogeneous sources; once signals are model-processed and QC’d, conventional scaling may be preferred. \\

\noindent\emph{Normalization Method Selection Considerations}: normalization aligns heterogeneous variables for meaningful comparison, fusion, and modeling. The optimal method depends on distributional shape, outlier presence, and downstream model requirements.   All natively supported normalizations are invertible mappings, but new normalizations may be readily integrated into the dFL GUI, as detailed in the \href{http://dfl.sophelio.io/documentation}{dFL documentation}.  Additionally, normalization may be performed at ingestion using the Fetch Data approach discussed in \ref{AppE}.
 
When data quality is high and distributions are near-normal, $z$-score normalization remains a strong default. For bounded output on clean data, Min-Max scaling integrates seamlessly with many machine learning models. Scalar multiplication is optimal for rapid, unit-preserving adjustments across already harmonized features. When variable skewness must be reduced---whether strictly positive or extended to non-positive ranges via a constant shift---the Box-Cox transformation offers a principled, tunable path toward Gaussianization and variance stabilization. In the presence of outliers or heavy-tailed distributions, robust scaling provides stability without distorting relationships within the bulk of the data. In practice, hybrid workflows often combine methods, e.g., robust scaling followed by Min-Max or Box-Cox followed by $z$-score, tailored to the statistical and physical structure of the dataset, and the requirements of the modeling target. Examples of how to easily implement serial normalizations into dFL are provided in the \href{http://dfl.sophelio.io/documentation}{dFL documentation}.

\section{A Note on Units and Axis Labeling}\label{units}

A subtle limitation of the present dFL implementation is the absence of explicit $x$ and $y$ axis unit labeling in its visualization components. While time is generally understood to be the abscissa (the independent variable, usually $t$, though not necessarily), the ordinate (dependent variable, often $y$ or $f(t)$) remains unlabeled, reflecting the fact that many harmonization operations---particularly normalization and standardization---transform raw measurements into dimensionless or non-intuitive scales. This practice is commonplace in machine learning and artificial intelligence pipelines, where preserving statistical comparability across heterogeneous feature spaces often supersedes the preservation of absolute units. By contrast, in traditional physics workflows, units are non-negotiable: they are essential for dimensional consistency, physical interpretation, and reproducibility, and errors in units can propagate catastrophically. 

A concrete illustration arises when considering electron temperature $T_e$ measured in electronvolts (eV) alongside electron density $n_e$ measured in $\mathrm{m}^{-3}$. Typical core tokamak values might span $T_e \sim 1$--$10\,\mathrm{keV}$ and $n_e \sim 10^{19}$--$10^{20}\,\mathrm{m}^{-3}$. If these raw values are ingested directly into an ML pipeline without normalization, the density feature dominates numerically by more than fifteen orders of magnitude relative to temperature. In effect, the units themselves act as an implicit statistical weight, biasing learning algorithms toward density-driven structure while suppressing temperature variations that may be equally or more relevant for physical inference. Such imbalances reduce the utility of otherwise powerful algorithms, as the optimization landscape becomes warped by unit-induced scaling disparities rather than by underlying correlations or dynamics. Normalization or variance-based scaling thus serves not only as a numerical convenience, but as an essential step in mitigating hidden biases introduced by standard (or conventional) units.  In addition, dFL supports multimodal visualizations (in part to further develop intuition and understanding of how data-driven algorithms, for example, see and operate on the underlying data spaces) where signals with disparate base units (e.g., temperatures, magnetic fields, and densities) are shown together on a common axis. In such cases, plotting in absolute units is rarely practical, since signals must be rescaled or normalized to avoid one variable’s dynamic range obscuring the others, making normalization a prerequisite for multimodal visualization. 

\textit{Transformation-aware units:} not all preprocessing operations treat units equally. Resampling or trimming typically preserves units, while $z$-scoring or PCA whitening explicitly destroys them. Transformations such as logarithms or Box-Cox mappings produce re-scaled units that remain interpretable, but only if metadata tracking the mappings. Without automated unit propagation, these subtleties can be lost.

\textit{Operational and human factors:} in practice, units often act as guardrails. Operators and diagnosticians often catch errors (e.g., kiloamps vs.\ megaamps, keV vs.\ eV) by their units alone. When stripped away, silent misinterpretations become more likely, especially in archival datasets where future users may not recall the original dimensionality of a signal. Unit metadata thus provides a natural audit trail.

\textit{Implications for PCS and control:} in real-time plasma control, units are frequently inseparable from actuator guardrails. For example, actuator constraints such as ``do not exceed 5~MW of injected ECH power'' cannot be expressed in normalized space. Feeding unitless features directly into a plasma control system (PCS) therefore introduces risk unless a unit-aware mapping layer exists. In principle, dFL can serve as such an intermediate layer; where signals may be normalized for ML pipelines while retaining the ability to ``round-trip'' back into physical units when interfaced with PCS, MPC, or physics validation.

\textit{Relation to data fusion mathematics:} Bayesian or statistical data fusion explicitly requires noise covariance matrices $R_i$ expressed in physical units, since weights are determined by inverse variances. If signals are unitless or inconsistently scaled, these weights lose physical meaning and produce biased or non-optimal estimators. Correct unit handling is therefore not simply cosmetic, but mathematically essential for heteroscedastic fusion across modalities.

\textit{Provenance and metadata standards:} the International Atomic Energy Agency's IMAS framework and tools such as OMAS already mandate explicit unit metadata, but these are rarely surfaced into harmonization and visualization layers. For dFL, the lack of unit propagation is thus not a theoretical impossibility but an implementation gap. Proper unit handling requires metadata-rich datasets where units are explicitly stored, propagated, and updated under transformation, including transparent annotations of when units are lost or remapped.

\textit{Potential solutions:} future dFL extensions plan to integrate unit-handling libraries (e.g., schema-aware extensions of \texttt{pint} \citep{Grecco2013Pint}) to track dimensional consistency automatically. This would enable hybrid visualization modes: (\emph{i}) \emph{absolute mode}, where raw units are preserved; (\emph{ii}) \emph{relative mode}, where all signals are normalized; and (\emph{iii}) \emph{hybrid mode}, where multiple axes or dual scales expose both. Planned unit-aware metadata will also support heteroscedastic data fusion where per-modality variances $R_i$ are interpreted in physical units, while still allowing unit-stripped versions for ML workflows requiring dimensionless feature vectors and multimodal visualization.

For now, if units must be explicitly preserved, the easiest workaround is to simply rename input features to include their base units (e.g., \texttt{density\_in\_m\string^-3}); or, if needing to preserve unit structures through transformations, to use a custom implementation of \texttt{pint} through the \texttt{Fetch Data} field of the dFL \texttt{data coordinator}. Howveer, while future dFL releases plan to include automated unit propagation and guardrails against common errors, ultimate responsibility for correct unit management will remain with the practitioner. Even the most sophisticated metadata systems cannot substitute for principled scientific use. Users must remain vigilant in tracking and validating transformations, ensuring dimensional consistency, and interpreting outputs in physically meaningful terms. In this sense, dFL can provide scaffolding and reminders, but it will never be able to fully insulate users from mistakes introduced by careless or unprincipled handling of units.

\bibliographystyle{elsarticle-num}
\bibliography{labeler}

\end{document}